\documentclass{nature3}

\usepackage{graphicx}%
\usepackage{multirow}%

\usepackage{lineno}

\usepackage{titlesec}
\titleformat{\section}{\normalfont\large\bfseries}{\thesection.}{3pt}{\space}[]
\titlespacing*{\section}{0em}{1ex}{1em}[0em]
\titleformat{\subsection}{\bfseries}{\thesubsection.}{3pt}{\space}[]
\titlespacing*{\subsection}{0em}{1ex}{1em}[0em]

\usepackage{xcolor}
\usepackage[colorlinks,breaklinks=true,plainpages=false,citecolor=blue,linkcolor=blue,urlcolor=blue,bookmarksopen=true,bookmarksnumbered=false,bookmarksdepth=5]{hyperref}
\usepackage[symbol]{footmisc}
\renewcommand{\thefootnote}{\fnsymbol{footnote}}
\usepackage{xurl}

\usepackage{afterpage}

\usepackage{rotating}

\usepackage{amsmath,amssymb,amsfonts}%
\usepackage{amsthm}%
\usepackage{mathrsfs}%
\usepackage[title]{appendix}%
\usepackage{xcolor}%
\usepackage{textcomp}%
\usepackage{manyfoot}%
\usepackage{booktabs}%
\usepackage{algorithm}%
\usepackage{algorithmicx}%
\usepackage{algpseudocode}%
\usepackage{listings}%
\usepackage{mhchem}

\def\la{\mathrel{\mathchoice {\vcenter{\offinterlineskip\halign{\hfil$\displaystyle##$\hfil\cr<\cr\sim\cr}}}
{\vcenter{\offinterlineskip\halign{\hfil$\textstyle##$\hfil\cr<\cr\sim\cr}}}
{\vcenter{\offinterlineskip\halign{\hfil$\scriptstyle##$\hfil\cr
<\cr\sim\cr}}}
{\vcenter{\offinterlineskip\halign{\hfil$\scriptscriptstyle##$\hfil\cr><cr\sim\cr}}}}}

\def\ga{\mathrel{\mathchoice {\vcenter{\offinterlineskip\halign{\hfil$\displaystyle##$\hfil\cr>\cr\sim\cr}}}
{\vcenter{\offinterlineskip\halign{\hfil$\textstyle##$\hfil\cr>\cr\sim\cr}}}
{\vcenter{\offinterlineskip\halign{\hfil$\scriptstyle##$\hfil\cr
<\cr\sim\cr}}}
{\vcenter{\offinterlineskip\halign{\hfil$\scriptscriptstyle##$\hfil\cr><cr\sim\cr}}}}}

\usepackage{multibib}
\newcites{Supp}{Supplementary References}

\newcount\longrefs
\def\aap{\ifnum\longrefs=1 {Astron.\ Astrophys.}\else 
                           {A\hbox{\rm \&}A}\fi}
\def\aapl{\ifnum\longrefs=1 {Astron.\ Astrophys.\ Lett.}\else 
                           {A\hbox{\rm \&}A}\fi}
\def\aapr{\ifnum\longrefs=1 {Astron.\ Astrophys.\ Rev.}\else 
                            {A\hbox{\rm \&}AR}\fi}
\def\aaps{\ifnum\longrefs=1 {Astron.\ Astrophys.\ Suppl.}\else 
                            {A\hbox{\rm \&}AS}\fi}
\def\aj{\ifnum\longrefs=1 {Astron.\ J.}\else 
                          {AJ}\fi} 
\def\ao{\ifnum\longrefs=1 {Applied Optics}\else 
                           {Appl.\ Opt.}\fi} 
\def\aspcs{\ifnum\longrefs=1 {Astron.\ Soc.\ Pacific Conf. Series}\else 
                           {ASP Conf.\ Ser.}\fi} 
\def\apj{\ifnum\longrefs=1 {Astrophys.\ J.}\else 
                           {ApJ}\fi} 
\def\apjl{\ifnum\longrefs=1 {Astrophys.\ J.\ Lett.}\else 
                            {ApJ}\fi} 
\def\aplett{\ifnum\longrefs=1 {Astrophys.\ J.\ Lett.}\else 
                            {ApJ}\fi} 
\def\apjs{\ifnum\longrefs=1 {Astrophys.\ J.\ Suppl.}\else 
                            {ApJS}\fi}
\def\apss{\ifnum\longrefs=1 {Astrophys.\ and Space Science}\else 
                            {Ap\hbox{\rm \&}SS}\fi}
\def\araa{\ifnum\longrefs=1 {Ann.\ Rev.\ Astron.\ Astrophys.}\else 
                            {ARA\hbox{\rm \&}A}\fi}
\def\azh{\ifnum\longrefs=1 {Astronomicheskii Zhurnal}\else 
                            {Astron.\ Zhur.}\fi}
\def\baas{\ifnum\longrefs=1 {Bull.\ Am.\ Astron.\ Soc.}\else 
                            {BAAS}\fi}
\def\bain{\ifnum\longrefs=1 {Bull.\ Astronom.\ Institutes Netherlands}\else
                            {Bull.\ Astr.\ Inst.\ Neth.}\fi}
\def\gca{\ifnum\longrefs=1 {Geochim.\ Cosmochim.\ Acta}\else 
                           {Geochim.\ Cosmochim.\ Acta}\fi}
\def\grl{\ifnum\longrefs=1 {Geophys.\ Res.\ Lett.}\else 
                           {Geoph.\ Res.\ Lett.}\fi}
\def\iaucirc{\ifnum\longrefs=1 {IAU Circulars}\else 
                          {IAU Circ.}\fi}
\def\ip{\ifnum\longrefs=1 {in press}\else 
                          {in press}\fi}
\def\jchemp{\ifnum\longrefs=1 {J.\ Chem.\ Phys.}\else 
                           {J.\ Chem.\ Phys.}\fi}  
\def\jcp{\ifnum\longrefs=1 {J.\ Chem.\ Phys.}\else 
                           {J.\ Chem.\ Phys.}\fi}  
\def\jgr{\ifnum\longrefs=1 {J.\ Geophys.\ Res.}\else 
                           {J.\ Geophys.\ Res.}\fi}  
\def\jmolspec{\ifnum\longrefs=1 {J.\ Mol.\ Spectrosc.}\else 
                           {J.\ Mol.\ Spectrosc.}\fi}  
\def\jqsrt{\ifnum\longrefs=1 {J.\ Quant.\ Spectrosc.\ Radiat.\ Transfer}\else 
                           {J.\ Quant.\ Spectrosc.\ Radiat.\ Transfer}\fi}  
\def\jrasc{\ifnum\longrefs=1 {J.\ Royal Astron.\ Soc.\ Canada}\else 
                           {JRAS Can.}\fi}  
\def\mnras{\ifnum\longrefs=1 {Mon.\ Not.\ Roy.\ Astron.\ Soc.}\else 
                             {MNRAS}\fi} 
\def\nat{\ifnum\longrefs=1 {Nature}\else 
                           {Nat}\fi}
\def\pasj{\ifnum\longrefs=1 {Pub.\ Astron.\ Soc.\ Japan}\else 
                            {PASJ}\fi} 
\def\pasp{\ifnum\longrefs=1 {Pub.\ Astron.\ Soc.\ Pacific}\else 
                            {PASP}\fi} 
\def\physscr{\ifnum\longrefs=1 {Physica Scripta}\else 
                            {Phys.\ Scrip.}\fi} 
\def\planss{\ifnum\longrefs=1 {Planetary \& Space Science}\else 
                            {Plan. \& Space Sci.}\fi} 
\def\procspie{\ifnum\longrefs=1 {Proc.\ SPIE}\else 
                            {Proc.\ SPIE}\fi} 
\def\qjras{\ifnum\longrefs=1 {Quarterly J.\ Royal Astron.\ Soc.}\else 
                            {QJRAS}\fi} 
\def\sa{\ifnum\longrefs=1 {Soviet Astron..}\else 
                               {Sov.\ Astron.}\fi}
\def\skytel{\ifnum\longrefs=1 {Sky \& Telescope}\else 
                            {Sky \& Tel.}\fi} 
\def\solphys{\ifnum\longrefs=1 {Solar Phys.}\else 
                               {Solar Phys.}\fi}
\def\ssr{\ifnum\longrefs=1 {Space Science Rev.}\else 
                               {Space\ Sci.\ Rev.}\fi}

\def\dutch{\def\refname{Referenties}\def\abstractname{Samenvatting}%
  \def\bibname{Bibliografie}\def\chaptername{Hoofdstuk}%
  \def\appendixname{Bijlage}\def\contentsname{Inhoudsopgave}%
  \def\listfigurename{Lijst van figuren}\def\listtablename{Lijst van tabellen}%
  \def\indexname{Index}\def\figurename{Figuur}\def\tablename{Tabel}%
  \def\partname{Deel}\def\enclname{Bijlage(n)}\def\ccname{Ter attentie van}%
  \def\headtoname{Aan}\def\headpagename{Pagina}%
  \def\today{\number\day\space\ifcase\month\or januari\or februari\or maart\or%
     april\or mei\or juni\or juli\or augustus\or september\or oktober\or%
     november\or december\fi \space\number\year}%
  \typeout{
              >>>>> use hlatex209 for Dutch hyphenation <<<<< 
         }}
\newcounter{onefig} \newcounter{fignumber}
\newcount\nocaptions \newcount\nofigures \newcount\figwidth
\newcount\viewgraphs
  \def\paper{}  \def\figlabel{} 
\long\def\nextfig#1{\setcounter{figure}{\value{fignumber}}
  \addtocounter{fignumber}{1}
  \ifnum \viewgraphs=1 \newpage \pagestyle{empty} \fi 
  \ifnum\value{onefig}=0 #1 \fi                 
  \ifnum\value{onefig}=\value{fignumber} #1 \fi}
\def\figwidths#1#2{\ifnum \nocaptions=1 #2mm \else #1mm \fi}  
\def\paper#1{}  
\long\def\plotfig#1#2{\ifnum \nofigures=1 \else #2 \fi}
\long\def\captiontext#1{\ifnum \nofigures=1 \raggedright \fi 
   \ifnum \nocaptions=1 \paper
     \ifnum \viewgraphs=0 
       \newline  \mbox{}\hrulefill\mbox{} \newline 
       \newline label:~\{\figlabel\} 
     \fi 
     \else \ifnum \nofigures=0 \fi 
   #1 \fi}

\newcount\panelwidth \newcount\panelheight 
\newcount\bxmin \newcount\bymin \newcount\bxmax \newcount\bymax
\newcount\tbxmin \newcount\tbymin
\newcount\tpanelwidth \newcount\tpanelheight \newcount\tpdif
\def\panelsize #1,#2;{\panelwidth=#1 \panelheight=#2}  
\def\setbb #1,#2;#3,#4;#5,#6;{
  \tbxmin=#1 \tbymin=#2    
  \bxmin=#3 \bymin=#4      
  \bxmax=#5 \bymax=#6}     
\def\barepanel #1{%
  \ifnum\panelheight=0 
    \tpdif=\bymax \advance\tpdif by -\bymin
    \multiply \tpdif by \panelwidth
    \tpanelheight=\tpdif
    \tpdif=\bxmax \advance\tpdif by -\bxmin
    \divide \tpanelheight by \tpdif
  \else \tpanelheight=\panelheight \fi
  \epsfig{file=#1,%
     bbllx=\bxmin bp,bblly=\bymin bp,bburx=\bxmax bp,bbury=\bymax bp,clip=,%
     width=\panelwidth mm,height=\tpanelheight mm}}
\def\labelypanel #1{
  \ifnum\panelheight=0 
    \tpdif=\bymax \advance\tpdif by -\bymin
    \multiply \tpdif by \panelwidth
    \tpanelheight=\tpdif
    \tpdif=\bxmax \advance\tpdif by -\bxmin
    \divide \tpanelheight by \tpdif
  \else \tpanelheight=\panelheight \fi
  \tpdif=\bxmax \advance\tpdif by -\tbxmin
  \tpanelwidth=\panelwidth \multiply \tpanelwidth by \tpdif
  \tpdif=\bxmax \advance\tpdif by -\bxmin
  \divide \tpanelwidth by \tpdif
  \epsfig{file=#1,%
    bbllx=\tbxmin bp,bblly=\bymin bp,bburx=\bxmax bp,bbury=\bymax bp,%
    clip=,width=\tpanelwidth mm,height=\tpanelheight mm}}
\def\labelxpanel #1{%
  \ifnum\panelheight=0 
    \tpdif=\bymax \advance\tpdif by -\bymin
    \multiply \tpdif by \panelwidth
    \tpanelheight=\tpdif
    \tpdif=\bxmax \advance\tpdif by -\bxmin
    \divide \tpanelheight by \tpdif
  \else \tpanelheight=\panelheight \fi
  \tpdif=\bymax \advance\tpdif by -\tbymin
  \multiply \tpanelheight by \tpdif
  \tpdif=\bymax \advance\tpdif by -\bymin
  \divide \tpanelheight by \tpdif
  \epsfig{file=#1,%
    bbllx=\bxmin bp,bblly=\tbymin bp,bburx=\bxmax bp,bbury=\bymax bp,%
    clip=,width=\panelwidth mm,height=\tpanelheight mm}}
\def\labelxypanel #1{%
  \ifnum\panelheight=0 
    \tpdif=\bymax \advance\tpdif by -\bymin
    \multiply \tpdif by \panelwidth
    \tpanelheight=\tpdif
    \tpdif=\bxmax \advance\tpdif by -\bxmin
    \divide \tpanelheight by \tpdif
  \else \tpanelheight=\panelheight \fi
  \tpdif=\bxmax \advance\tpdif by -\tbxmin
  \tpanelwidth=\panelwidth \multiply \tpanelwidth by \tpdif
  \tpdif=\bxmax \advance\tpdif by -\bxmin
  \divide \tpanelwidth by \tpdif 
  \tpdif=\bymax \advance\tpdif by -\tbymin 
  \multiply \tpanelheight by \tpdif
  \tpdif=\bymax \advance\tpdif by -\bymin
  \divide \tpanelheight by \tpdif
  \epsfig{file=#1,%
    bbllx=\tbxmin bp,bblly=\tbymin bp,bburx=\bxmax bp,bbury=\bymax bp,%
    clip=,width=\tpanelwidth mm,height=\tpanelheight mm}}

\def\CC{\par \vspace*{-2ex} \footnotesize \baselineskip=8pt \begin{verbatim}}

\long\def\startignore #1\stopignore{}   

\def\setlistparams{         
  \topsep=0.7ex                 
  \itemsep=0.7ex                
  \leftmargini=3ex}             
\setlistparams                  

\newcounter{alistindex}       

\def\spacing #1{\small \renewcommand{\baselinestretch}{#1}
                \normalsize \setlistparams}             

\newcounter{romenumnr}

\newlength{\minipagewidth}

\newsavebox{\boxcontent}
\newcommand{\ovalhead}[1]{
  \unitlength=1cm
  \sbox{\boxcontent}{\mbox{~~{#1}~~}}
  \begin{center}
    \ifdim\wd\boxcontent>6ex 
    \ifdim\wd\boxcontent<8cm 
    \begin{picture}(8,3) \thicklines     
      \put(4.0,0.8){\oval(8,1.6)} 
      \put(0.0,0.7){\parbox{8cm}{
         \begin{center} \usebox{\boxcontent} \end{center}}}
    \end{picture}
    \else \ifdim\wd\boxcontent<12cm 
    \begin{picture}(12,3) \thicklines     
        \put(6.0,0.8){\oval(12,1.6)} 
        \put(0.0,0.7){\parbox{12cm}{
           \begin{center} \usebox{\boxcontent} \end{center}}}
    \end{picture}
    \else
    \begin{picture}(16,3) \thicklines     
        \put(8.0,0.8){\oval(16,1.6)} 
        \put(0.0,0.7){\parbox{16cm}{
           \begin{center} \usebox{\boxcontent} \end{center}}}
    \end{picture}
    \fi \fi \fi
  \end{center}} 

\newcounter{headnr}            
\newcounter{subheadnr}[headnr]
\newcounter{subsubheadnr}[subheadnr]
\def\head #1\par{
  \stepcounter{headnr}                          
  \vspace{2ex} \noindent                        
  {\bf \theheadnr~~~~#1}\\[1ex] \noindent}      
\def\subhead #1\par{  
  \stepcounter{subheadnr}
  \vspace{1.3ex} \noindent
  {\bf \theheadnr.\arabic{subheadnr}~~~#1}\\[0.3ex] \noindent}
\def\subsubhead #1\par{
  \stepcounter{subsubheadnr}
  \vspace{1.0ex} \noindent
  {\bf \theheadnr.\arabic{subheadnr}.\arabic{subsubheadnr}~~~#1}\\ \noindent}

\font\dropfont= cmr12 scaled \magstep5
\def\dropcap#1#2{{\noindent
    \setbox0\hbox{\dropfont #1}\setbox1\hbox{#2}\setbox2\hbox{(}%
    \count0=\ht0\advance\count0 by\dp0\count1\baselineskip
    \advance\count0 by-\ht1\advance\count0by\ht2
    \dimen1=.5ex\advance\count0by\dimen1\divide\count0 by\count1
    \advance\count0 by1\dimen0\wd0
    \advance\dimen0 by.25em\dimen1=\ht0\advance\dimen1 by-\ht1
    \global\hangindent\dimen0\global\hangafter-\count0
    \hskip-\dimen0\setbox0\hbox to\dimen0{\raise-\dimen1\box0\hss}%
    \dp0=0in\ht0=0in\box0}#2}

\def\level #1 #2#3#4{$#1 \: ^{#2} \mbox{#3} ^{#4}$}   

\def\mathstacksym#1#2#3#4#5{\def#1{\mathrel{\hbox to 0pt{\lower 
    #5\hbox{#3}\hss} \raise #4\hbox{#2}}}}

\mathstacksym\lta{$<$}{$\sim$}{1.5pt}{3.5pt} 
\mathstacksym\gta{$>$}{$\sim$}{1.5pt}{3.5pt} 
\mathstacksym\lrarrow{$\leftarrow$}{$\rightarrow$}{2pt}{1pt} 
\mathstacksym\lessgreat{$>$}{$<$}{3pt}{3pt} 

\title{SO$_2$, silicate clouds, but no CH$_4$ detected in a warm Neptune}

\usepackage{comment}

\begin{document}

\renewcommand{\thefootnote}{\alph{footnote}}

\maketitle

\noindent\author{
Achr\`ene Dyrek$^{1,\star,\dagger}$,
Michiel Min$^{2,\dagger}$,
Leen Decin$^{3,\dagger}$,
Jeroen Bouwman$^{4}$,
Nicolas Crouzet$^{5}$,
Paul Molli\`ere$^{4}$,
Pierre-Olivier Lagage$^{6}$,
Thomas Konings$^{3}$,
Pascal Tremblin$^{7}$, 
Manuel G\"udel$^{8,4,9}$,
John Pye$^{10}$,
Rens Waters$^{11,12,2}$,
Thomas Henning$^{4}$, 
Bart Vandenbussche$^{3}$,
Francisco Ardevol Martinez$^{13,2,14,15}$,
Ioannis Argyriou$^{3}$,
Elsa Ducrot$^{6}$,
Linus Heinke$^{3,14,15}$,
Gwenael Van Looveren$^{8}$,
Olivier Absil$^{16}$,
David Barrado$^{17}$,
Pierre Baudoz$^{18}$,
Anthony Boccaletti$^{18}$,
Christophe Cossou$^{19}$,
Alain Coulais$^{6,20}$,
Billy Edwards$^{2}$,
Ren\'e Gastaud$^{19}$,
Alistair Glasse$^{21}$,
Adrian Glauser$^{9}$,
Thomas P.\ Greene$^{22}$,
Sarah Kendrew$^{23}$,
Oliver Krause$^{4}$,
Fred Lahuis$^{2}$,
Michael Mueller$^{13}$,
Goran Olofsson$^{24}$,
Polychronis Patapis$^{9}$,
Daniel Rouan$^{17}$,
Pierre Royer$^{3}$,
Silvia Scheithauer$^{4}$,
Ingo Wald\-mann$^{25}$,
Niall White\-ford$^{26}$,
Luis Colina$^{17}$,
Ewine F.\ van Dishoeck$^{5}$,
G\"oran Ostlin$^{27}$,
Tom P.\ Ray$^{28}$,
Gillian Wright$^{29}$
}

\noindent$^\dagger$Equal contribution is indicated by shared first-authorship. \\

\begin{affiliations}
	\item Universit\'e Paris Cit\'e, Universit\'e Paris-Saclay, CEA, CNRS, AIM, F-91191 Gif-sur-Yvette, France
 	\item
	SRON Netherlands Institute for Space Research, Niels Bohrweg 4, 2333 CA Leiden, the Netherlands

    \item 
    Institute of Astronomy, KU Leuven, Celestijnenlaan 200D, 3001 Leuven, Belgium
    \item 
    Max-Planck-Institut f\"ur Astronomie (MPIA), K\"onigstuhl 17, 69117 Heidelberg, Germany
    \item
    Leiden Observatory, Leiden University, P.O. Box 9513, 2300 RA Leiden, the Netherlands
    \item 
    Universit\'e Paris-Saclay, Universit\'e Paris Cit\'e, CEA, CNRS, AIM, F-91191 Gif-sur-Yvette, France
    \item 
    Universit\'e Paris-Saclay, UVSQ, CNRS, CEA, Maison de la Simulation, 91191, Gif-sur-Yvette, France.    
 \item 
 Department of Astrophysics, University of Vienna, T\"urkenschanzstrasse 17, 1180 Vienna, Austria
 \item
 ETH Z\"urich, Institute for Particle Physics and Astrophysics, Wolfgang-Pauli-Strasse 27, 8093 Z\"urich, Switzerland
 \item
 School of Physics \& Astronomy, Space Research Centre, Space Park Leicester, University of Leicester, 92 Corporation Road, Leicester, LE4 5SP, UK
 \item 
 Department of Astrophysics/IMAPP, Radboud University, PO Box 9010, 6500 GL Nijmegen, the Netherlands
 \item
 HFML - FELIX. Radboud University PO box 9010, 6500 GL Nijmegen, the Netherlands
 \item
 Kapteyn Institute of Astronomy, University of Groningen, Landleven 12, 9747 AD Groningen, the Netherlands 
\item
Centre for Exoplanet Science, University of Edinburgh, Edinburgh, EH9 3FD, UK
\item
School of GeoSciences, University of Edinburgh, Edinburgh, EH9 3FF, UK
\item 
STAR Institute, Universit\'e de Li\`ege, All\'ee du Six Ao\^ut 19c, 4000 Li\`ege, Belgium
\item 
Centro de Astrobiología (CAB), CSIC-INTA, ESAC Campus, Camino Bajo del Castillo s/n, 28692 Villanueva de la
Ca\~nada, Madrid, Spain
\item 
LESIA, Observatoire de Paris, CNRS, Universit\'e Paris Diderot, Universit\'e Pierre et Marie Curie, 5 place Jules Janssen, 92190 Meudon, France
\item 
Universit\'e Paris-Saclay, CEA, D\'epartement d'Electronique des D\'etecteurs et d'Informatique pour la Physique, 91191, Gif-sur-Yvette, France.
\item 
LERMA, Observatoire de Paris, Universit\'e PSL, Sorbonne Universit\'e, CNRS, Paris, France
\item 
UK Astronomy Technology Centre, Royal Observatory, Blackford Hill, Edinburgh EH9 3HJ, UK
\item 
Space Science and Astrobiology Division, NASA’s Ames Research Center, M.S.\ 245-6, Moffett Field, 94035, California, USA
\item 
European Space Agency, Space Telescope Science Institute, Baltimore, Maryland, USA
\item 
Department of Astronomy, Stockholm University, AlbaNova University Center, 10691 Stockholm, Sweden
\item 
Department of Physics and Astronomy, University College London, Gower Street, WC1E 6BT, UK
\item 
Department of Astrophysics, American Museum of Natural History, New York, NY 10024, USA
\item 
Department of Astronomy, Oskar Klein Centre, Stockholm University, 106 91 Stockholm, Sweden
\item 
School of Cosmic Physics, Dublin Institute for Advanced Studies, 31 Fitzwilliam Place, Dublin, D02 XF86, Ireland
\item 
UK Astronomy Technology Centre, Royal Observatory Edinburgh, Blackford Hill, Edinburgh EH9 3HJ, UK
\end{affiliations}

\maketitle

\setlength{\parskip}{10pt}

\bigskip

\begin{abstract} 
WASP-107b is a warm ($\sim$740\,K) transiting planet with a Neptune-like mass of $\sim$30.5\,$M_{\oplus}$ and Jupiter-like radius of $\sim$0.94\,$R_{\rm J}$\,\cite{anderson_2017, piaulet_2021} whose extended atmosphere is eroding\,\cite{spake_2018}. 
Previous observations showed evidence for water vapour and a thick high-altitude condensate layer in WASP-107b's atmosphere\,\cite{kreidberg_2018, 2022arXiv221100649E}. 
Recently, photochemically produced sulphur dioxide (SO$_2$) was detected in the atmosphere of a  hot ($\sim$1,200\,K) Saturn-mass planet from transmission spectroscopy near 4.05\,$\mu$m\,\cite{rustamkulov_early_2023, alderson_early_2023}, but for temperatures below $\sim$1,000\,K sulphur is predicted to preferably form sulphur allotropes instead of SO$_2$\,\cite{Zahnle2016ApJ...824..137Z, tsai_2021,tsai_photochemically_2023}. 
Here we report the 9$\sigma$-detection of two fundamental vibration bands of SO$_2$, at 7.35\,$\mu$m and 8.69\,$\mu$m, in the transmission spectrum of WASP-107b using the Mid-Infrared Instrument (MIRI) of the JWST. This discovery establishes WASP-107b as the second irradiated exoplanet with confirmed photochemistry, extending the temperature range of exoplanets exhibiting detected photochemistry from $\sim$1,200\,K down to $\sim$740\,K. Additionally, our spectral analysis reveals the presence of silicate clouds, which are strongly favoured ($\sim$7$\sigma$) over simpler cloud setups. Furthermore, water is detected ($\sim$12$\sigma$), but methane  is not. 
 These findings provide evidence of disequilibrium chemistry and indicate a dynamically active atmosphere with a super-solar metallicity. 
 \end{abstract}

\bigskip

WASP-107b was observed with JWST MIRI on 19\,--\,20 January 2023. The \texttt{SLIT\-LESS\-PRISM} subarray of the low-resolution spectrometer was used, offering a spectral resolution ranging from 30 and 100 over a wavelength span of 4.61 to 11.83\,$\mu$m. We performed three independent data reductions using the \texttt{CASCADe}\,\cite{bouwman_2023}, \texttt{Eureka!}\,\cite{bell_eureka_2022}, and \texttt{TEATRO} packages; see Supplementary Information (SI).  Each method extracted 51 spectroscopic light curves. For all channels, we obtained a minimal level of correlated noise in the residuals, consistent with normally distributed noise. The 1$\sigma$ error displayed a minimum of 80\,ppm at 7.5\,$\mu$m. The transmission spectra derived from the different reductions, shown in Figure~\ref{fig:wasp107_spectrum} and tabulated in Extended Data Table~\ref{tab:jwst_all_spectra}, are within 3$\sigma$ agreement and 95\% of the points within 2$\sigma$; see SI. 

We performed atmospheric retrievals using two independent frameworks, \texttt{ARCiS}\,\cite{min_2020} and \texttt{petit\-RADTRANS} \texttt{(pRT)}\,\cite{molliere_wardenier_2019}, including both our JWST data and previous near-infrared (1.121\,--\,1.629 $\mu$m) HST data\,\cite{kreidberg_2018}. Free abundance retrievals were run including the following species: H$_{2}$O, CO, CO$_2$, CH$_4$, C$_2$H$_2$, SO$_2$, SO, H$_{2}$S, SiO, HCN, NH$_3$, and PH$_3$. The remaining atmosphere consisted of H$_2$ and He. A variety of cloud models were tested, ranging from cloud-free to more complex models, the latter focusing mostly on silicate clouds (MgSiO$_3$, SiO$_2$, and SiO); see SI.

Figure~\ref{fig:wasp107_retrieval} shows the best fit to the data, including main contributions from molecular species and clouds to the spectrum. The figure presents the results based on the \texttt{CASCADe} package, but our conclusions are consistent across the three data reductions. Both retrieval codes detect SO$_2$ at $\sim$9$\sigma$, H$_2$O at $\sim$12$\sigma$, and the presence of high-altitude clouds at $\sim$9$\sigma$, with a $\sim$7$\sigma$ preference for silicate clouds over more simple cloud setups. 
We also tentatively detect H$_2$S ($\sim$4$\sigma$), NH$_3$ ($\sim$2\,--\,3$\sigma$), and CO ($\sim$2\,--\,3$\sigma$), although the CO detection relies on the first three spectral points and requires confirmation at shorter wavelengths. CH$_4$ is not detected, with an upper limit of its volume mixing ratio (VMR) being a few times 10$^{-6}$. Table~\ref{tab:retrieval_result} presents the detection significance and VMR for each species. Due to differences in cloud structure setups, the absolute VMRs are different between the two retrieval codes. We therefore focus the discussion on the detection significance and relative abundances.

The MIRI data of WASP-107b presents the mid-infrared discovery of SO$_2$ in an exoplanet atmosphere. Both the $\nu_1$ symmetric stretch and the $\nu_3$ asymmetric stretch vibration bands  of SO$_2$ (with fundamental frequency at 8.69\,$\mu$m and 7.35\,$\mu$m, respectively\,\cite{Herzberg1966msms.book.....H}) are detected (Figure~\ref{fig:wasp107_spectrum} and Figure~\ref{fig:wasp107_retrieval}).  
Recently, the ($\nu_1+\nu_3$) combination band of SO$_2$ was assigned as carrier of a spectral feature near 4.05\,$\mu$m detected in the JWST NIRspec spectrum of WASP-39b\cite{rustamkulov_early_2023, jakobsen_2022}, a hot ($\sim$1,200\,K) irradiated Saturn-mass exoplanet.
The MIRI detection of SO$_2$ in WASP-107b extends the range of exoplanet temperatures with detected SO$_2$ from $\sim$1,200\,K down to $\sim$740\,K.

The measured VMR of SO$_2$ in WASP-107b (see Table~\ref{tab:retrieval_result}) is several orders of magnitude higher than expected for chemical equilibrium, which predicts H$_2$S to be the dominant sulphur-bearing molecule for Neptune-like planets\,\cite{tsai_2021}. Disequilibrium processes can drive abundances considerably away from chemical equilibrium,  with photochemistry and atmospheric transport being the dominant mechanisms\,\cite{Polman2023A&A...670A.161P}. Notably, the SO$_2$ feature observed in WASP-39b has been attributed to photochemical processes occurring within its atmosphere\,\cite{tsai_photochemically_2023}. Indeed, UV irradiation initiates the photodissociation of H$_2$O, yielding H and OH radicals. These OH radicals are key for oxidising sulphur that is liberated from H$_2$S\,\cite{tsai_photochemically_2023}. However, these models\cite{tsai_photochemically_2023} predict that SO$_2$ would not be detectable using JWST MIRI for a planet with an equilibrium temperature of $\sim$740\,K. This is in contrast with our detection of SO$_2$ in the atmosphere of WASP-107b. 

To unravel the production paths of SO$_2$ in WASP-107b, we computed a grid of disequilibrium models (see SI). To ensure an accurate consideration of the upper-atmosphere chemistry, we observed -- contemporaneously with the JWST observations -- the Near-Ultraviolet (NUV) emission of the host star WASP-107 with {\it Neil Gehrels Swift}. Additionally, we reanalysed the 2018 X-ray emission observed with \textit{XMM-Newton} (see SI). Figure~\ref{fig:wasp107_model} provides evidence that only models incorporating photochemistry in combination with a super-solar metallicity predict a detectable level of SO$_2$ in WASP-107b. Key disparities from prior predictions\,\cite{tsai_photochemically_2023, Polman2023A&A...670A.161P} stem from UV radiation and gravity. Previous models adopted a gravity of 1,000\,cm/s$^2$\,\cite{tsai_photochemically_2023} and 2,140\,cm/s$^2$\,\cite{Polman2023A&A...670A.161P}, while WASP-107b's gravity is $\sim$260\,cm/s$^2$. Moreover, the NUV flux is a factor $\sim$200 lower for WASP-107b than for WASP-39b, and a factor $\sim$100\,--\,1,000 lower in the FUV 
(see SI). A lower gravity, an overall decreased UV flux, and a low FUV/NUV ratio collectively contribute to the increased formation of SO$_2$; see Extended Data Figure~\ref{fig:chemistry_UV_g}.

The overarching scenario that unfolds is that the initiating pathways for SO$_2$ formation in WASP-107b are twofold. First, H$_2$O photodissociation in upper layers at pressures $\la$10$^{-5}$\,bar generates atomic H and OH radicals, leading to sequential oxidation of sulphur liberated from H$_2$S. Second, in the pressure range of $\sim$10$^{-5}$\,--\,1\,bar, photolysis of various abundant molecules -- beyond just H$_2$O -- provides free atoms and radicals, partially redistributed through eddy diffusion. This initiates a cascade of thermochemical reactions that progressively yield sufficient OH radicals for SO$_2$ oxidation. Given the fact that a large fraction of these reactions are temperature-independent and barrierless, the equilibrium temperature is not the sole determinant for SO$_2$ formation. As long as the UV irradiation and FUV/NUV ratio remain moderate and the gravity is low, these processes lead to detectable SO$_2$ levels even at the low equilibrium temperature of WASP-107b.  

The sensitivity to metallicity can be attributed to the larger abundance of sulphur and OH radicals 
at higher metallicities. At a metallicity of 6$\times$ solar, the SO$_2$ spectral features contribute partially to the 7.8\,$\mu$m feature and dominate at $\ga$10$\times$ solar (see Extended Data Figure~\ref{fig:metallicities}). 
For metallicities $\ge$6$\times$ solar, the gas-phase photochemical models predict a SO$_2$ VMR above $\sim$5$\times$10$^{-7}$ at pressures between 10$^{-7}$\,--\,10$^{-4}$\,bar (see Figure~\ref{fig:wasp107_model}). This result is corroborated by the retrieval outcomes, which show that the spectral contribution of SO$_2$ predominantly comes from the region above the high-altitude cloud layer (situated at pressures of a few times 10$^{-5}$\,bar; see Extended Data Figure~\ref{fig:contribution_functions_ARCiS}). This establishes SO$_2$ as a key diagnostic for the atmospheric metallicity in exoplanet atmospheres.

At super-solar metallicity, our models also predict a detectable CH$_4$ feature (see Extended Data Figure~\ref{fig:metallicities}). The confirmation of the HST CH$_4$ non-detection\,\cite{kreidberg_2018} in the MIRI wavelength range raises questions regarding the predicted predominance of CH$_4$ in the atmosphere\,\cite{madhusudhan_co_2012, benneke_sub-neptune_2019}. 
One potential reason for the non-detection of CH$_4$ is that it is hidden by the high-altitude cloud layer. In contrast to SO$_2$, CH$_4$ is photochemically destroyed at high altitudes, i.e.\ above the cloud layer (see Extended Data Figure~\ref{fig:chemistry_Tint_COratios}). Another explanation for the non-detection of CH$_4$ is the influence of the strong irradiation experienced by WASP-107b, which can significantly impact the thermodynamical properties of its atmosphere\,\cite{Cooper2006ApJ...649.1048C, Showman2009ApJ...699..564S,Drummond2020A&A...636A..68D}. 
WASP-107b is highly inflated, pointing to a deep atmosphere that has significantly higher temperatures\,\cite{Showman2002A&A...385..166S, Sainsbury2019A&A...632A.114S, Sarkis2021A&A...645A..79S, Schneider2022A&A...666L..11S} than those predicted by conventional models. The latter suggest an intrinsic 
heat flux ($\sigma T_{\rm{int}}^4$, with $\sigma$ the Boltzmann constant and $T_{\rm{int}}$ the intrinsic temperature) with corresponding $T_{\rm{int}}$ of 150\,K\,\cite{Baraffe_2010}. 
We computed chemical models with different intrinsic temperatures ranging from 250\,--\,600\,K to reproduce a hot deep atmosphere (see Extended Data Figure~\ref{fig:chemistry_Tint_COratios}). 
Our results show that the CH$_4$ molar fraction in the atmospheric layers probed by our MIRI observations is reduced by more than 3 orders-of-magnitude when we increase $T_{\rm{int}}$ from 250 to 600\,K. Interestingly, the SO$_2$ abundance in the detectable upper atmosphere is almost unaffected since sulphur species do not quench in deep regions\,\cite{tsai_2021} (see Extended Data Figure~\ref{fig:chemistry_Tint_COratios}).  
Lowering the C/O ratio from solar (\,=\,0.55) to 0.10 yields an additional decrease of the CH$_4$ abundance by a factor $\sim$15 (see Extended Data Figure~\ref{fig:chemistry_Tint_COratios}).

The presence of silicate clouds in the atmospheres of strongly irradiated planets has been predicted for a long time\,\cite{seager_2000}. Although a hint of silicate emission has been claimed for one single object\,\cite{2007Natur.445..892R}, clear observational evidence has been lacking. Recently, silicate clouds were detected in a young, self-luminous planet at large separation from its host star\,\cite{miles_2023}. The MIRI data of WASP-107b present the detection of the 10$\,\mu$m Si-O stretching mode of solid silicate particles. The retrieval models favour a high-altitude cloud layer  composed of small (sub-micron) amorphous silicate particles.
At the pressure levels and temperatures where we find the cloud layer, gas phase SiO can nucleate and condensate efficiently to form small solid state mineral particles\,\cite{nuth_2006}. However, according to the traditional picture on cloud formation, these particles would eventually rain down to deeper, hotter layers of the atmosphere, depleting the relatively cold upper atmosphere from gas phase SiO. In the deeper, hotter layers, the particles evaporate, allowing the gas phase material to mix and serve as building blocks for new nucleation/condensation cycles. Thus, we can conclude that the presence of silicate clouds at such high altitudes in a relatively cold part of the atmosphere is another indication of strong mixing from either a hot inner atmosphere (similar to the finding related to CH$_4$) or potentially from the intensely irradiated hot dayside. 

\afterpage{\clearpage}
\newpage

\begin{figure*}[!htp]
\centering
	\includegraphics[width=12truecm]{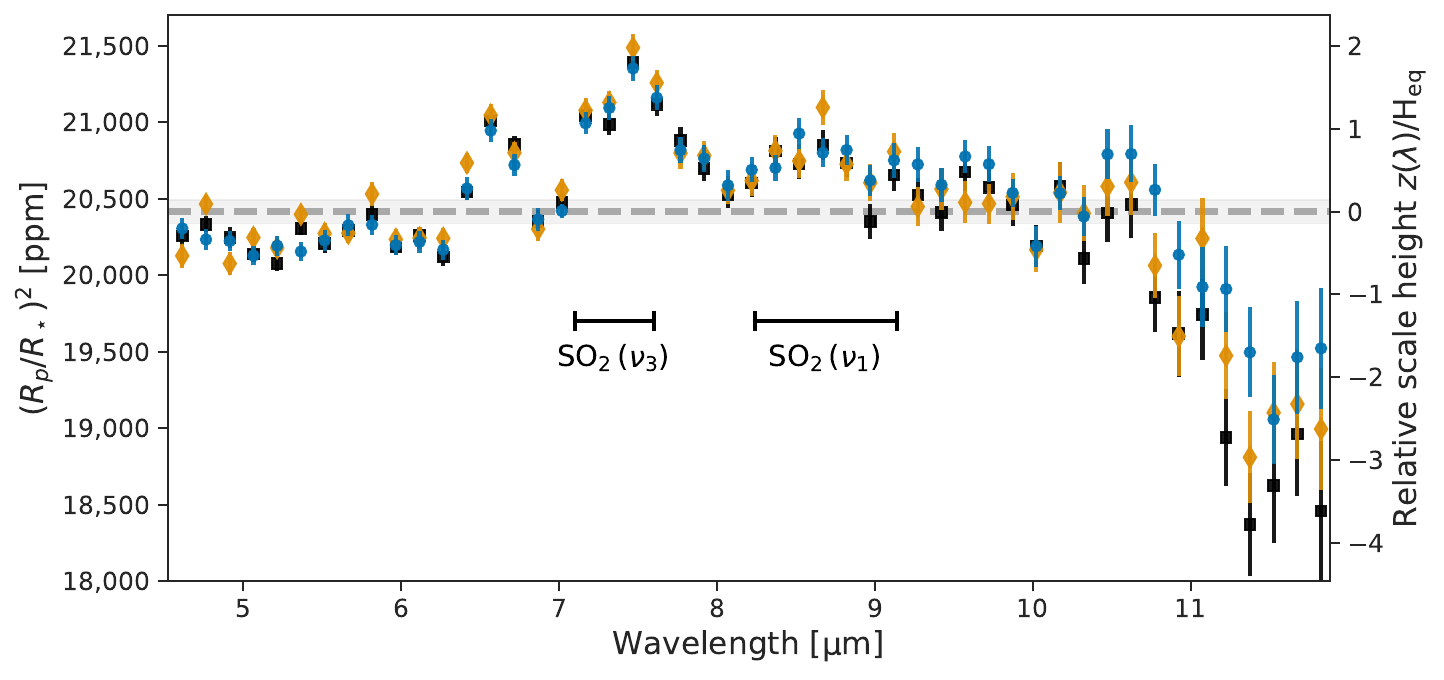}
	\caption{\textbf{JWST MIRI transmission spectrum of WASP-107b.} Comparison of the JWST MIRI transmission spectra obtained from the three independent reductions considered in this work (coloured points, with 1-$\sigma$ error bars). The blue dots show the results from the \texttt{CASCADe} code, the red dots using the \texttt{Eureka!} package and the black dots are from the \texttt{TEATRO} routines. The thick dashed grey line indicates the band-averaged transit depth from the \texttt{CASCADe} analysis at 20,463\,ppm and the shaded area the 95\% confidence interval of 39\,ppm. The right y-axis gives the planetary spectrum in units of atmospheric scale height of the planetary atmosphere, assuming a hydrogen dominated atmosphere. The spectra have a constant wavelength bin width, corresponding to a spectral resolution of 50 at 7.5~$\mu$m. The two horizontal bars indicate the $\nu_1$ symmetric stretch and $\nu_3$ asymmetric stretch vibration bands of SO$_2$.
 } 
		\label{fig:wasp107_spectrum}
\end{figure*}

\afterpage{\clearpage}
\newpage

\begin{figure}[ht!]
	\centering
	\includegraphics[width=12truecm]{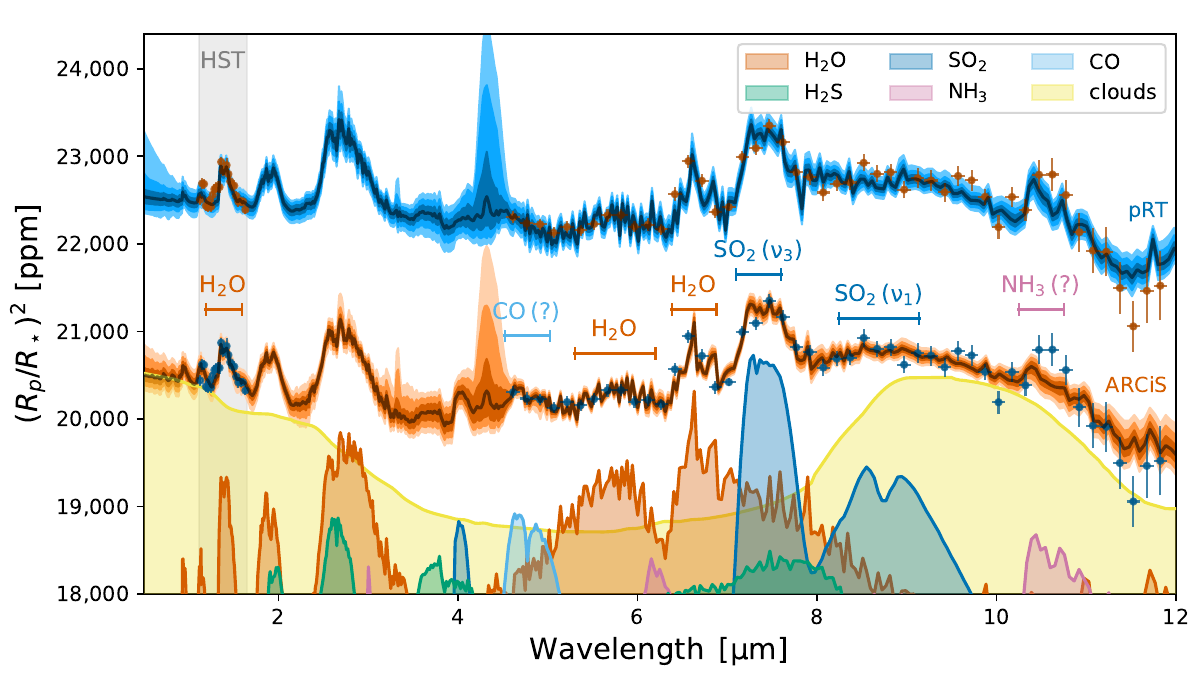}
	\caption{\textbf{Transmission spectrum of WASP-107b with key contributions. }
 The (red and pink) points with 1-$\sigma$ error bars correspond to the measured JWST MIRI \texttt{CASCAde} transit depths of the spectrophotometric light curves at different wavelengths. The (red and pink) points in the area indicated by the grey band are from HST. The median spectrum model predicted from  \texttt{ARCiS}\,\cite{min_2020} retrievals is shown in blue and from \texttt{petitRADTRANS} \texttt{(pRT)}\,\cite{molliere_wardenier_2019} in green (offset by 2,000\,ppm for clarity). The shaded regions of the model spectra correspond to the 1-,2-,3-$\sigma$ credibility envelopes predicted by the retrievals. The bottom part of the figure shows the silicate cloud and molecular contributions for those gases inferred by our analysis of WASP-107b's spectrum.}
	\label{fig:wasp107_retrieval}
\end{figure}

\afterpage{\clearpage}
\newpage

\begin{figure}[ht!]
	\centering
	\includegraphics[width=12truecm]{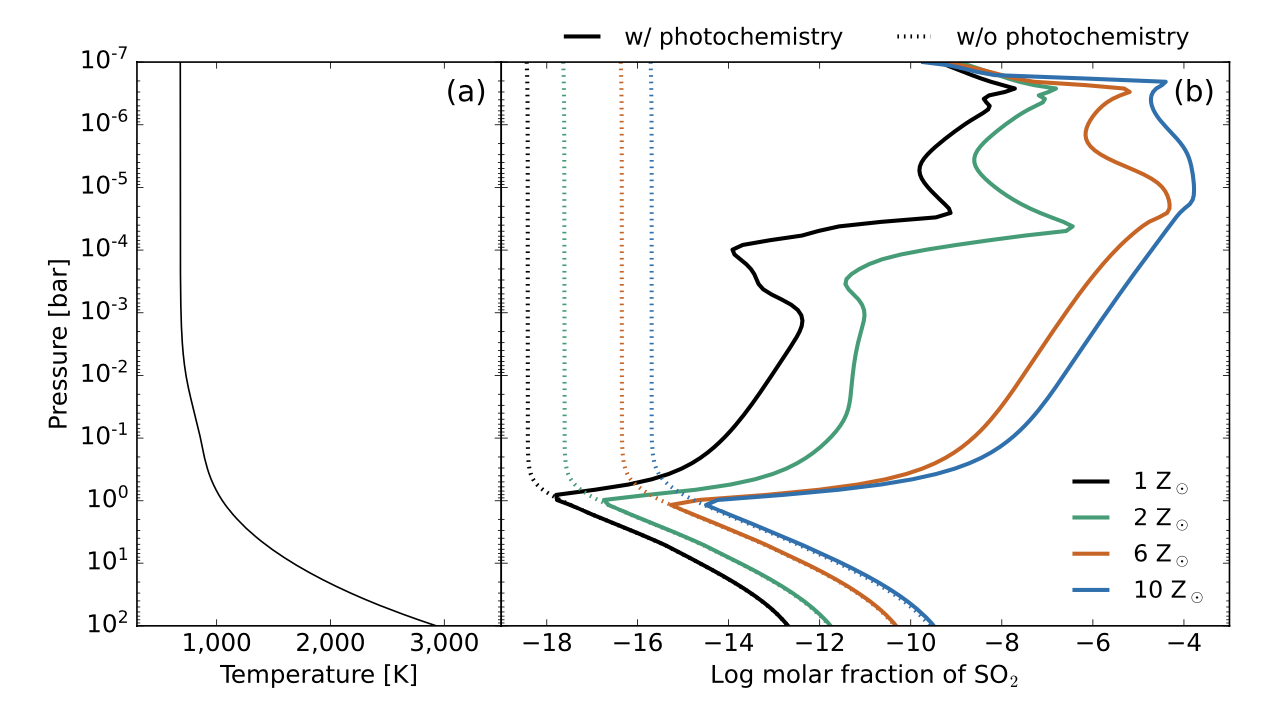}
\caption{\textbf{Predicted SO$_2$ molar fraction.}
Panel~(a): Adopted temperature-pressure ($T$-$P$) profile for WASP-107b following the analytical equation of ref.~\cite{Guillot2010A&A...520A..27G}; see SI. Panel~(b): Predicted SO$_2$ molar fraction for four values of the metallicity (1, 2, 6, and 10 Z$_\odot$; in black, green, red, and blue, respectively) for models with (full line) and without (dotted line) the inclusion of photochemistry. 
Predictions are calculated for a model with an intrinsic temperature of 400\,K, a solar C/O ratio, and a $\log_{10}$($K_{\rm{zz}}$, \rm{cgs})\,=\,10. }
	\label{fig:wasp107_model}
\end{figure}

\afterpage{\clearpage}
\newpage

\begin{table*}[!htp]
	\caption{\textbf{Outcome of the retrieval analyses.} Given are the detection significance and volume mixing ratios (VMR) of the various components in the \texttt{ARCiS} and \texttt{petitRADTRANS} (\texttt{pRT}) retrieval analysis.}
	\label{tab:retrieval_result}
	\begin{center}
	\begin{tabular}{@{\extracolsep\fill}lcccc}
		\toprule%
		& \multicolumn{2}{@{}c@{}}{Significance [$\sigma$]}
  		& \multicolumn{2}{@{}c@{}}{$\log_{10}(\rm{VMR})$}
        \\\cmidrule{2-3}\cmidrule{4-5}%
		Component & \texttt{ARCiS} & \texttt{pRT} & \texttt{ARCiS} & \texttt{pRT} \\
		\midrule
		H$_2$O      & 13.2 &  12.5 & $-2.19_{-0.26}^{+0.42}$ & $-3.81_{-0.29}^{+0.38}$ \\
		SO$_2$      & 8.8 &  9.1 & $-5.03_{-0.18}^{+0.33}$ & $-6.72_{-0.23}^{+0.30}$ \\
		H$_2$S      & 4.7 &  3.5 & $-2.65_{-0.38}^{+0.49}$ & $-3.88_{-0.33}^{+0.41}$ \\
		NH$_3$      & 2.3 & 3.4 & $-5.47_{-0.66}^{+0.34}$ & $-6.04_{-0.25}^{+0.30}$ \\
		CO          & 2.8 & 2.3 & $-2.41_{-0.28}^{+0.29}$ & $-4.58_{-0.59}^{+0.40}$ \\
		PH$_3$      & --$^{(b)}$ & --$^{(a)}$ & $-6.29_{-2.24}^{+1.29}$ & $-7.60_{-1.39}^{+0.89}$ \\
		HCN         & --$^{(b)}$ & --$^{(a)}$ & $-9.26_{-1.63}^{+1.73}$ & $-9.19_{-1.15}^{+1.20}$ \\
		C$_2$H$_2$  & --$^{(b)}$ & --$^{(a)}$ & $-9.08_{-1.73}^{+1.73}$ & $-9.19_{-1.14}^{+1.19}$ \\
		SiO         & --$^{(b)}$ & --$^{(a)}$ & $-6.08_{-3.34}^{+1.90}$ & $-9.03_{-1.40}^{+1.47}$ \\
		CH$_4$      & --$^{(b)}$ & --$^{(b)}$ & $-8.52_{-2.09}^{+2.09}$ & $-8.83_{-1.23}^{+1.25}$ \\
		CO$_2$      & --$^{(b)}$ & --$^{(b)}$ & $-8.05_{-2.37}^{+2.46}$ & $-8.49_{-1.70}^{+1.79}$ \\
		SO          & --$^{(b)}$ & --$^{(c)}$ & $-7.38_{-2.74}^{+2.76}$ & \\
		Silicate Clouds & 7.2 & 7.1 \\
		\bottomrule
	\end{tabular}\\
\end{center}
{$^{(a)}$~included in retrieval, but not tested for detection significance because posterior indicates an upper limit. \\
	$^{(b)}$~not favoured (Bayes factor $<$ 2). \\ $^{(c)}$~not included.}
\end{table*}

\clearpage
\newpage
\mbox{~}

\begin{addendum}

\item[Supplementary Information] is linked to the online version of the paper at www.nature.com/nature.

\item[Data availability]
The JWST MIRI data presented in this paper are part of the JWST MIRI GTO program (Program identifier (PID) 1280; P.I.\  P.O.\ Lagage). The JWST data will be publicly available in the Barbara A.\ Mikulski Archive for Space Telescopes (MAST; \url{https://archive.stsci.edu/}) after January 20, 2024, and can be found either using the program identifier or using the Data Object Identifier (DOI): \url{https://doi.org/10.17909/as3s-x893}. The DOI link can also be used to retrieve the publicly available HST WFC3 data used in this paper from MAST. The NUV data are in an online publicly-accessible archive: \url{https://heasarc.gsfc.nasa.gov/cgi-bin/W3Browse/swift.pl}. The HST data are available in the MAST archive and can be found using DOI: \url{https://doi.org/10.17909/as3s-x893}. The XMM X-ray data archive is available at \url{https://nxsa.esac.esa.int/nxsa-web/#search}.

\item[Code availability]
The codes used in this publication to extract, reduce, and analyse the data are as follows. The data reduction pipeline \texttt{jwst} can be found at \url{https://jwst-pipeline.readthedocs.io/en/latest/}. The data analysis codes \texttt{Eureka!}, \texttt{CASCADe}, and \texttt{TEATRO}  can be found respectively at \url{https://eurekadocs.readthedocs.io/en/latest/},
\url{https://gitlab.com/jbouwman/CASCADe}, and \url{https://github.com/ncrouzet/TEATRO}. The \texttt{CASCADe-filtering}, and \texttt{CAS\-CADe-\-jitter} sub-packages can be found, respectively, at \url{https://gitlab.com/jbouwman/CASCADe-filtering}, and
\url{https://gitlab.com/jbouwman/CASCADe-jitter}. The atmospheric model codes used to fit the data can be found at \url{https://www.exoclouds.com/} for the 
\texttt{ARCiS}-code\,\cite{min_2020} and at
\url{https://petitradtrans.readthedocs.io/en/latest/} for the \texttt{petitRADTRANS}-code\,\cite{molliere_wardenier_2019}. The XSPEC package\,\cite{arnaud_xspec_1996}  is available at \url{https://heasarc.gsfc.nasa.gov/xanadu/xspec/}.
The HEASoft package (including the Swift UVOT tools) is available at \url{https://heasarc.gsfc.nasa.gov/lheasoft/download.html} (version 6.31.1 was used in our work). The NUV {\it Swift}-project pipeline is available at \url{https://swift.gsfc.nasa.gov/quicklook/swift_process_overview.html}. The VULCAN chemical network can be found at
\url{https://github.com/exoclime/VULCAN/blob/master/thermo/SNCHO_photo_network.txt}.

\item[Inclusion \& Ethics] All authors have committed to upholding the principles of research ethics \& inclusion as advocated by the Nature Portfolio journals.

\item [Acknowledgments] This work is based on observations made with the NASA/ESA/CSA JWST. The data were obtained from the Mikulski Archive for Space Telescopes at the Space Telescope Science Institute, which is operated by the Association of Universities for Research in Astronomy, Inc., under NASA contract NAS 5-03127 for JWST. These observations are associated with program 1280. MIRI draws on the scientific and technical expertise of the following organisations: Ames Research Center, USA; Airbus Defence and Space, UK; CEA-Irfu, Saclay, France; Centre Spatial de Li\`ege, Belgium; Consejo Superior de Investigaciones Cient\'ificas, Spain; Carl Zeiss Optronics, Germany; Chalmers University of Technology, Sweden; Danish Space Research Institute, Denmark; Dublin Institute for Advanced Studies, Ireland; European Space Agency, Netherlands; ETCA, Belgium; ETH Zurich, Switzerland; Goddard Space Flight Center, USA; Institut d’Astrophysique Spatiale, France; Instituto Nacional de T\'ecnica Aeroespacial, Spain; Institute for Astronomy, Edinburgh, UK; Jet Propulsion Laboratory, USA; Laboratoire d’Astrophysique de Marseille (LAM), France; Leiden University, Netherlands; Lockheed Advanced Technology Center (USA); NOVA Opt-IR group at Dwingeloo, Netherlands; Northrop Grumman, USA; Max-Planck Institut für Astronomie (MPIA), Heidelberg, Germany; Laboratoire d’Etudes Spatiales et d’Instrumentation en Astrophysique (LESIA), France; Paul Scherrer Institut, Switzerland; Raytheon Vision Systems, USA; RUAG Aerospace, Switzerland; Rutherford Appleton Laboratory (RAL Space), UK; Space Telescope Science Institute, USA; Toegepast Natuurwetenschappelijk Onderzoek (TNO-TPD), Netherlands; UK Astronomy Technology Centre, UK; University College London, UK; University of Amsterdam, Netherlands; University of Arizona, USA; University of Bern, Switzerland; University of Cardiff, UK; University of Cologne, Germany; University of Ghent; University of Groningen, Netherlands; University of Leicester, UK; University of Leuven, Belgium; University of Stockholm, Sweden; Utah. The following National and International Funding Agencies
funded and supported the MIRI development:
NASA; ESA; Belgian Science Policy Office (BELSPO);
Centre Nationale d’Etudes Spatiales (CNES); Danish
National Space Centre; Deutsches Zentrum fur Luftund
Raumfahrt (DLR); Enterprise Ireland; Ministerio
De Economalia y Competividad; Netherlands Research
School for Astronomy (NOVA); Netherlands Organisation
for Scientific Research (NWO); Science and Technology
Facilities Council; Swiss Space Office; Swedish
National Space Agency; and UK Space Agency. 
C.C., A.D.\, P.-O.L.\, R.G.\, A.C.\ acknowledge funding support from CNES. 
O.A.\, I.A.\, B.V.\ and P.R.\ thank the European Space Agency (ESA) and the Belgian Federal Science Policy Office (BELSPO) for their support in the framework of the PRODEX Programme.
D.B.\ is supported by Spanish MCIN/AEI/10.13039/501100011033 grant PID2019-107061GB-C61 and and No. MDM-2017-0737. 
L.D.\ acknowledges funding from the KU Leuven Interdisciplinary Grant (IDN/19/028), the European Union H2020-MSCA-ITN-2019 under Grant no. 860470 (CHAMELEON) and the FWO research grant G086217N. 
I.K.\ acknowledges support from grant TOP-1 614.001.751 from the Dutch Research Council (NWO). 
O.K.\ acknowledges support from the Federal Ministery of Economy (BMWi) through the German Space Agency (DLR).
J.P.P.\ acknowledges financial support from the UK Science and Technology Facilities Council, and the UK Space Agency, and acknowledges the advice of the {\it Swift} project team, and especially Kim Page (UK Swift Science Data Centre at University of Leicester) in planning and analysis of the {\it Swift} observations.
G.O.\ acknowledge support from the Swedish National Space Board and the Knut and Alice Wallenberg Foundation. 
P.T.\ acknowledges support by the European Research Council under Grant Agreement ATMO 757858. 
I.P.W.\ acknowledges funding from the European Research Council (ERC) under the European Union’s Horizon 2020 research and innovation programme (grant agreement No 758892, ExoAI), from the Science and Technology Funding Council grants ST/S002634/1 and ST/T001836/1 and from the UK Space Agency grant ST/W00254X/1.
F.A.M.\ has received funding from the European Union’s Horizon 2020 research and innovation programme under the Marie Sk\l{}odowska-Curie grant agreement no.\ 860470. 
E.D.\ has received funding from the European Union’s Horizon 2020 research and innovation programme under the Marie Skłodowska-Curie actions Grant Agreement no 945298-ParisRegionFP. 
G.V.L.\ acknowledges that some results of this work were partially achieved at the Vienna Scientific Cluster (VSC).
L.H.\ has received funding from the European Union’s Horizon 2020 research and innovation program under the Marie Sk\l{}odowska-Curie grant agreement no. 860470.
T.K.\ acknowledges funding from the KU Leuven Interdisciplinary Grant (IDN/19/028). 
L.C.\ acknowledges support by grant PIB2021-127718NB-100 from the Spanish Ministry of Science and Innovation/State Agency of Research MCIN/AEI/10.13039/501100011033. 
E.vD.\ acknowledges support from A-ERC grant 101019751 MOLDISK. 
T.P.R.\ acknowledges support from the ERC 743029 EASY. 
G.O.\ acknowledges support from SNSA.
P.P.\ thanks the Swiss National Science Foundation (SNSF) for financial support under grant number 200020\_200399.
O.A.\ is a Senior Research Associate of the Fonds de la Recherche Scientifique - FNRS.
We thank Luis Welbanks for a fruitful discussion on the significance of the retrieval results.
We thank Olivia Venot for sharing with us the new sulphur photo-absorption cross-sections and Robin Baeyens for implementing them and the VULCAN thermo-chemical network in our full chemical network. We thank the MIRI instrument team and the many other people who contributed to the success of JWST.

\item[Author Contributions]
All authors played a significant role in one or more of the following: development of the original proposal, management of the project, definition of the target list and observation plan, analysis of the data, theoretical modelling and preparation of this paper. Some specific contributions are listed as follows. 
P.-O.L.\ is PI of the JWST MIRI GTO European consortium program dedicated to JWST observations of exoplanet atmospheres; R.W.\ is co-lead of this JWST MIRI GTO European consortium. 
L.D.\ and N.C.\ provided overall program leadership and management of the WASP-107b working group.
P.-O.L., J.B., T.H., R.W., T.G.\ and L.D. made significant contributions to the design of the observational program and contributed to the setting of the observing parameters. 
A.D., J.B.\ and N.C. generated simulated data for prelaunch testing of the data reduction methods. 
J.B., A.D., and N.C.\ reduced the data, modelled the light curves and produced the planetary spectrum. 
P.T., T.K., and L.D.\ generated theoretical model grids for comparison with the data. 
M.Mi.\ and P.M.\ fitted the generated spectrum with retrieval models. 
J.P.\ led the associated {\it Swift} observing programme and performed the analysis of the {\it Swift} data. M.G.\ led the data reduction and the analysis of the {\it XMM-Newton} X-ray data. 
L.D., M.Mi, J.B., and A.D.\ led the writing of the manuscript. 
L.D., A.D., M.Mi., P.M., J.B., N.C., T.K., J.P., M.G., R.W., P.T., and P.-O.L.\ made significant contributions to the writing of this paper. 
T.K., M.Mi., J.B., and P.M.\ generated figures for this paper.
G.W.\ is the European PI of the JWST MIRI instrument,
P.-O.L., T.H., M.G, B.V., L.C., E.vD., T.R., and G.O.\ are European co-PI, and
L.D., R.W., O.A., I.K., O.K., J.P., G.O.\ and D.B.\ are European co-I of the JWST MIRI instrument. A.G.\ led the MIRI instrument testing and commissioning effort.

\item[Rights and permissions]
Open Access This article is licensed under a Creative Commons Attribution 4.0 International Licence, which permits use, sharing, adaptation, distribution and reproduction in any medium or format, as long as you give appropriate credit to the original author(s) and the source, provide a link to the Creative Commons licence, and indicate if changes were made. The images or other third party material in this article are included in the article’s Creative Commons licence, unless indicated otherwise in a credit line to the material. If material is not included in the article’s Creative Commons licence and your intended use is not permitted by statutory regulation or exceeds the permitted use, you will need to obtain permission directly from the copyright holder. To view a copy of this licence, visit \url{http://creativecommons.org/licenses/by/4.0/}.

\item[Author Information]
$^\star$Correspondence and requests for materials should be addressed to achrene.dyrek@cea.fr. 

\item[Competing Interests] The authors declare that they have no competing financial interests.

\end{addendum}

\section*{References}
\longrefs=1
\bibliographystyle{naturemag}
\bibliography{sn-bibliography}

\afterpage{\clearpage}
\newpage
\setcounter{table}{0} 
\renewcommand{\tablename}{Extended Data -- Table}

\afterpage{\clearpage}
\newpage

\setcounter{figure}{0} 
\renewcommand{\figurename}{Extended Data -- Figure}
\section*{Extended Data}

\begin{table*}[!htp]
\caption{\textbf{Measured transit depth of WASP-107b.} First and fifth column list the wavelength, and the other columns the transit depths as obtained with \texttt{Eureka!}, \texttt{CASCADe} and \texttt{TEATRO}. Each wavelength bin has a full width of 0.15~$\mu$m.}
\label{tab:jwst_all_spectra}
\begin{center}
\setlength{\tabcolsep}{0.8mm}
\begin{tabular}{rllllrlll}
    \toprule
    {\small{$\lambda$}} & \multicolumn{3}{c}{{\small{Transit depth}}} & \multicolumn{1}{c}{} & {\small{$\lambda$}} & \multicolumn{3}{c}{{\small{Transit depth}}}\\
    \cline{2-4}
    \cline{7-9}
    & {\small{\texttt{Eureka!}}} &  {\small{\texttt{CASCAde}}} & {\small{\texttt{TEATRO}}} & \multicolumn{1}{c}{} & & {\small{\texttt{Eureka!}}} & {\small{\texttt{CASCADe}}} & {\small{\texttt{TEATRO}}} \\
	   ($\mu$m) & (ppm) & (ppm)& (ppm) & \multicolumn{1}{c}{} & ($\mu$m) & (ppm) & (ppm)& (ppm) \\
    \midrule
    4.61 & 20,126 $\pm$ 80 & 20,305 $\pm$  69 & 20,257 $\pm$ 73  & \multicolumn{1}{c}{} & 8.37 & 20,813 $\pm$ 100 &  20,701 $\pm$  87 &   20,811 $\pm$ 93\\
	4.76 & 20,464 $\pm$ 60 & 20,233 $\pm$  70 & 20,331 $\pm$ 52 & \multicolumn{1}{c}{} &  8.52 & 20,745 $\pm$ 106 &  20,925 $\pm$  99 &   20,727 $\pm$ 98   \\ 
	4.91 & 20,075 $\pm$ 73 & 20,223 $\pm$  63 &20,245 $\pm$ 70  & \multicolumn{1}{c}{} &  8.67 & 21,095 $\pm$ 112  &  20,800 $\pm$  96  &   20,850 $\pm$ 99 \\ 
	5.06 & 20,245 $\pm$ 60 & 20,128 $\pm$  64 & 20,139 $\pm$ 53  & \multicolumn{1}{c}{} &  8.82 & 20,725 $\pm$ 108 &  20,818 $\pm$  96 &   20,733 $\pm$ 95  \\ 
	5.21 & 20,178 $\pm$ 60 & 20,192 $\pm$  64 &20,077 $\pm$ 52  & \multicolumn{1}{c}{} &   8.97 & 20,603 $\pm$ 118 & 20,620 $\pm$ 100  &   20,350 $\pm$ 112 \\ 
	5.36 & 20,396 $\pm$ 55 & 20,154 $\pm$  63 & 20,306 $\pm$ 46  & \multicolumn{1}{c}{} &    9.12 & 20,807 $\pm$ 118 &  20,750 $\pm$ 114  &  20,654 $\pm$ 107\\ 
	5.51 & 20,272 $\pm$ 69 & 20,227 $\pm$  70 & 20,206 $\pm$ 58 & \multicolumn{1}{c}{} &   9.27 & 20,447 $\pm$ 125 & 20,724 $\pm$ 114 &   20,523 $\pm$ 114 \\ 
	5.67 & 20,274 $\pm$ 65 & 20,326 $\pm$  72 & 20,292 $\pm$ 53  & \multicolumn{1}{c}{} &   9.42 & 20,562 $\pm$ 132 &  20,590 $\pm$ 110 &  20,411 $\pm$ 122 \\ 
	5.82 & 20,530 $\pm$ 76 & 20,329 $\pm$  67 & 20,396 $\pm$ 65  & \multicolumn{1}{c}{} & 9.57 & 20,475 $\pm$ 137 &  20,775 $\pm$ 111 &   20,676 $\pm$ 127 \\ 
	5.97 & 20,234 $\pm$ 64 & 20,196 $\pm$  66 & 20,186 $\pm$ 55  & \multicolumn{1}{c}{} & 9.72 & 20,466 $\pm$ 127  &  20,725 $\pm$ 117 &   20,574 $\pm$ 122\\ 
	6.12 & 20,243 $\pm$ 68 & 20,217 $\pm$  72 & 20,258 $\pm$ 59  & \multicolumn{1}{c}{} &  9.87 & 20,512 $\pm$ 149 &  20,537 $\pm$  94 &  20,460 $\pm$ 141 \\ 
	6.27 & 20,240 $\pm$ 66 & 20,166 $\pm$  64 & 20,118 $\pm$ 61  & \multicolumn{1}{c}{} &   10.02 & 20,166 $\pm$ 148 &  20,190 $\pm$ 134  &   20,190 $\pm$ 139\\ 
	6.42 & 20,733 $\pm$ 66 & 20,567 $\pm$  78 & 20,544 $\pm$ 55  & \multicolumn{1}{c}{} &    10.17 & 20,539 $\pm$ 174 &  20,535 $\pm$ 110 &   20,584 $\pm$ 154\\  
    6.57 & 21,045 $\pm$ 77 & 20,945 $\pm$  73 & 21,006 $\pm$ 65 & \multicolumn{1}{c}{} &  10.32 & 20,404 $\pm$ 186 &  20,384 $\pm$ 125 &  20,109 $\pm$ 165 \\    
    6.72 & 20,799 $\pm$ 67 & 20,720 $\pm$  74  & 20,853 $\pm$ 58 & \multicolumn{1}{c}{} &    10.47 & 20,580 $\pm$ 192 &   20,789 $\pm$ 164 &  20,406 $\pm$ 189\\
    6.87 & 20,301 $\pm$ 79 & 20,364 $\pm$  74 & 20,351 $\pm$ 73  & \multicolumn{1}{c}{} &  10.62 & 20,606 $\pm$ 213 & 20,792 $\pm$ 186 &    20,462 $\pm$ 219\\
    7.02 & 20,555 $\pm$ 72 & 20,420 $\pm$  47 & 20,475 $\pm$ 65 & \multicolumn{1}{c}{} &   10.77 & 20,062 $\pm$ 208 &  20,558 $\pm$ 172 &  19,853 $\pm$ 226\\
    7.17 & 21,076 $\pm$ 78 & 20,993 $\pm$  73 & 21,044 $\pm$ 69  & \multicolumn{1}{c}{} &   10.92 & 19,600 $\pm$ 266 &  20,133 $\pm$ 223 &   19,618 $\pm$ 282\\
    7.32 & 21,128 $\pm$ 76 & 21,093 $\pm$  77 & 20,985 $\pm$ 71  & \multicolumn{1}{c}{} &    11.07 & 20,239 $\pm$ 262 &  19,923 $\pm$ 260 &   19,742 $\pm$ 294\\
    7.47 & 21,486 $\pm$ 88 & 21,351 $\pm$  79 & 21,390 $\pm$ 77  & \multicolumn{1}{c}{} &  11.22 & 19,474 $\pm$ 278 &  19,909 $\pm$ 282 &   18,938 $\pm$ 316\\
    7.62 & 21,257 $\pm$ 83 & 21,160 $\pm$  83 & 21,117 $\pm$ 73  & \multicolumn{1}{c}{} &    11.37 & 18,810 $\pm$ 295 &  19,496 $\pm$ 295 &   18,371 $\pm$ 339\\
    7.77 & 20,797 $\pm$ 100 & 20,817 $\pm$  86 & 20,880 $\pm$ 85 & \multicolumn{1}{c}{} &  11.52 & 19,101 $\pm$ 328 &  19,057 $\pm$ 291 &   18,625 $\pm$ 378\\
    7.92 & 20,782 $\pm$ 93 & 20,762 $\pm$  87 & 20,697 $\pm$ 81  & \multicolumn{1}{c}{} &  11.68 & 19,157 $\pm$ 363 &  19,464 $\pm$ 369  &  18,963 $\pm$ 404\\
     8.07 & 20,554 $\pm$ 90 & 20,587 $\pm$ 101 & 20,524 $\pm$ 83 & \multicolumn{1}{c}{} &    11.83 & 18,994 $\pm$ 397 &  19,522 $\pm$ 396 &   18,459 $\pm$ 457 \\
    8.22 & 20,617 $\pm$ 98 & 20,688 $\pm$  82 & 20,601 $\pm$ 89  & \multicolumn{1}{c}{} &  & -- &  -- & -- \\     
\bottomrule
\end{tabular}
\end{center}
\end{table*}

\begin{figure}[htp]
    \centering
    \includegraphics[width=12truecm]{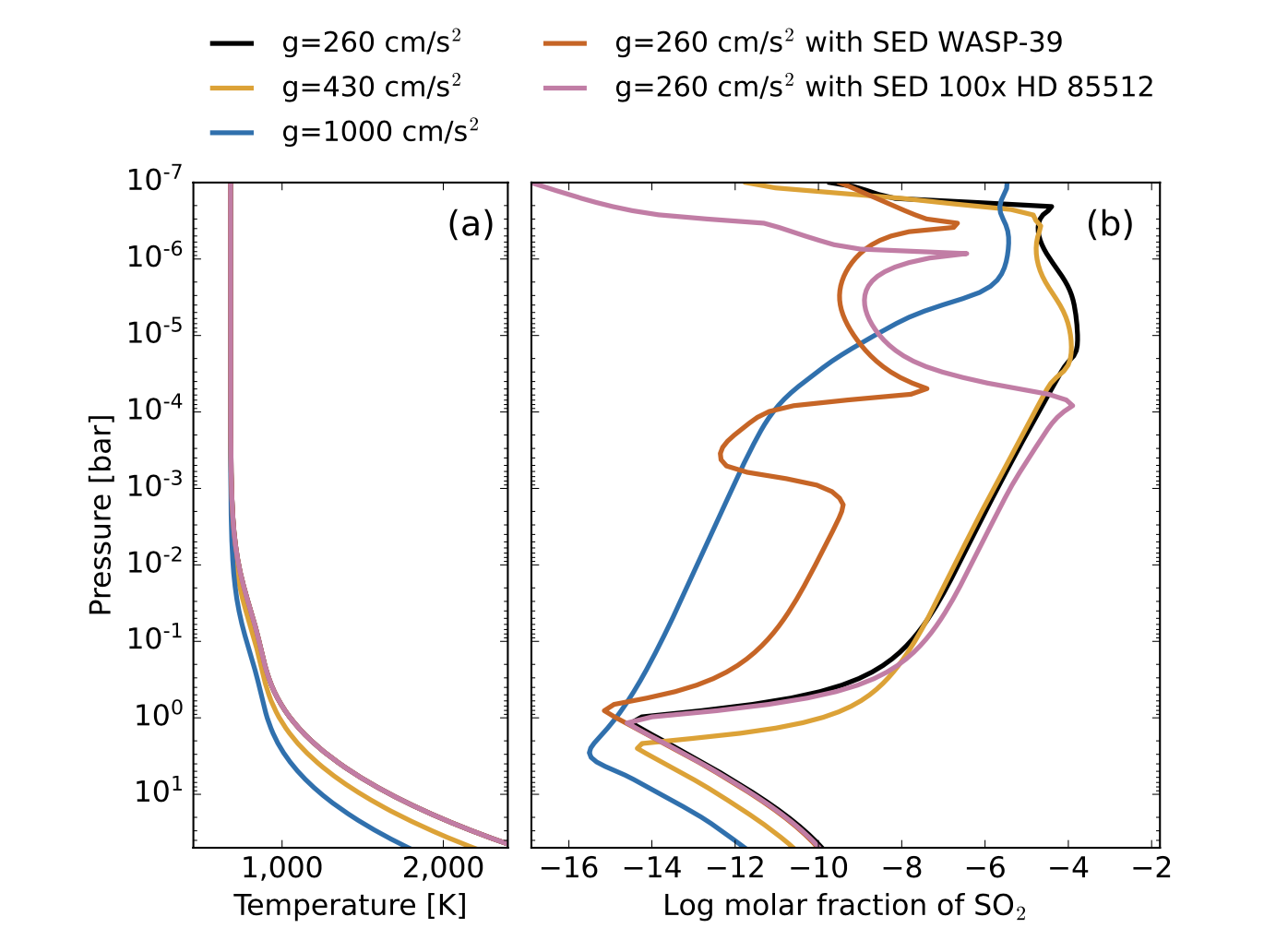}
    \caption{\textbf{Impact of gravity and UV irradiation on predicted SO$_2$ molar fraction.} 
    The base model (shown in black in each panel) has an intrinsic temperature of 400\,K, a solar C/O ratio, a metallicity of 10$\times$ solar, a $\log_{10}(K_{\rm{zz}}, \rm{cgs})$\,=\,10, and uses the SED of HD~85512\,\cite{Loyd2016ApJ...824..102L} -- used as a proxy for WASP-107 -- as input stellar spectrum (see SI). Panel~(a): Temperature-pressure ($T$-$P$) profile for a gravity $g$ of 2.6\,m/s$^2$ (black, purple and brown), 4.3\,m/s$^2$ (blue), and 10\,m/s$^2$ (orange). Panel~(b): Predicted SO$_2$ molar fractions for different gravity values with colours corresponding to panel~(a). While the black curve uses the HD~85512 SED as input spectrum, the purple curve uses the WASP-39 SED instead (see SI), and the brown curve the HD~85512 SED scaled with a factor 100.}
    \label{fig:chemistry_UV_g}
\end{figure}

\begin{figure}[htp]
    \centering
    \includegraphics[width=12truecm]{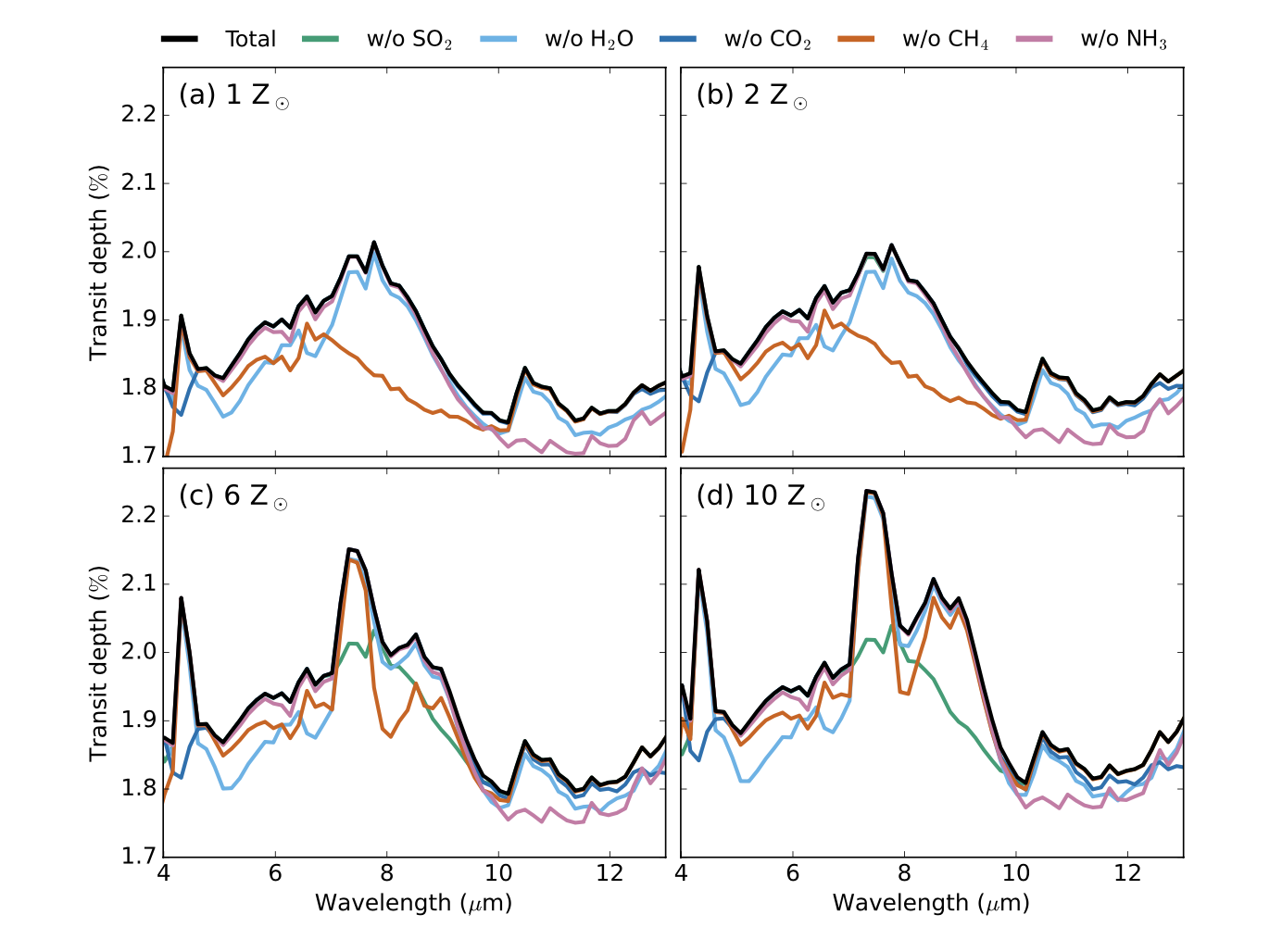}
    \caption{\textbf{Predicted transit depth for WASP-107b assuming a cloudless atmosphere.} The four panels display for a metallicity of 1, 2, 6, and 10 Z$_\odot$, the total transit depth (in black) and the transit depth without relative contributions of SO$_2$ (light blue), H$_2$O (dark blue), CO$_2$ (light green), CH$_4$ (pink) and NH$_3$ (orange). Predictions are calculated for a model with an intrinsic temperature of 400\,K, a solar C/O ratio, and a $\log_{10}$($K_{\rm{zz}}$, \rm{cgs})\,=\,10.
    }
    \label{fig:metallicities}
\end{figure}

\begin{figure}[htp]
    \centering
    \includegraphics[width=12truecm]{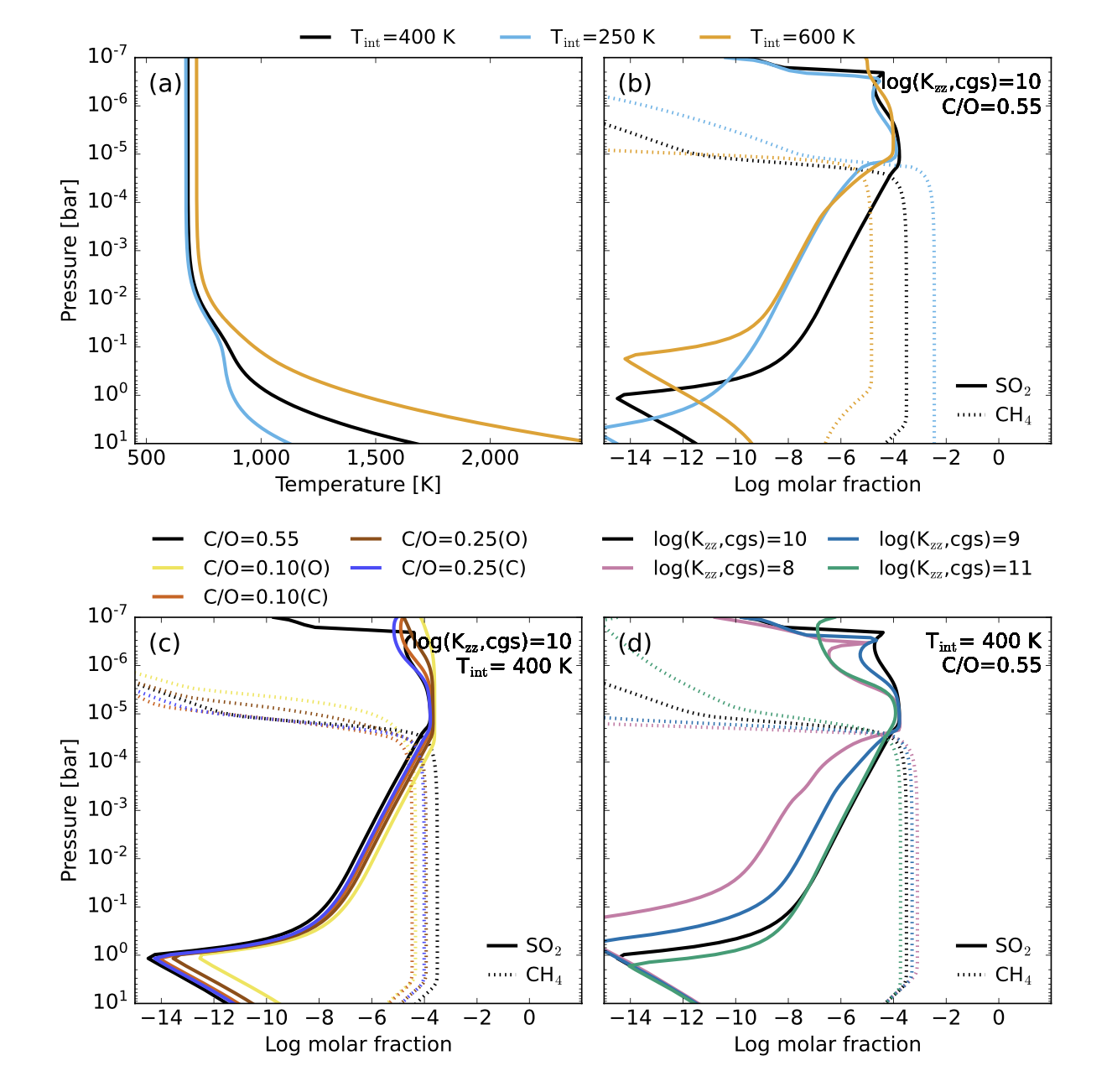}
    \caption{\textbf{Sensitivity of SO$_2$ and CH$_4$ molar fractions to various input parameters.} Shown are the impact of the intrinsic temperature, C/O ratio and eddy diffusion coefficient on predicted SO$_2$ and CH$_4$ molar fractions. The base model (shown in black in each panel) has an intrinsic temperature of 400\,K, a solar C/O ratio, a metallicity of 10$\times$ solar, and a $\log_{10}$($K_{\rm{zz}}$, \rm{cgs})\,=\,10.
    Panel~(a): Temperature-pressure ($T$-$P$) profile for intrinsic temperatures of 250, 400, and 600\,K (light blue, black, and orange, respectively). Panel~(b): Predicted SO$_2$ (full line) and CH$_4$ (dotted line) molar fractions for different $T$-$P$ structures (and hence intrinsic temperatures), with colours corresponding to panel~(a). Panel~(c): Predicted SO$_2$ and CH$_4$ molar fraction for different C/O ratios. Comparison with predictions for a solar C/O ratio (of 0.55, black line) for which either the carbon or oxygen atomic abundance has been adapted (indicated by `C' or `O' between parenthesis, respectively). Panel (d): Predicted SO$_2$ and CH$_4$ molar fraction for different values of the eddy diffusion coefficient $K_{\rm{zz}}$.}
    \label{fig:chemistry_Tint_COratios}
\end{figure}

\begin{figure}[htp]
    \centering
    \includegraphics[width=8.9truecm]{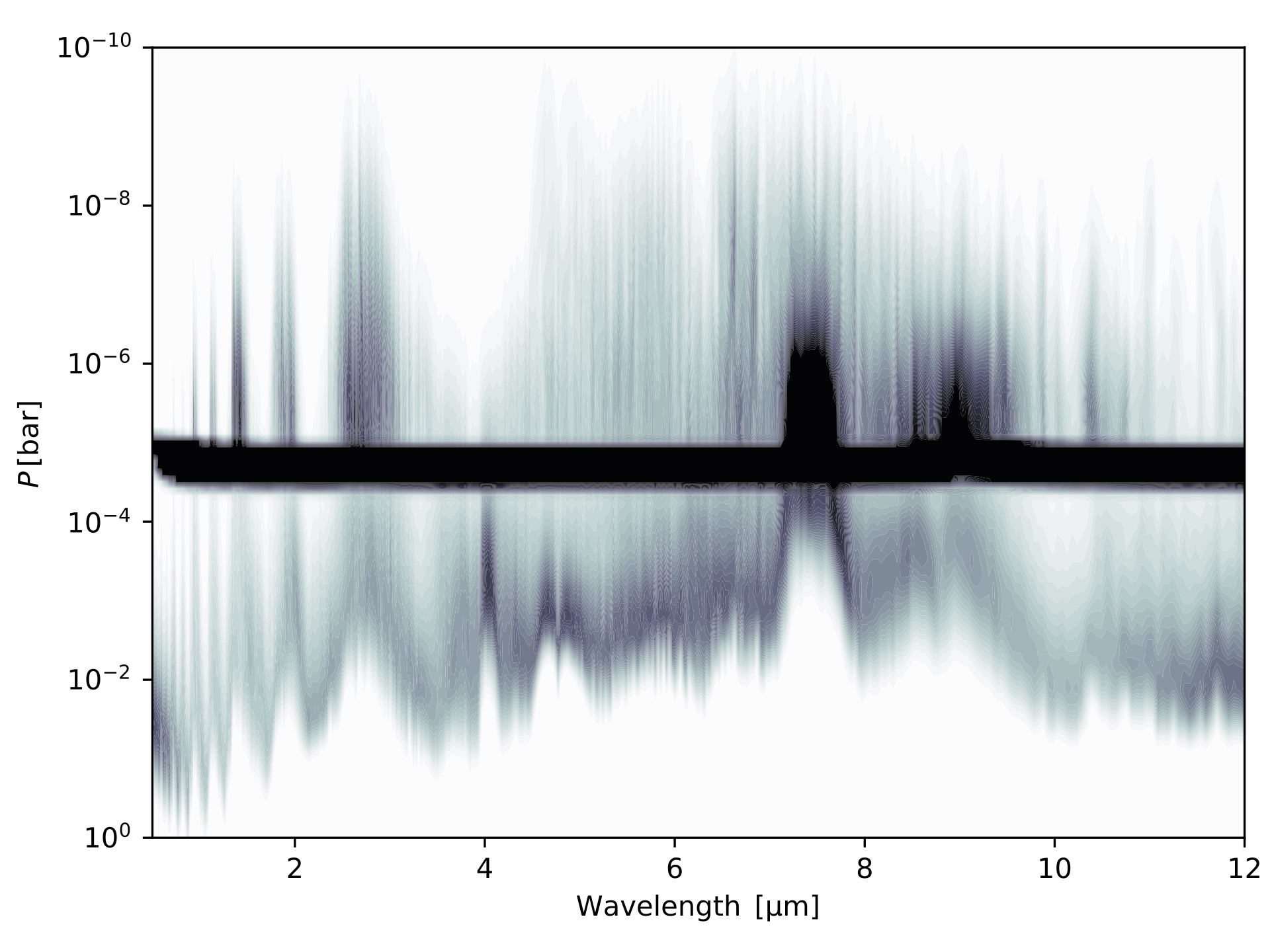}
    \caption{\textbf{Contribution function for the atmosphere as retrieved by \texttt{ARCiS}.} Shown in the contour colours are the relative contributions of various layers in the atmosphere to the transit spectrum as a function of wavelength. The dark horizontal line in this plot is located at the geometrically thin cloud layer, dominating the spectrum at all wavelengths. The spectral variation on top of this baseline predominantly comes from below the cloud layer at the wavelengths where water features are present, while it is dominated by regions above the cloud layer for the spectral features of SO$_2$. This is consistent with the pressures where we expect SO$_2$ to be abundant in the atmosphere from our photochemical modelling (see Figure~\ref{fig:wasp107_model}). The molecular contribution to the spectral variation in the 9.5\,--\,11\,$\mu$m region is significantly fainter than at other wavelengths. This is the region where the silicate feature is most prominent and thus also the spectral variation is dominated by the cloud layer.}
    \label{fig:contribution_functions_ARCiS}
\end{figure}

\begin{figure}[htp]
    \centering
    \includegraphics[width=8.9truecm]{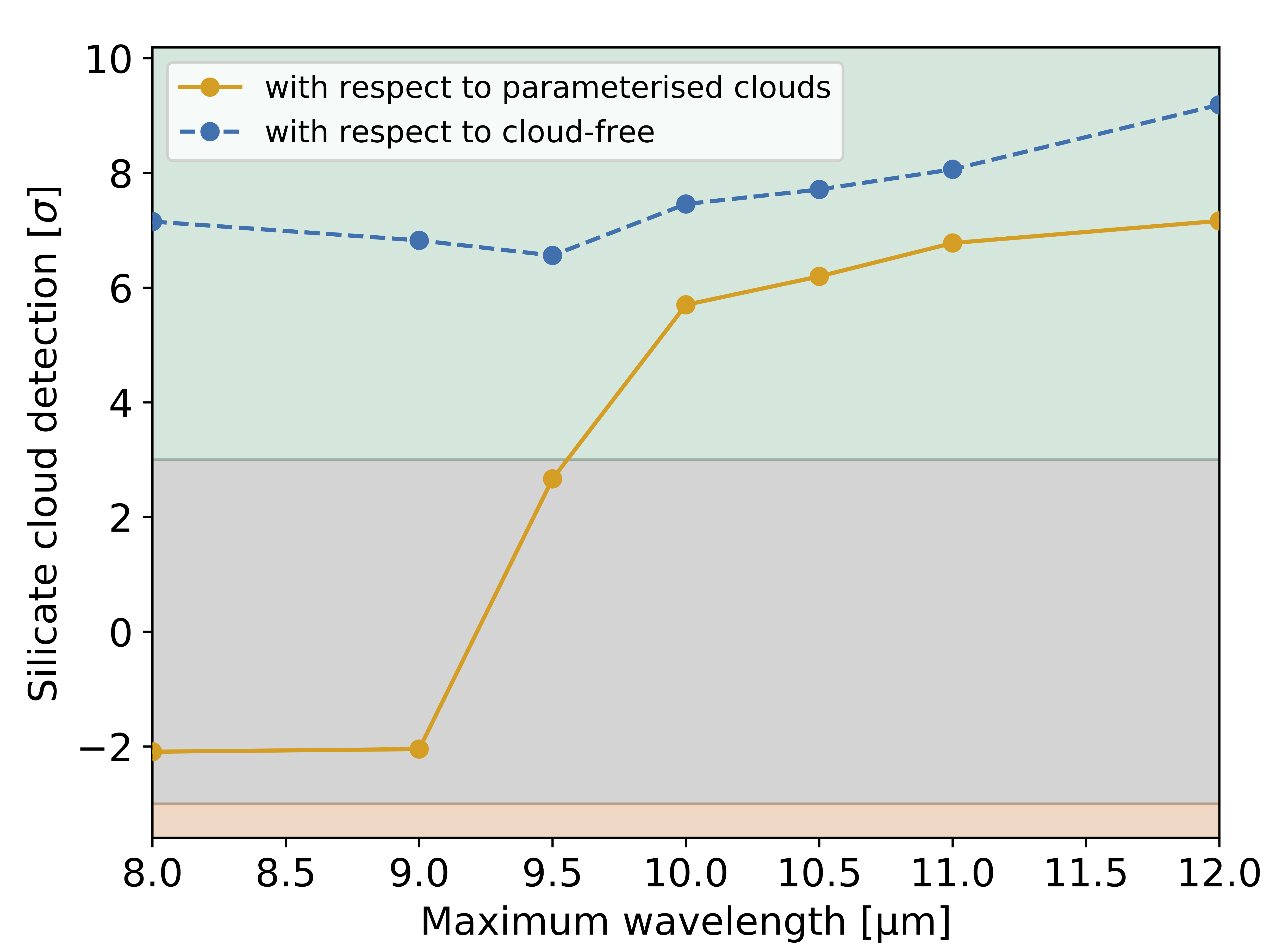}
    \caption{\textbf{Silicate cloud detection significance.} The detection significance, measured in $\sigma$, of the silicate cloud model is presented in comparison to both the cloud-free model (indicated by the blue dashed curve) and the parameterised cloud model (represented by the red solid line) as a function of the maximum wavelength used in the analysis. Even when limiting the analysis to wavelengths below 10\,$\mu$m, the silicate cloud remains favoured, at a significance of 5.7$\sigma$, over the parameterised cloud configuration.}
    \label{fig:silicate_detection}
\end{figure}


\afterpage{\clearpage}
\newpage
   
\setcounter{page}{1}


\noindent{\Large{\textbf{Supplementary information}}\label{sec:supp_information}}

\setcounter{figure}{0} 
\renewcommand{\figurename}{Suppl.\ Inf.\ -- Figure}
\setcounter{table}{0} 
\renewcommand{\tablename}{Suppl.\ Inf.\  -- Table}

\section{JWST MIRI observations and data processing}

WASP-107b was observed with the Low-Resolution Spectrometer (LRS \citeSupp{kendrew_2015}) of the Mid-Infrared Instrument (MIRI \citeSupp{rieke_2015}) on board the JWST on the 19\,--\,20 January 2023. The data is part of the GTO program under program identifier (PID) 1280 (P.I.\  P.O.\ Lagage). The observation started on 19 January at 18:25 UT, in a time-series of 4546 integrations lasting 8h14m, starting approximately 4h50m before the centre of the WASP-107b transit. This total time duration includes the out-of-transit time, $\sim$30 min of detector settling time, and additional time to accommodate scheduling flexibility. 

The data was acquired using the \texttt{SLITLESSPRISM} subarray and the \texttt{FASTR1} readout mode\,\citeSupp{kendrew_2015, ressler_2015}. The integrations consisted of 40 groups (or frames as the MIRI instrument uses 1 frame per group). With this particular number of frames, a maximum signal level of about 75\% of the saturation level is reached, ensuring that a photon-noise limited signal is measured while still avoiding the strongest non-linearity effects occurring for signals approaching saturation.

The data processing began with the uncalibrated raw data products retrieved from the Barbara A.\ Mikulski Archive for Space Telescopes (MAST; \url{https://archive.stsci.edu/}). In order to ensure results that are not influenced by the calibration or potential uncorrected instrumental systematics, we performed three independent data reductions and light curve analyses. In short, our reductions are based on the \texttt{CASCADe} reduction package\,\citeSupp{bouwman_2023}, that was used both for the JWST MIRI and the HST/WFC3 data (Sect.~\ref{method:cascade}),  the \texttt{Eureka!} package\,\citeSupp{bell_eureka_2022} (Sect.~\ref{Sec:Eureka}) and the \texttt{TEATRO} package (see Sect.~\ref{Sec:TEATRO}). To ensure a correct relative flux calibration, we derived and applied a specific non-linearity correction of the ramps (see Sect.~\ref{method:non_linearity}). 
The outcomes of the three data reduction methods are compared in Sect.~\ref{Sec:comparison_reduction}.

\begin{figure}[htp]
    \centering
	\includegraphics[width=0.95\linewidth]{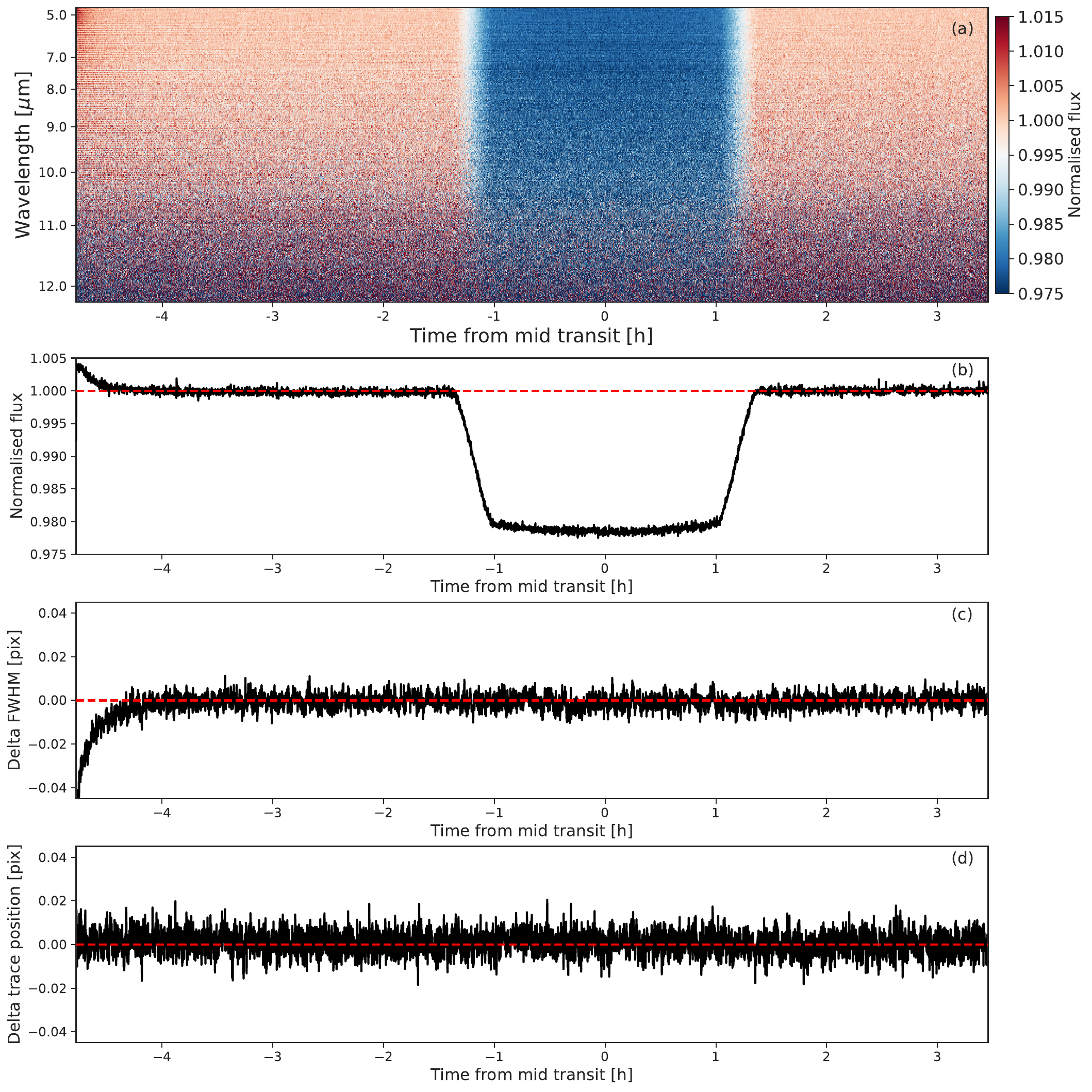}
	\caption{\textbf{Band-average time series of the JWST MIRI/LRS observations of the WASP-107b transit.} Panel~(a): 
    Normalised spectral time series data. Panel~(b): Normalised light curve of the WASP-107b transit integrated between 4.61~$\mu$m to 11.83~$\mu$m. Panel~(c): Change in full-width half maximum (FWHM) in the spectral trace between detector rows 280 and 390, corresponding to the shortest wavelengths. Panel~(d): Change in cross-dispersion position of the spectral trace.  The dashed lines are drawn to indicate the value to expect in case of no variations in values of the plotted data.}
	\label{fig:fig_timeseries}
  \end{figure}

Each method extracted 51 spectroscopic light curves between 4.61 and 11.83\,$\mu$m with a 0.15\,$\mu$m bin width. Suppl.\ Inf.\ Figure~\ref{fig:fig_timeseries} shows the \texttt{CASCADe} MIRI/LRS transit observation of WASP-107b. The shorter wavelength channels ($<$7\,$\mu$m) show the strongest (downward) drift at the start of the spectral time series, consistent with the behaviour observed in the MIRI/LRS data of the transit of L168-9b\,\citeSupp{bouwman_2023}. At the longest wavelengths ($>$11\,$\mu$m), a slight upward drift can be observed, although at a much lower amplitude compared to the short wavelength channels. Note that longward of 10\,$\mu$m, the noise substantially increases due to a decreasing response of the instrument. 

The \texttt{SLITLESSPRSIM} subarray covers parts of 3 distinctive regions on the MIRI imager detector. For wavelengths shorter than 10.5\,$\mu$m the spectra fall within the Lyot coronographic subarray. For wavelengths between 10.5\,$\mu$m and about 11\,$\mu$m the LRS spectra fall on an area of the detector covered by the focal plane mask, and at longer wavelengths in the subarray of one of the 4 quadrant phase mask coronographs. Prior operations of the MIRI imager detector (i.e.\ Exposures with different duration and filter wheel position or idling operation), may impact the detector in the specific regions differently\citeSupp{bell_first_2023}, and could create calibration offsets and extra noise. However, we find no such effects in our data. 

Each spectroscopic light-curve as well as the band-averaged light-curve were fitted by a transit model. We fixed the orbital period $\rm P = 5.7214$ days\,\citeSupp{anderson_2017} and derived the semi-major axis, inclination and mid-transit time from fitting the band-averaged light curve. The free parameters for the spectral light curve fitting were the ratio of the planet radius over the stellar radius, $R_p/R_\star$, the instrumental systematics parameters and the limb-darkening coefficients. The transit ephemeris were taken from ref.~\citeSupp{ivshina_2022}, other system parameters from ref.~\citeSupp{dai_2017}, and quadratic limb-darkening coefficients computed with the \texttt{ExoTETHyS} package\,\citeSupp{morello_2020}, using the parameterised quadratic parameters from ref.~\citeSupp{kipping_2013}. The parameters retrieved from the band-averaged light curve fitting are presented in Suppl.\ Inf.\ Table~\ref{tab:white_fitting_result}. 

\begin{table*}[!htp]
	\caption{\textbf{Parameters retrieved from the band-averaged light curve fitting.} Listed are the values retrieved with the \texttt{Eureka!} and \texttt{TEATRO} reduction methods. \texttt{CASCADe} uses the \texttt{TEATRO} output parameters. The mid-transit timing $T_0$ is the Barycentric Modified Julian date / Temps Dynamique Barycentrique (BMJD\_TDB) time system.}
	\label{tab:white_fitting_result}
	\begin{center}
 \renewcommand{\arraystretch}{1.5}

 \begin{tabular}{lcc}
		\toprule%
 	  Parameter & \texttt{Eureka!} &  \texttt{TEATRO} \\
		\midrule
        Orbital period [d] & $5.7214742\,^{(a)}$ & $5.7214904\,^{(b)}$ \\
        Planetary radius $R_p$ [$R_{\star}]$  & $0.14336^{+6.55154 \times 10^{-5}}_{-6.75549 \times 10^{-5}} $ & $0.14341 \pm 0.00011$ \\
    	Semi-major axis $a$ [$R_{\star}$]  & $18.10815^{+0.00710}_{-0.00712}$ & $18.0249 \,^{(c)}$  \\
        Inclination $i$ [deg]  & $89.59059^{+0.00837}_{-0.00851}$ & $89.516 \pm 0.016$ \\
        Mid-transit timing $T_0$ [d]  & $59963.9687968^{+1.30 \times 10^{-5}}_{-1.27 \times 10^{-5}}$ & $59963.968763 \pm 1.6 \times 10^{-5}$ \\
        Limb-darkening coefficient $u_1$ & $0.095^{+0.0017}_{-0.0015}$  & $0.089 \pm 0.014$ \\
        Limb-darkening coefficient $u_2$ & $0.017^{+0.059}_{ -0.053}$ & $0.042 \pm 0.029$ \\     
		\bottomrule
 \end{tabular}\\
\end{center}
        $^{(a)}$ Fixed\citeSupp{piaulet_2021}. \\
        $^{(b)}$ Fixed\citeSupp{ivshina_2022}. \\
        $^{(c)}$ Fixed, computed from the orbital period\citeSupp{ivshina_2022} and the stellar mass and radius\citeSupp{anderson_2017}. \\
\renewcommand{\arraystretch}{1}
\end{table*}

\subsection{CASCADe data reduction setup}\label{method:cascade}

The first method for spectral extraction and time series analysis was based on the Calibration of trAnsit Spectroscopy using CAusal Data (\texttt{CASCADe}) data reduction package developed within the \emph{Exoplanet Atmosphere New Emission Transmission Spectra Analysis} (\texttt{ExoplANETS-A}) Horizon-2020 program and described in detail in ref.~\citeSupp{bouwman_2023}. For the basic data calibration and spectral extraction, we used the \texttt{jwst} calibration pipeline version 1.9.4 and reference files from the JWST Calibration Reference Data System (CRDS) using context 1030. We followed the procedure described in ref.\,\citeSupp{bouwman_2023} with a few exceptions. We found that the dark correction applied in context version 1030 was not optimal, as it introduced an excess scatter on the detector ramps. We, therefore, overrode the dark reference file with a custom one, which we derived by taking the standard CRDS dark file and running a median smoothing (or running median) to remove the observed excess scatter in the dark estimate. For a complete discussion on our linearity correction we refer to Sect.~\ref{method:non_linearity}. Secondly,  we used the reset switch charge decay (RSCD) step in the Detector1 pipeline stage that flags the first 4 groups of each integration as `do not use'. Though this decreases the effective integration time, the linearity and stability of the detector signals are improved in a substantial way, resulting in increased signal-to-noise ratios of the final extracted spectra. Note that also the last group of each integration is standard flagged as `do not use', as this group is strongly affected by the detector reset (see also ref.\,\citeSupp{morrison_2023}).
The infrared background emission was removed by determining a median background per detector row and integration using detector columns 12 to 19 (starting from 0) and 52 to 59. We used the \texttt{CASCADe-filtering} package version 1.0.2 to identify any bad pixels or cosmic ray hits not identified in the Detector1 pipeline stage. We then used this package to clean all pixels flagged as `do not use' before spectral extraction. We used the \texttt{CASCADe-jitter} package version 0.9.5 to determine the spectral trace to be able to precisely position the extraction aperture. The time averaged polynomial coefficients of the spectral trace are $35.29$, $4.313 \times 10^{-3}$,  $5.947 \times 10^{-6}$ and $-9.484 \times 10^{-8}$ from zero to third order, respectively.
We extracted the 1D spectral time series data from the spectral images using the \texttt{extract1d} pipeline step. In this step we used the polynomial coefficients from the trace fit listed above to centre a constant width extraction aperture of 8 pixels at the exact source position for all wavelengths. The spectral flux values are calculated by summing the signal on the detector within the region defined by the extraction aperture and wavelength bins. Suppl.\ Inf.\ Figure~\ref{fig:fig_timeseries} shows the time series of the extracted LRS spectra. Also shown in that figure is the derived movement of the spectral trace in the cross dispersion direction and the full-width at half maximum (FWHM) of the spectral trace at the shortest wavelengths. Apart from the first half an hour, no substantial photometric or positional drifts can be observed, showing the exquisite stability of the MIRI instrument.

\begin{figure}[htp]
	\centering
	\includegraphics[width=0.95\columnwidth]{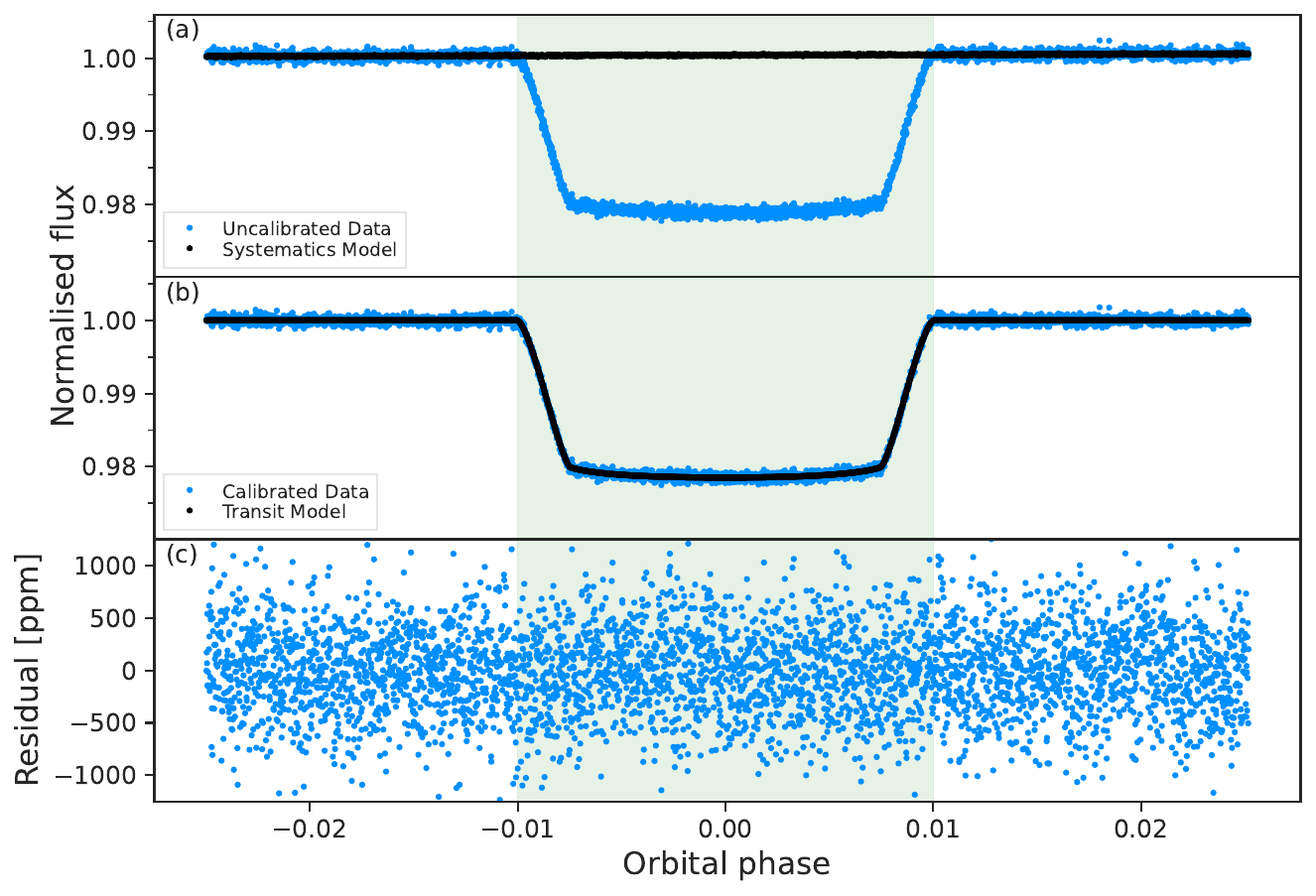}
	\caption{\textbf{\texttt{CASCADe} band average light curve analysis.} Panel~(a): The band-averaged JWST MIRI/LRS light curve data of the transit of WASP-107b and the fitted systematics model. Panel~(b): The systematics corrected band-averaged light curve with the fitted transit model. Panel~(c): The band-averaged residuals after subtracting the best-fit light curve model. The shaded area indicates the orbital phases during which the transit occurs. Note that the first 744 integrations have been removed from the light curves shown in this figure.}
    \label{fig:fig_residuals}
\end{figure}

For the light curve fitting, we used the identical procedure as described in ref.~\citeSupp{bouwman_2023} using the \texttt{CASCADe}-package version 1.2.2. We omitted the first 744 integrations (about 1.3 hours) to avoid the response drifts seen in Suppl.\ Inf.\ Figure~\ref{fig:fig_timeseries}. Before the spectral light curve fitting, we binned the spectra to a uniform wavelength grid with a 0.15~$\mu$m bin width.
For the systematics model (see ref.\,\citeSupp{carone_2021} for details), we used as additional regression parameters the time, the FWHM of the spectral trace, and the trace position as plotted in Suppl.\ Inf.\ Figure~\ref{fig:fig_timeseries}. The orbital parameters of WASP-107b were fixed to the values derived in the band-averaged light curve analysis from the \texttt{TEATRO} data reduction (see Suppl.\ Inf.\ Table~\ref{tab:white_fitting_result}). Limb-darkening coefficients for each spectral channel were calculated using the \texttt{ExoTETHyS}-package~\,\citeSupp{morello_2020} (see Suppl.\ Inf.\ Table~\ref{tab:white_fitting_result}).
The \texttt{CASCADe} band-averaged results from our spectral light curve analysis are presented in Suppl.\ Inf.\ Figure~\ref{fig:fig_residuals} and the \texttt{CASCADe} transmission spectroscopy results are provided in Extended Data Table~\ref{tab:jwst_all_spectra} and shown in Figure~\ref{fig:wasp107_spectrum}. The error estimates on the transit depths were derived by performing a bootstrap analysis.

\subsection{Eureka! data reduction setup}\label{Sec:Eureka}

Data reduction was conducted using the STScI \texttt{jwst} pipeline version 1.8.5 under CRDS context 1030. At the ramp and pixel scale, the first four frames corresponding to the ones affected by the RSCD effect\,\citeSupp{argyriou_calibration_2021} were flagged and ramp non-linearity correction was performed using a custom correction file ( Suppl.\ Inf.\ ~\ref{method:non_linearity}). Contrary to the \emph{CASCADe} and \emph{TEATRO} data reductions,  no custom dark file is needed as pipeline version 1.8.5 uses different correction which did not show an excess scatter.
Cosmic rays were flagged with a rejection threshold of 5$\sigma$ and the ramps were fitted using a least-squared minimisation algorithm. To comply with the JWST MIRI spectroscopic performances of time-series observations\,\citeSupp{bouwman_2023}, the electronic gain value was lowered from 5.5 to 3.1 $\rm e^- \; \rm DN^{-1}$. The background was subtracted following the same method as in ref.~\citeSupp{bouwman_2023}. In particular, 7 columns on the left and 7 on the right sides of the trace (column 36) were selected, a median value was taken and then subtracted from the spectral image. A spatial filter of outlier detection was then applied to remove any hot pixel that would have been left in the subarray. An optimal spectral extraction with a half-width extraction aperture of 4 pixels was then performed using the \texttt{Eureka!} package\,\citeSupp{bell_eureka_2022}. We extracted 51 spectroscopic light curves between 4.61 and 11.83 $\mu$m with a 0.15  $\mu$m bin width, ran a sigma-clipping of 20 integrations with a rejection threshold of 5$\sigma$, and trimmed the 250 first integrations to get rid of strong persistence effects. Light curves were then fitted using the MCMC \texttt{emcee} sampler\,\citeSupp{foreman_2013}, the \texttt{batman} transit model\,\citeSupp{kreidberg_2015} and both an exponential and a second order polynomial model for systematics. We used the band-averaged light curve fit to refine the mid-transit timing, the ratio of the semi-major axis over the stellar radius, and inclination parameters (see Suppl.\ Inf.\ Table~\ref{tab:white_fitting_result}). The ratio of the planet radius over the stellar radius, the limb-darkening coefficients and the systematics parameters were then used as free parameters for all spectroscopic channels (see Suppl.\ Inf.\ Table~\ref{tab:white_fitting_result}). The \texttt{Eureka!} transmission spectroscopy results are provided in Extended Data Table~\ref{tab:jwst_all_spectra} and shown in Figure~\ref{fig:wasp107_spectrum}.

\subsection{TEATRO data reduction setup}\label{Sec:TEATRO}

We processed the data using the Transiting Exoplanet Atmosphere Tool for Reduction of Observations (\texttt{TEATRO}) that runs the \texttt{jwst} package, extracts and cleans the stellar spectra and light curves, and runs light curve fits. In the \texttt{jwst} Detector1 pipeline, we use the same dark and linearity corrections as in \texttt{CASCADe} by overriding the default reference files. We subtracted the background per integration and per detector row and corrected for flagged pixels. We extracted the stellar spectra by summing the flux in a 12 pixel wide aperture, summed them between 4.61\,--\,11.83 $\mu$m to obtain the band-averaged light flux, and binned them in 51 wavelength bins from 4.61 to 11.83\,$\mu$m (bin width of 0.15\,$\mu$m) to obtain the spectroscopic light curves. We discarded the first 1.4\,hr that show a decay caused by persistence effects, normalised the light curves by the out-of-transit flux, and removed outliers. We fitted the light curves by a transit light curve model computed with the \texttt{exoplanet} package\,\citeSupp{kumar_2019, exoplanet_2021} and a linear trend. We fitted that model to the data using a MCMC procedure based on the \texttt{PyMC3} package as implemented in \texttt{exoplanet}\citeSupp{kumar_2019, exoplanet_2021}. 
We refined the mid-transit time, planet-to-star radius ratio, and inclination from a band-averaged light curve fit (Suppl.\ Inf.\ Table \ref{tab:white_fitting_result}), and let only the planet-to-star radius ratio and a linear trend as free parameters for the spectroscopic light curve fits. The limb-darkening coefficients for each spectral channel were fixed to the values used in the \texttt{CASCADe} reduction. We used the median of the posterior distributions as final parameters, and computed the transit depth uncertainties by a quadratic sum of the standard deviations of the residuals of the in- and out-of-transit points divided by the square root of their respective number of points, because it gives more conservative uncertainties than those obtained from the MCMC posterior distributions. The \texttt{TEATRO} transmission spectroscopy results are provided in Extended Data Table~\ref{tab:jwst_all_spectra} and shown in Figure~\ref{fig:wasp107_spectrum}.

\subsection{Data non-linearity correction} \label{method:non_linearity}

The adopted readout pattern for all JWST instruments, including those of the MIRI instrument, is the so-called \texttt{MULTIACCUM} readout pattern. 
The MIRI pixels are read non-destructively (charges are read but
not reset) at a constant rate until a final read followed by two resets to clear the accumulated charges. An integration thus consists of a number of samples of the accumulating detector signal, resulting in a ramp, that, when fitted, yields a measure of the flux per pixel. For a detailed discussion of the MIRI focal plane arrays and read out patterns we refer to ref.\,\citeSupp{ressler_2015}. 

The MIRI detector ramps show several non-ideal behaviours, influencing the slope derivation and thus the flux estimates. We refer to ref.\,\citeSupp{morrison_2023} for a review of all detector effects influencing the sampling of the detector ramps and their mitigation in the JWST data reduction pipeline. The two main non-linearity effects which are important for transit observations are the reset switch charge decay\,\citeSupp{morrison_2023} and the debiasing effect in combination with a diffusion of electrons to neighbouring pixels\,\citeSupp{morrison_2023, argyriou_2023}. While the former affects mainly the first few reads of the detector ramps, and can be mitigated by not using the affected reads when determining the slope of the detector ramps, the latter effects need to be corrected before the slope of the detector ramps can be correctly measured.  
For a detailed discussion on the detector voltage debiasing and related effects, see ref.~\citeSupp{argyriou_2023}. In brief, a detector circuit as used in MIRI can be seen as a resistor-capacitor circuit. Charge accumulation at the integration capacitors reduces the net bias voltage, which in turn leads to a lower response of the detector as it causes the width of the depletion region to shrink below the active layer width, and
a smaller fraction of the produced photoelectrons are guided to the pixels. The diffusion of photo-excited electrons in the undepleted region of a (near) saturated pixel to the depleted region at neighbouring pixels -- dubbed the brighter-fatter effect\,\citeSupp{argyriou_2023} -- can be observed in the WASP-107b data but only at a low level, as the maximum observed signal level of the detector ramps remains well below the saturation limit.
The main effect of the electron diffusion in the WASP-107b data is an additional loss of electrons in the central pixels of the spectral point spread function (PSF), in combination with a small gain of electrons in the neighbouring pixels in the wing of the PSF. {\bf As we will show in the following, parametric 
model can still be used for this data set to derive an \emph{effective} debiasing of the detector pixels and properly linearize the detector ramps, mitigating the combined effects of detector debiasing plus electron diffusion.} We, therefore, ignored the electron diffusion effect in our analysis, and focused on correcting the main detector ramp non-linearity due to debiasing.

The standard correction for the non-linearity of the detector ramps due to the debiasing effect, implemented in the \texttt{linearity} step of the JWST data reduction pipeline, is derived by fitting a cubic polynomial to the detector ramps of dedicated calibration data, and using the linear term as an estimate of the linearised signal of the detector ramp. A functional relation is then determined between linearised signal and observed signal using a fourth-order polynomial. The polynomial coefficients from this fit are stored in the CRDS calibration file for the \texttt{linearity} pipeline step. Note that the standard linearity correction implements an identical correction for all detector pixels in the MIRI/LRS subarray. Also note that the standard correction was derived using data from an spatially extended illumination source, which results in data not influenced by electron diffusion as there is no significant electrical field differences between neighbouring pixels.

To test the default linearity correction (pmap version 1030), we checked the behaviour of the detector ramps by creating pair-wise differences of the readouts (frames or groups in case of MIRI). In case of a perfect linear ramp, the pair-wise differences of a detector ramp for a single detector pixel should have a constant value. Panels a, d, g, j and m (the left column) of Suppl.\ Inf.\ Figure~\ref{fig:fig_linearity} displays the pair-wise differences of the uncalibrated data (\texttt{uncal} data product), clearly showing non-constant values for those pixels receiving the highest photon flux. Note that the slope change of the first few differences is dominated by the RSCD, and the last pair by the last-frame effect. Applying the default linearity correction substantially improves the linearity of the ramps but a slope can still be seen when plotting the pair-wise differences in panels b, e, h, k and n (the second column ) of Suppl.\ Inf.\ Figure~\ref{fig:fig_linearity}, indicating that the default correction is not yet optimal. As non-linearity effects can have a substantial impact on the derived transit depth, we derived an alternative linearity correction based on the data itself. We fitted the following {\bf parametric model} to the detector ramps

\begin{multline*}
	\mathrm{S_{ij}}(t) = a_{ij,0} + \tau_{ik,1} \cdot a_{ij,1} \cdot \left(1 - \mathrm{e}^{ \frac{-t} {\tau_{ij,1}} }\right) - \tau_{ij,2} \cdot a_{ij,2} \cdot \mathrm{e}^{ \frac{-t}{\tau_{ij,2}} } \\ \quad i \in \{0, \dots, 415\},
	\quad j \in \{0, \dots, 72\},  \quad  0 \le t \le \mathrm{T}_{\mathrm{int}}\label{eq:ramp}
\end{multline*}

In this equation, $t$ is the time between $0$ and the duration of a single integration $\mathrm{T}_{\mathrm{int}}$. The first term represents the debiasing effect, with $a_{ij,0}$ the reset level for a single pixel with detector row index $i$ and column index $j$, $a_{ij,1}$ and $\tau_{ik,1}$ the linearised slope of the detector ramp and {\bf the time constant combined effects of the detector debiasing plus electron diffusion}, respectively. The second term models the RSCD effects with $a_{ij,2}$ and $\tau_{ij,2}$ the amplitude and time constant for the estimate of the RSCD effect. Though we will not use the fitted contribution of the RSCD effect in this study, we included the term in the fit to ensure we obtained an unbiased estimate of the {\bf combined debiasing and electron diffusion effects}. Using this model, we fitted the detector ramps after applying the \texttt{reset} pipeline step, for all integrations after the transit. Using the fitted estimate of the linearised signal, we followed the procedure described in  ref.~\citeSupp{morrison_2023} to derive a custom non-linearity correction used in the \texttt{linearity} pipeline step. Panels c, f, i, l and o (the right column) of Suppl.\ Inf.\  Figure~\ref{fig:fig_linearity} show the slope estimates of the linearised ramps using our custom non-linearity procedure. 
One can see that our custom linearisation improves the linearity of the detector pixels in the detector column at the centre of the spectral trace. 
To check the linearisation in the of the detector signals in the direction across the spectral trace, we show in Suppl.\ Inf.\  Figure~\ref{fig:fig_linearity_cross_cut} our results for several detector columns across the spectral PSF for detector row 385, which corresponds to the shortest wavelengths in our spectra.  Comparing our results shown in this latter figure to the linearised ramps using the standard calibration, one can observe again a substantial improvement in the linearity of the ramps. 
Note that for detector pixels with a row number below 305 (equivalent to wavelengths beyond approximately 8~$\mu$m), which see a sufficiently low signal, no differences can be observed between our calibration and the standard CRDS linearisation. This is expected, as the detector ramps for those pixels are expected to be (near) linear.

\begin{figure}[!htp]
\begin{center}
	\includegraphics[width=0.9\textwidth]{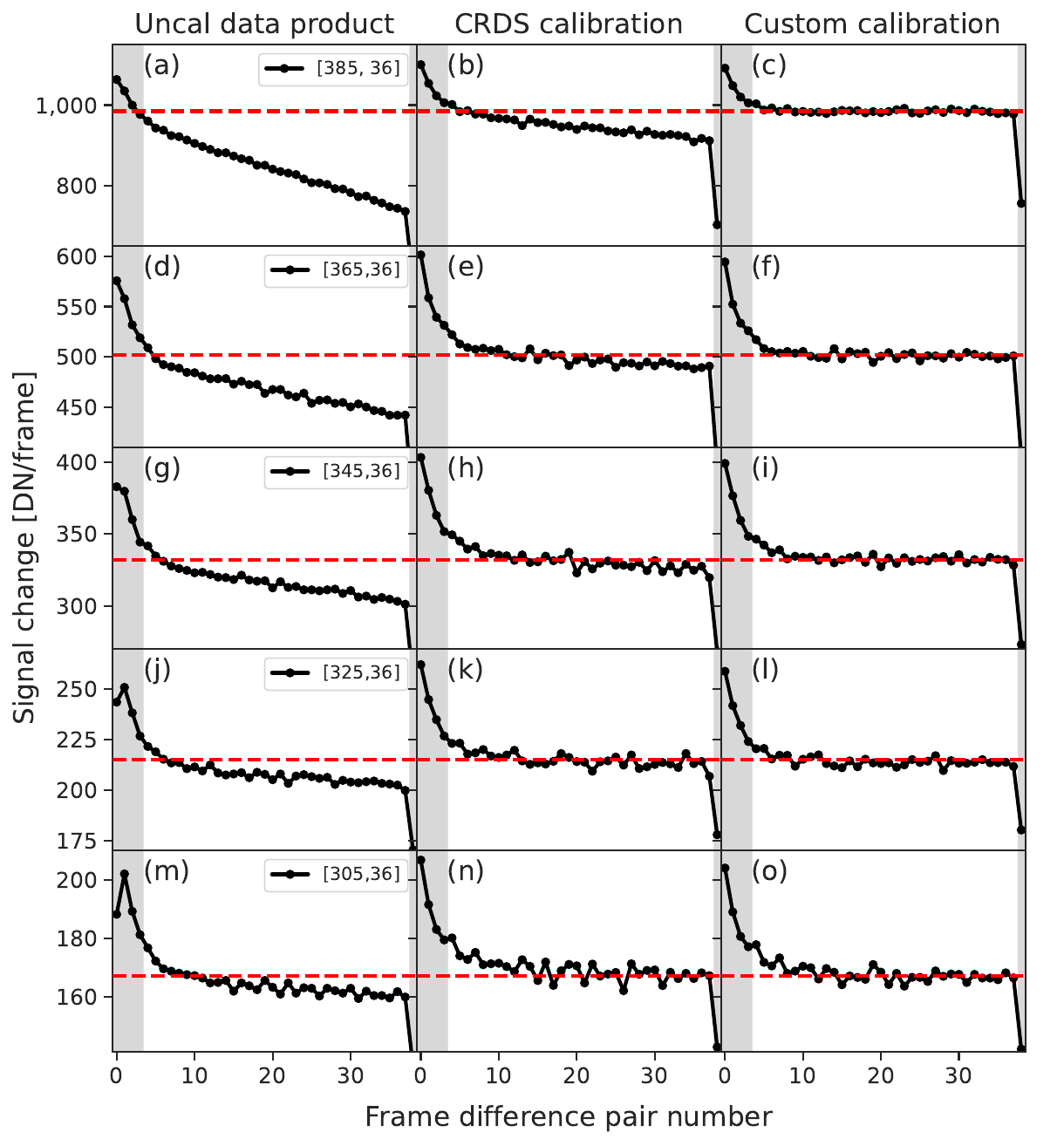}
 \end{center}
	\caption{\textbf{Linearity of the detector ramps for selected detector pixels along the spectral trace.} Shown are the pair-wise differences of the samples of the detector ramps for a number of detector pixels. From left to right are shown the ramp gradients for the \texttt{uncal} data product, the standard \texttt{reset}, \texttt{dark} and \texttt{linearise} processed data using the calibration files from CRDS with pmap version 1030, and the data product using a custom calibration for the \texttt{linearise} and \texttt{dark} calibration steps. From top to bottom are shown the data for 5 detector pixels corresponding to the maximum signal in the spectral trace of WASP-107 at different wavelengths. The pixel indices are indicated in the legends shown in the left column. The shaded regions indicate the data not used in the final determination of the slopes of the detector ramps. The dashed lines are plotted to guide the eye and represent the average linear slope after applying our custom calibration.}
	\label{fig:fig_linearity}
\end{figure}

\begin{figure}[!htp]
\begin{center}
	\includegraphics[width=0.9\textwidth]{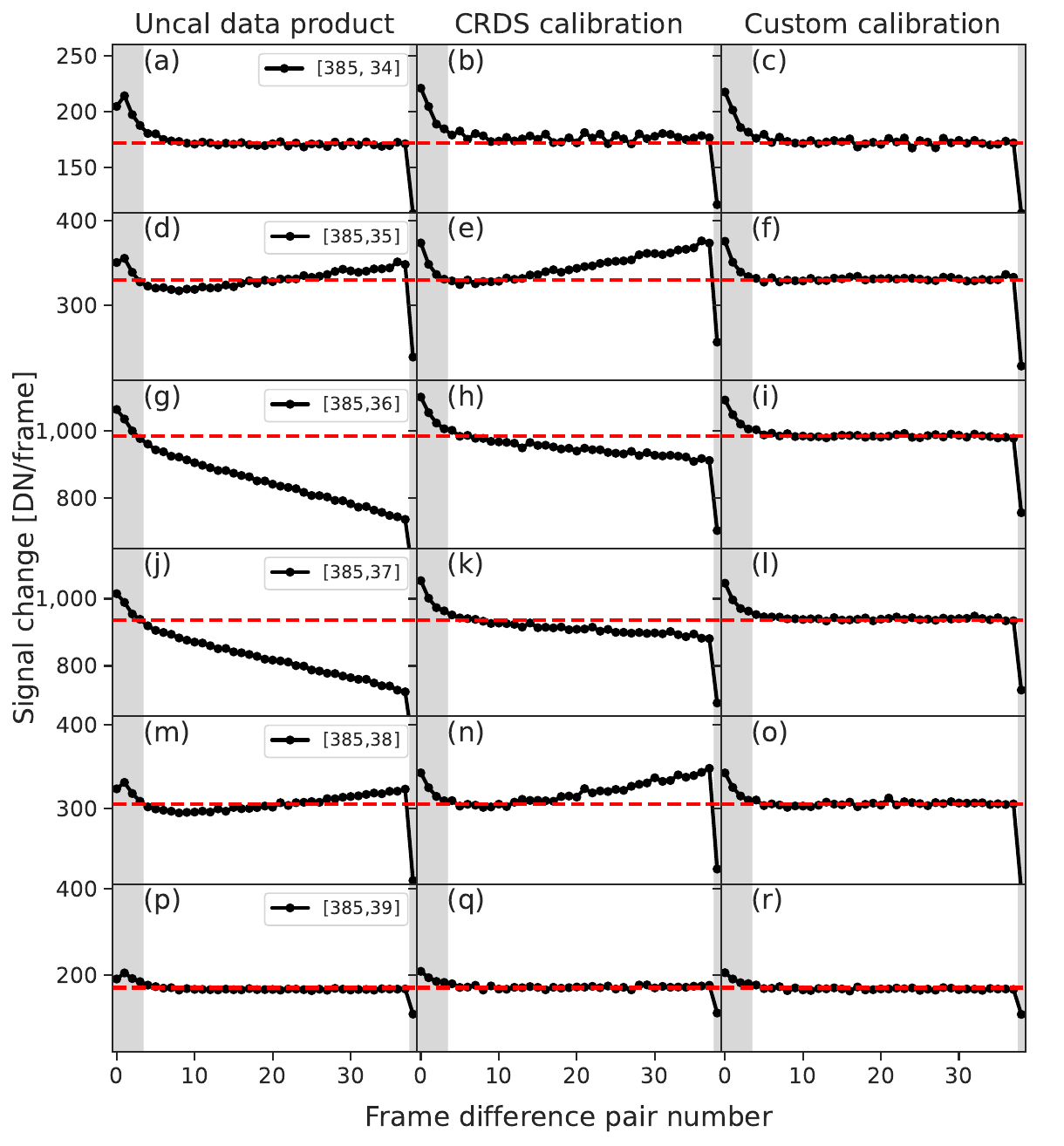}
 \end{center}
	\caption{\textbf{Linearity of the detector ramps for selected pixels across the spectral trace.} 
    The data shown in this figure is similar to Suppl.\ Inf.\  Figure~\ref{fig:fig_linearity} but now for 6 detector pixels corresponding to a cross-section (from top to bottom, detector columns 34 to 39) of the spectral trace of the dispersed light at detector row 385, the latter corresponding to the shortest wavelength in our spectra.  Note that panels g,h,and i correspond to panels a,b, and c of Suppl.\ Inf.\  Figure~\ref{fig:fig_linearity}.}
    \label{fig:fig_linearity_cross_cut}
\end{figure}

Another test to check the non-linearity correction of the detector ramps is to look at the FWHM of the spectral trace. As the central detector pixels in the spectral trace see a stronger signal, they will be subject to a stronger non-linearity, leading to a broadening of the point spread function of the individual readouts of the detector ramps during an integration\,\citeSupp{argyriou_2023}.  Suppl.\ Inf.\  Figure~\ref{fig:fig_FWHM} shows our estimates of the FWHM of the spectral trace for frame difference pairs along the detector ramp. Panels~(a) and (c) show the average FWHM of the spectral trace for frame difference pairs 5 to 10, which are the first samples not substantially influenced by the RSCD effects, and the frame difference pairs 34 to 38, respectively. The data calibrated using the standard calibration (panel~(a)) clearly shows a broadening of the point spread function (PSF) during an integration. The custom calibrated data, however, shows no such effect (lower left panels). Panels (b) and (d) of Suppl.\ Inf.\  Figure~\ref{fig:fig_FWHM} show the average FWHM as a function of frame difference pair for the detector rows 382 to 386, which sample the shortest wavelengths and receive the highest photon flux from the target. Again, the detector data calibrated with the standard calibration shows a broadening of the PSF during the sampling up the ramp (panel~(b)) while no such effect can be observed for the data calibrated with our custom calibration (panel~(d)). The shaded grey regions in the right panels indicate the data not used in the final determination of the slopes of the detector ramps, as those points are strongly affected by the RSCD and last-frame effects.

\begin{figure}[htp]
    \begin{center}
        \includegraphics[width=0.95\linewidth]{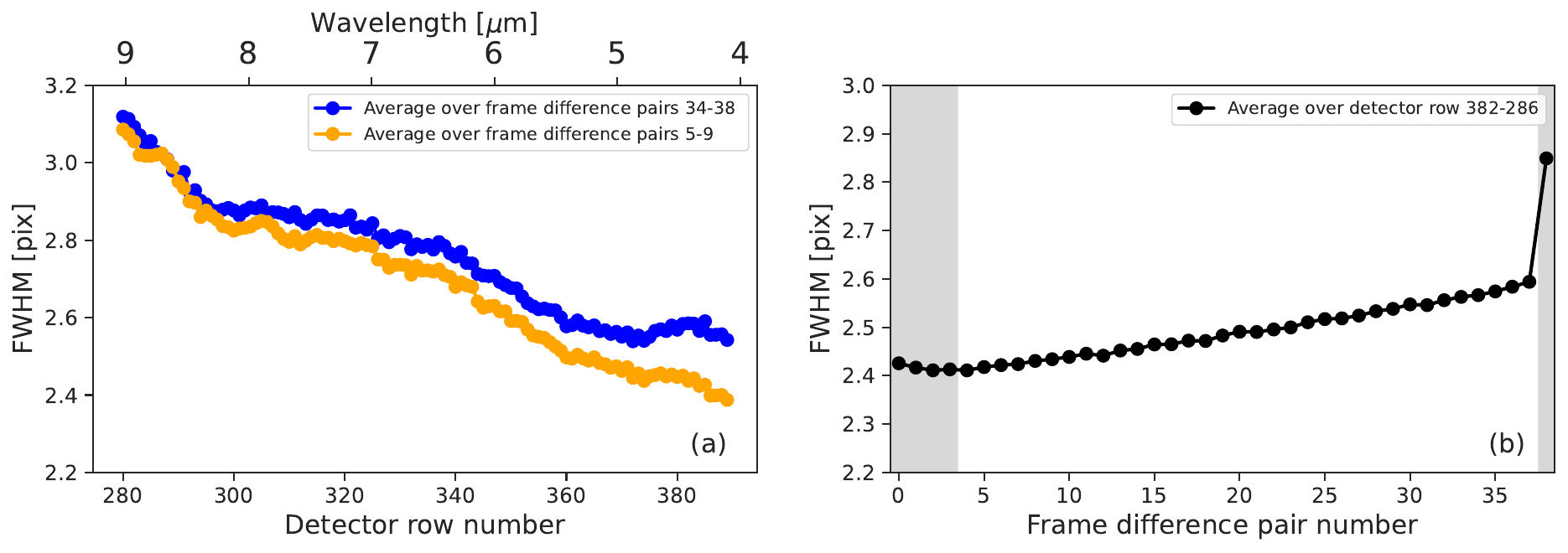} \\
		\includegraphics[width=0.95\linewidth]{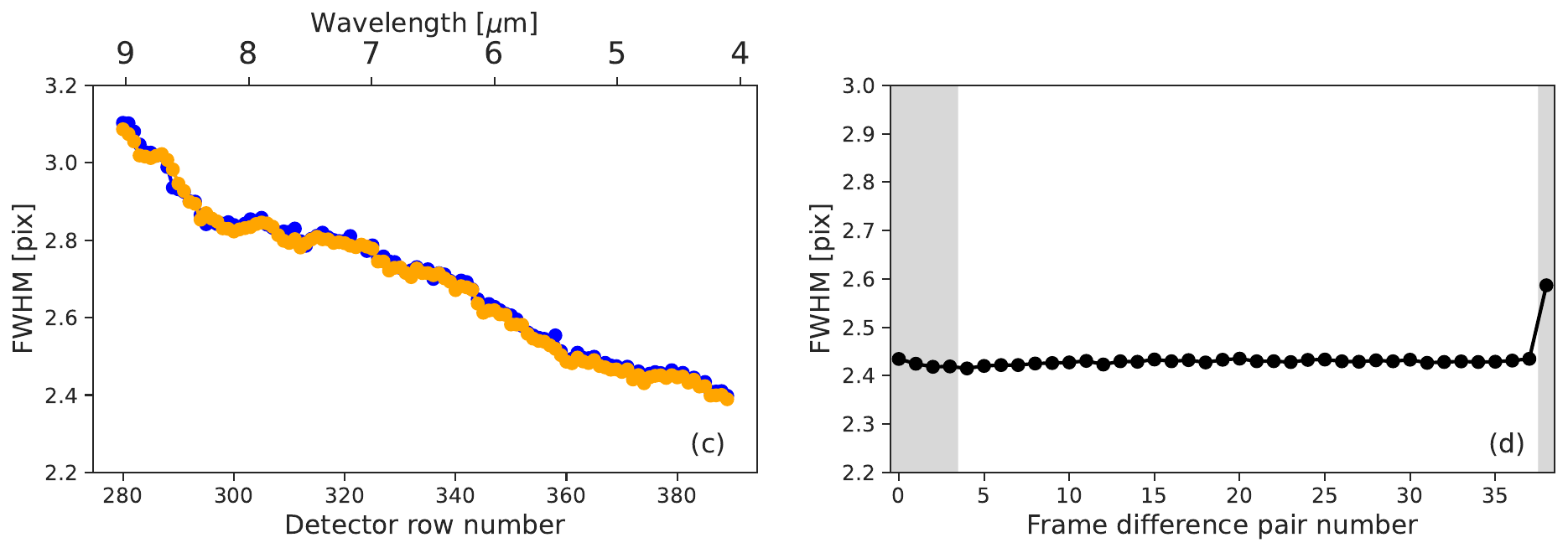}
	\end{center}
	\caption{\textbf{FWHM estimates of the spectral trace for different detector ramp frames.} 
 Panels (a) and (b) show the results for the standard calibrated detector ramps while panels (c) and (d) show the results from our custom calibrated data. Panels (a) and (c) show the average FWHM of the spectral trace for detector rows 280 to 390, for frame difference pairs 5 to 10, and 34 to 38, respectively. Panels (b) and (d) show the average FWHM as a function of frame difference pair number for detector rows with the highest signal. The shaded regions in the right panels indicate the data not used in the final determination of the slopes of the detector ramps.
	\label{fig:fig_FWHM}}
\end{figure}

Finally, Suppl.\ Inf.\  Figure~\ref{fig:fig_FWHM_comp} shows the FWHM of the brightest pixels as a function of time. As evident in that figure, the data calibrated using the standard CRDS calibration shows a drop of the derived FWHM during the transit. The drop in the observed signal during the transit of about 2\% is clearly enough to have a measurable effect on the photometric signal in case the non-linearity of the detector ramps is not properly corrected. Applying our custom calibration for this dataset, no significant effect of the transit on the FWHM estimate can be observed.   

\begin{figure}[htp]
	\begin{center}
		\includegraphics[width=.7\textwidth]{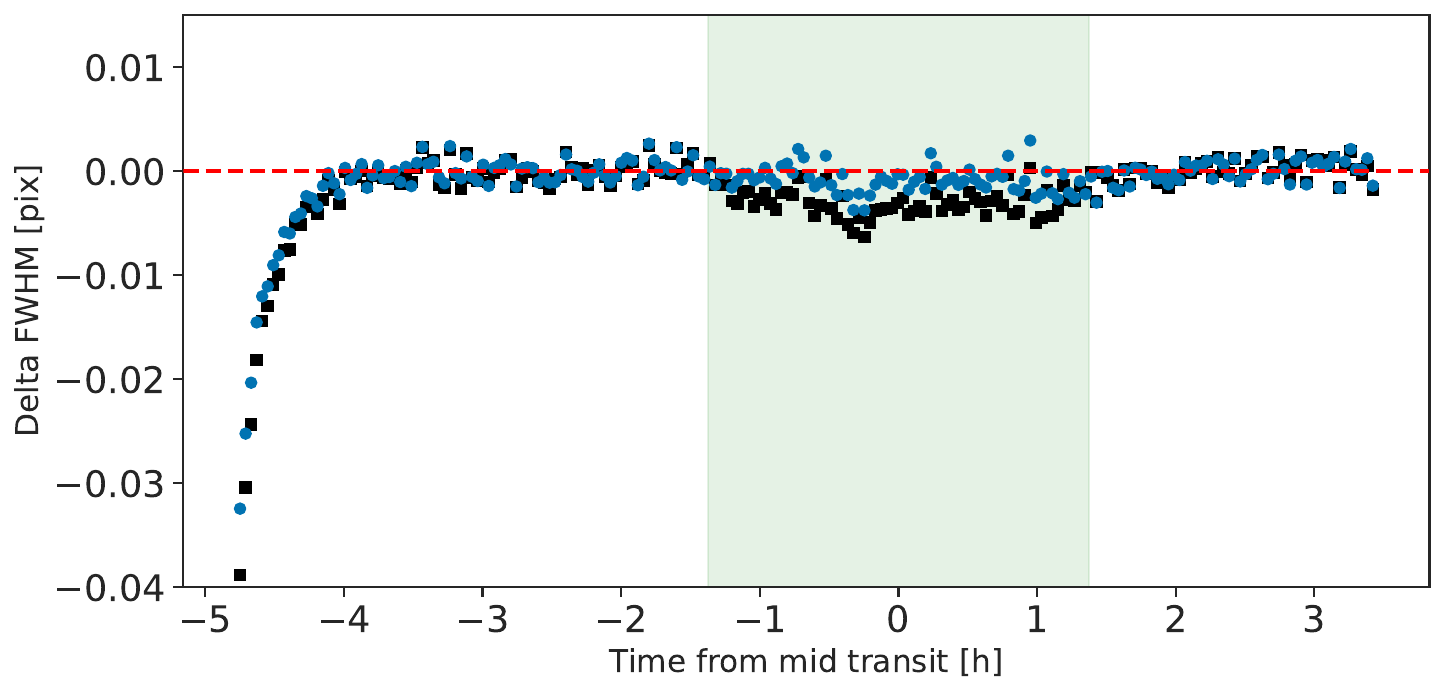}
	\end{center}
	\caption{\textbf{Mean FWHM of the spectral trace for the detector rows 380 to 390.} The blue squares show the FWHM as a function of time after applying the standard calibration from CRDS, while the black dots show the measured FWHM after using our custom non-linearity correction. To show the non-linearity effects more clearly, each data point represents an average of 22 integrations. The dashed line is plotted to guide the eye and represents zero variations. The shaded area indicates the time window where the transit occurs.
	\label{fig:fig_FWHM_comp}}
\end{figure}

\subsection{Comparison between the three JWST MIRI data reduction setups}\label{Sec:comparison_reduction}

To assess the quality of our data reductions and to identify possible biases between the 3 applied data reduction packages, we compared the uncertainty estimates and the differences in the derived transit depths.

\begin{figure}[htp]
    \centering
    \includegraphics[width=.7\textwidth]{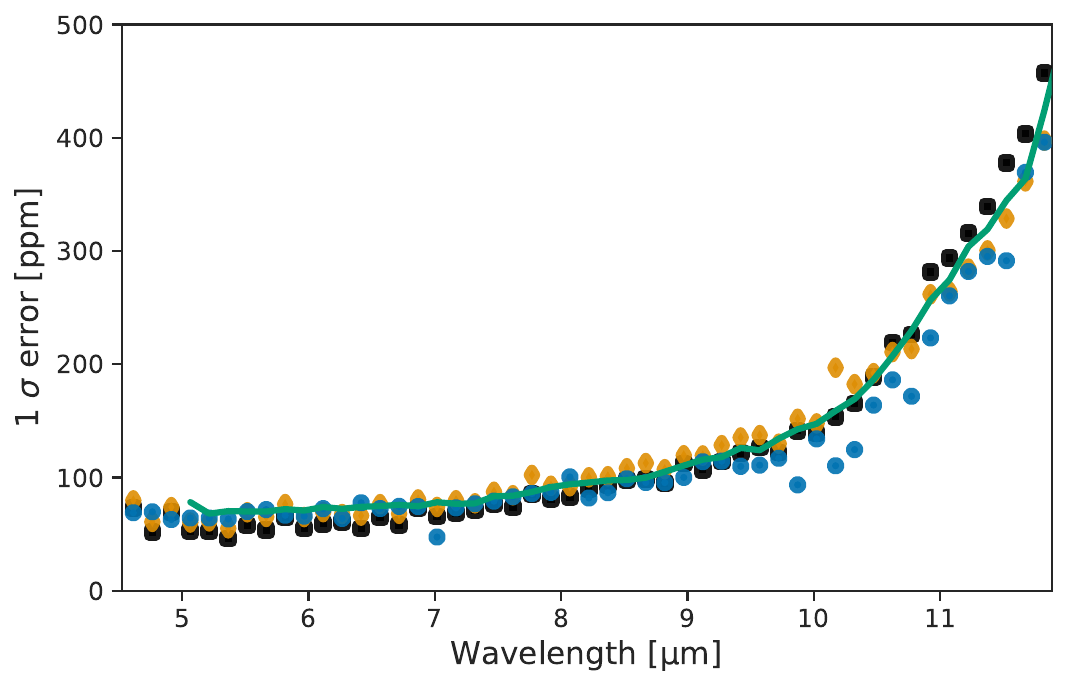}
    \caption{\textbf{1$\sigma$ uncertainties on the transit depths as a function of wavelength for the three data reductions.} The blue dots, orange diamonds and black squares show, respectively, the error estimates using the \texttt{CASCADe}, the \texttt{Eureka!}, and the \texttt{TEATRO} codes. The solid line shows the photon, dark and read noise limited performance estimate based on ETC calculations for comparison. This plot displays the performance of the MIRI/LRS instrument and the reliability of the three data reduction methods.}
    \label{fig:comparison_noise}
\end{figure}

We found that with a single-transit observation we reached a spectrophotometric precision of $\sim$80 ppm in the 7\,--\,8\,$\mu$m range at a spectral resolution $R$\,=\,50 (see Suppl.\ Inf.\ Figure~\ref{fig:comparison_noise}). We used the JWST Exposure Time Calculator (ETC)\,\citeSupp{pontoppidan_pandeia_2016} to estimate the signal-to-noise ratio on a single integration. Using this estimate,  we simulated the light curves per spectral channel assuming a constant transit depth equal to the observed band-averaged transit depth. The simulated light curves were then fitted using the \texttt{CASCADe} package to estimate the error of the simulated transit spectrum. This estimate is shown as the solid curve in Extended Data Figure~\ref{fig:comparison_noise}. All three data reductions are consistent with this estimate, indicating that our results are close to the photon noise limit of the instrument. Note that the noise limit estimate based on the ETC still contains uncertainties about the exact value of the detector gain and thus photon conversion efficiency and the level and modelling of the infrared background at longer wavelengths.
The largest differences between the error estimates are observed at wavelengths beyond 10~$\mu$m, which is expected as the signal to noise of the data rapidly drops beyond this wavelength.   

Our derived band-averaged transit depths are 20,463\ $\pm$\ 39\ ppm (see also Figure~\ref{fig:wasp107_spectrum}),  20,552\ $\pm$\ 17\ ppm, and 20,566\ $\pm$\ 33\ ppm, for \texttt{CASCADe}, \texttt{Eureka!}, and \texttt{TEATRO}, respectively. These values  are within 1$\sigma$ of the previously measured transit depth at near-infrared wavelengths with the HST (see Sect.~\ref{Sec:HST}), and well within 3$\sigma$ from each other, showing that all 3 methods give a consistent estimate of the overall transit depth.

For the comparison of the 3 derived transit spectra, we calculated 
the difference between pairs of data using a different reduction method as
\begin{equation}
\label{eq:sigma_differene}
\frac{\rm{TD}_1(\lambda) - \rm{TD}_2(\lambda)}{\sqrt{\rm{err}_1^2(\lambda) +  \rm{err}_2^2(\lambda)}},
\end{equation}
with TD$_1$ and TD$_2$ being the transit depth of reduction method 1, or 2, respectively, at wavelength $\lambda$ and err$_1$ and err$_2$ being the corresponding 1$\sigma$ errors shown in Suppl. Inf.\ Figure~\ref{fig:comparison_noise}. As can be seen in Suppl.\ Inf.\ Figure~\ref{fig:comparison_data_reduction}, the three data reduction methods are within 3$\sigma$ agreement, 96\% of the points being in 2$\sigma$ agreement.  For wavelengths shorter than 10\ $\mu$m, no significant systematic deviations between the data reductions can be observed. For the longer wavelengths, a small positive offset can be seen 
in Suppl.\ Inf.\ Figure~\ref{fig:comparison_data_reduction} for the three reductions that remains, however, within 1$\sigma$ difference.

The observed systematic trend in transit depth differences for the longer wavelengths can be attributed to the different systematics models employed by the \texttt{CASCADe}, \texttt{TEATRO} and \texttt{Eureka!} reduction codes. \texttt{TEATRO} cuts the initial ramp caused by persistence effects at the beginning of the observation and fits only a linear trend, \texttt{Eureka!} includes both an exponential model at the beginning of the observation and a polynomial one that fits any bending of the light curve, and in the \texttt{CASCADe} analysis, the initial response drift caused by persistence effects is also removed and the fitted systematics model is constructed from the data itself (see ref.\,\citeSupp{Carone2020MNRAS.496.3582C}) using the causal connection between the different wavelength channels in addition to the time, trace position and FWHM. Small differences in the curvature of the baseline will then translate in small differences of the fitted transit depth.
{\bf In general, however, these results demonstrate that all three reduction methods are compliant with each other.}

\begin{figure}[htp]
    \centering
    \includegraphics[width=\textwidth]{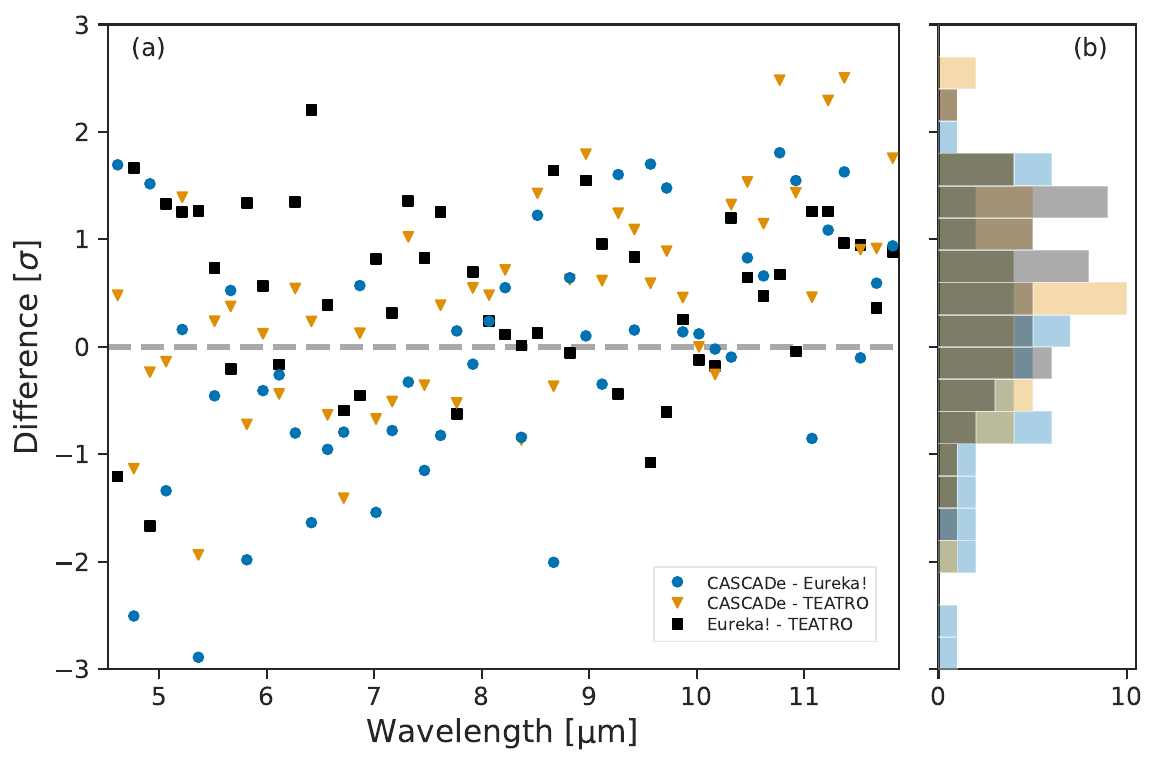}
    \caption{\textbf{Transit depth differences between the three data reduction methods as a function of wavelength.} In panel (a), the blue dots show the differences in units of $\sigma$ between the \texttt{CASCADe} and \texttt{Eureka!} reductions, the orange triangles the differences between the \texttt{CASCADe} and \texttt{TEATRO} reductions, and the black squares the differences between the \texttt{Eureka!} and \texttt{TEATRO} reductions as computed from Eq.~\ref{eq:sigma_differene}. The histograms in panel~(b) show the number of points in agreement within the different $\sigma$ ranges, with colours identical to those in panel (a).}
    \label{fig:comparison_data_reduction}
\end{figure}

\section{Ancillary data}\label{secA1}

\subsection{NUV data}\label{Sec:NUV}
Contemporaneously with the JWST observations, from 2023 January 5 to 29, {\it Swift} conducted a `Target of Opportunity' (ToO) observing campaign (Target Id.\ 15428) for WASP-107, with the UVOT\,\citeSupp{roming_swift_2005} as the primary instrument, and utilising the uvm2 filter to optimise the waveband definition and avoid redward `leaks' present in uvw2\,\citeSupp{Salz2019A&A...623A..57S}. The uvm2 filter has a central wavelength of 2246 \AA\ and a FWHM of 498 \AA\,\citeSupp{poole_photometric_2007}. The observing campaign consisted of 13 {\it observation segments} comprising a total of 20 {\it snapshots} (i.e.\ continuous exposure periods). Each segment was typically $\sim1.5$\,--\,2\,ks in duration; with snapshots ranging from the full segment length down to $\sim500$\,s. All observations were performed in {\it full imaging} mode, i.e.\ the snapshot duration was the maximum available time resolution.

Data from {\it Swift} observations are automatically processed by the {\it Swift}-project pipeline, and placed in an online publicly-accessible archive. The required data products, all FITS-format files, were downloaded from the archive, on 2023 February 22. These UVOT data products were, for each of the 13 observation segments, the segment image file summed over the snapshots in the segment (1 or 2 in the present case), the snapshot image file containing the individual snapshot images and the detected sources catalogue table. The photometry presented in the images is in units of recorded counts/pixel, where 1 pixel = $1 \times 1$ arcsec$^2$. The ancillary data and visual inspection of the snapshot-level images, indicated that one snapshot (segment-9, snapshot-1) had an aspect-solution problem. These data were excluded from the associated segment image and from further consideration in our analysis, and had been excluded from the automatic pipeline processing. All the following results reported here were based on the segment-level images, i.e.\ we have available 13 photometry values. We verified that, for the seven segments containing two snapshots, the photometry values were consistent within the statistical errors.

The information in the pipeline-generated source catalogue included, for each detected source, sky-coordinates and photometric values, the latter at successive levels of correction, from `raw' counts through to PSF-corrected isophotal flux densities. The pipeline source detection employs the {\it Swift} tool \texttt{uvotdetect}, which in turn invokes the SourceExtractor (SE) package\,\citeSupp{bertin_sextractor_1996} to perform source detection and characterisation, including isophotal signal extraction. For WASP-107, we identified, with no ambiguity, the relevant row of the source table based on an estimated epoch=J2023 position using coordinates and proper motions from CDS-SIMBAD. The UV coordinates for all segments lay within 1 arcsec of the estimated optical stellar location and within 0.5 arcsec of the mean UV position. The data were analysed interactively using the {\it Swift} software tools in \texttt{HEASoft} 6.31.1 and the latest available calibration files (CALDB dated 2021-11-08), with ds9 to display the images, and TOPCAT/STILTS\,\citeSupp{taylor_topcat_2005} to manipulate and view the source-catalogue tables. As recommended by the {\it Swift} project, we used the \texttt{uvotmaghist} tool, with a source-data extraction radius of 5 arcsec centred on the mean UV position, to perform aperture photometry for WASP-107 on the 13 segment images. We used an annular background region with the same centre, and inner and outer radii of 20 and 40 arcsec, respectively. We determined by inspection of the UVOT source detections and visually on the images, that the selected background region was free of contamination from nearby sources, and the inner radius was sufficiently removed from the target source to avoid significant contamination.

All 13 aperture-photometry values are consistent within the statistical errors (which dominate the overall errors, as reported by \texttt{uvotmaghist}), with a reduced chi-square $\chi^2/{\rm{dof}} \sim 1$ about the mean (with the degrees of freedom, dof, being 12); and at $\sim$10\%, the sample standard deviation was comparable with the 1$\sigma$ error on the individual data values. The source count rate from individual segments was $\sim$0.1$\pm$0.01\,ct/s. The mean flux density received at Earth distance was $1.08 \pm 0.03\ \rm erg\ cm^{-2}\ s^{-1}\ \AA^{-1}$, corresponding to a luminosity of $5.4\ \rm erg\ s^{-1}\ \AA^{-1}$ and a flux density incident on WASP-107b of $6.4\ \rm erg\ cm^{-2}\ s^{-1}\ \AA^{-1}$. We found good agreement between the flux values from \texttt{uvotmaghist} aperture photometry and \texttt{uvotdetect}/SE isophotal extraction. In making the conversion from instrumental count rate to calibrated flux values, \texttt{uvotmaghist} and \texttt{uvotdetect} assume a gamma-ray-burst-type spectrum, given the prime objective of the mission. However, the difference for a cool-star spectrum is expected to be no more than $\sim$15\%\,\citeSupp{breeveld_updated_2011}. Given the proximity of WASP-107 to Earth ($\sim$65\,pc) and relatively high galactic latitude ($\sim$52~deg), we have not attempted to make any allowance for extinction along the line-of-sight. We note that the NUV irradiance of WASP-107b by its host star is (by chance) comparable (within a factor $\sim2$) with that of the Earth by the Sun\,\citeSupp{woods_solar_2009}, the larger separation of the latter pair being offset by the Sun's hotter and larger-area photosphere (spectral type G2~V versus K6~V). 

\subsection{X-ray data}\label{Sec:Xray}

XMM-Newton has observed the host star WASP-107 on 2018-06-22 (ObsID 0830190901) with the EPIC X-ray telescope (pn, MOS1, MOS2 instruments; all utilising the \texttt{THIN} filter)\,\citeSupp{struder_european_2001, turner_european_2001}  yielding an exposure time of $\sim$ 60\,ks in a single, continuous observation. The host star was detected in X-rays\,\citeSupp{webb_xmm-newton_2020, nortmann2018, foster_exoplanet_2022, spinelli2023}, with an X-ray flux in the order of $1 \times 10^{-14}\ \rm erg\ cm^{-2}\ s^{-1}$ in the soft X-rays, equivalent to a luminosity of $\sim (4-7) \times 10^{27}\ \rm erg\ s^{-1}$ (depending on the adopted spectral energy range) for a distance of 64.7\,pc, yielding an X-ray flux incident on WASP-107b of $\sim$5$\times 10^2\ \rm erg\ cm^{-2}\ s^{-1}$ $^[$\citeSupp{foster_exoplanet_2022}$^]$. 

The flux and luminosity values in the cited literature have a wide range with differences of up to $\sim$40\%. Therefore, we have performed our own analysis of the XMM-Newton X-ray data, using the SAS data-analysis package, to extract source (and background) counts as a function of photon energy. We binned the spectra to bins with at least 25 source counts each to allow for proper application of $\chi^2$ fit statistics. The source count-rate was $\sim$0.01 ct/s, and the time-series showed no evidence for variability. The \texttt{XSPEC} package\,\citeSupp{arnaud_xspec_1996} was used for fitting optically-thin thermal models in collisional equilibrium (coronal models) to the extracted spectra, having two temperature components representing a wider, presumably continuous distribution of plasma, and a photoelectric absorption component to account for interstellar absorption. The data from all three EPIC instruments were fitted simultaneously after removing the notoriously difficult lowest-energy spectral bins below 0.2~keV. Following ref.\citeSupp{spinelli2023}, we adopted a fixed, interstellar photoelectric absorption component equivalent to a hydrogen column density of $N_{\rm H}=2\times 10^{19}$~cm$^{-2}$ given the distance to WASP-107.

Owing to the relatively modest signal-to-noise ratio (SNR) of the spectrum, the fits converged to two classes of solutions in very different temperature regimes. We discriminated between them by requiring that the solution  fulfils the general scaling law between average X-ray stellar surface flux and the logarithmically averaged coronal temperature, using the emission measures (EM $=\int n_en_i{\rm d}V$, where $n_e$ and $n_i$ are the coronal electron and ion number densities, respectively, and $V$ is the coronal volume occupied by the plasma) of the components as weights \citeSupp{johnstone2015}.

The coronal abundances are important quantities for such a fit but the limited SNR does not allow individual element abundances to be retrieved. We therefore used one common abundance factor $Z$ for all elements with respect to their solar photospheric values (relative to H). We then stepped through a grid of fixed $Z$ values, fitting the spectrum for each $Z$, and then searching for a solution that fulfils the coronal flux-temperature scaling relation while providing low $\chi^2$ value. Such a solution exists, with a reduced $\chi^2$ value of 0.94 for $Z = 0.22$.  The formal best-fit yielded temperatures of $T_{\rm 1} = 1.69$~MK (million K) and $ T_{\rm 2} = 8.6$~MK, with an emission-measure ratio $ {\rm EM}_{\rm 2} / {\rm EM}_{\rm 1} =0.54$. The EM-weighted logarithmic average of the coronal temperatures as defined in ref.\citeSupp{johnstone2015}  $\left( \log\bar{T} = \sum_i {\rm EM}_i \log T_i/\sum_i {\rm EM}_i \right)$ is $\bar{T} = 2.96$~MK, a relatively modest temperature as expected for a low-activity star. The corresponding absorption-corrected X-ray flux at Earth in the spectral range of 0.1\,--\,10~keV is $1.76\times 10^{-14}$~erg~cm$^{-2}$~s$^{-1}$,
equivalent to a luminosity of 
$L_{\rm X} \approx 8.8\times 10^{27}$~erg~s$^{-1}$
for a distance of 64.7 pc, yielding an X-ray flux incident on WASP-107b of $\sim 9.7 \times 10^2\ \rm erg\ cm^{-2}\ s^{-1}$. 

A rotation period of $17.5 \pm 1.5$ d was derived from {\it Kepler} K2 photometry\,\citeSupp{mocnik_starspots_2017},  while the WASP-107 photometry yields an estimate of $17 \pm 1$ d\,\citeSupp{anderson_2017}. From gyrochronology modelling and the rotation period derived from the WASP photometry, an age estimate of $3.4 \pm 0.3$ Gyr has been derived\,\citeSupp{piaulet_2021}. From recent studies of the activity-age-rotation relation for cool main-sequence stars\citeSupp{johnstone2021} we would expect an X-ray luminosity in the order of $10^{28}$~erg~s$^{-1}$ for a star with a mass of $0.68~M_{\odot}$ and an age of a few Gyr.
 This matches our derived X-ray luminosity very well.

\subsection{HST data}\label{Sec:HST}

A transit of WASP-107b was observed on June 5–6, 2017 with the Wide Field Camera 3 (WFC3) instrument onboard the \textit{Hubble Space Telescope} (HST) using the 1.41\,$\mu$m Grism (G141). The data were obtained as part of the general observer program 14915 (P.I.\ L.\ Kreidberg). We refer to ref.~\citeSupp{kreidberg_2018} for details on the observations and the initial data analysis. We performed an independent calibration and light curve fitting of the HST data using the \texttt{CASCADe} package. For details on the use of \texttt{CASCADe} on HST data, see ref.~\citeSupp{carone_2021}. We ran \texttt{CASCADe} using the same orbital and stellar parameters as used for the analysis of the JWST MIRI light curve data (see Methods), except for the ephemeris, for which we used the value published in ref.~\citeSupp{ivshina_2022}. This latter value gives a mid-transit time within 28~s of the value derived by ref.~\citeSupp{kreidberg_2018}. We choose to use the value of ref.~\citeSupp{ivshina_2022} as it resulted in slightly lower residuals after subtracting the best fit light curve model. 

Before fitting the spectral light curve data, we binned the original spectral resolution of the HST/WFC3 data to a uniform wavelength grid with a spectral bin width of $0.00757 \mu$m. Of the first HST orbit, the first 6 spatial scans were not used in our analysis as they showed a very strong initial drift. For the systematics model (see ref.~\citeSupp{carone_2021} for details), the additional regression parameters were the time variable and the trace position. The derived transit spectrum is plotted in the top panel of Figure~\ref{fig:fig_HST} (blue squares). We derived a band-averaged transit depth of 20,448$\pm$79\,ppm, consistent within 1$\sigma$ of the transit depth derived from the JWST MIRI observations. The errors in the transit spectrum and band-averaged depth were estimated by performing a bootstrap analysis. For the retrieval analysis, we binned the spectrum to a slightly lower spectral resolution, with a spectral bin width of about 0.02\,$\mu$m to increase the signal-to-noise ratio per spectral channel and to ensure that each spectral bin is independent. A comparison of the spectrum derived using the \texttt{CASCADe} package to the previous published spectrum of ref.~\citeSupp{kreidberg_2018} can be seen in the lower panel of Suppl. Inf.\ Figure~\ref{fig:fig_HST}.  Both spectra are in excellent agreement with each other.  The band-averaged transit depth of ref.~\citeSupp{kreidberg_2018} is 145\,ppm, less than $2\sigma$, larger than the averaged depth we derived. This difference is consistent with the quoted error bars and can easily be explained by the large systematics and sparse time sampling of the data, in combination with the different methods used to fit the baselines of the spectral light curves.

\begin{figure}[htp]
	\begin{center}
		\includegraphics[width=.7\textwidth]{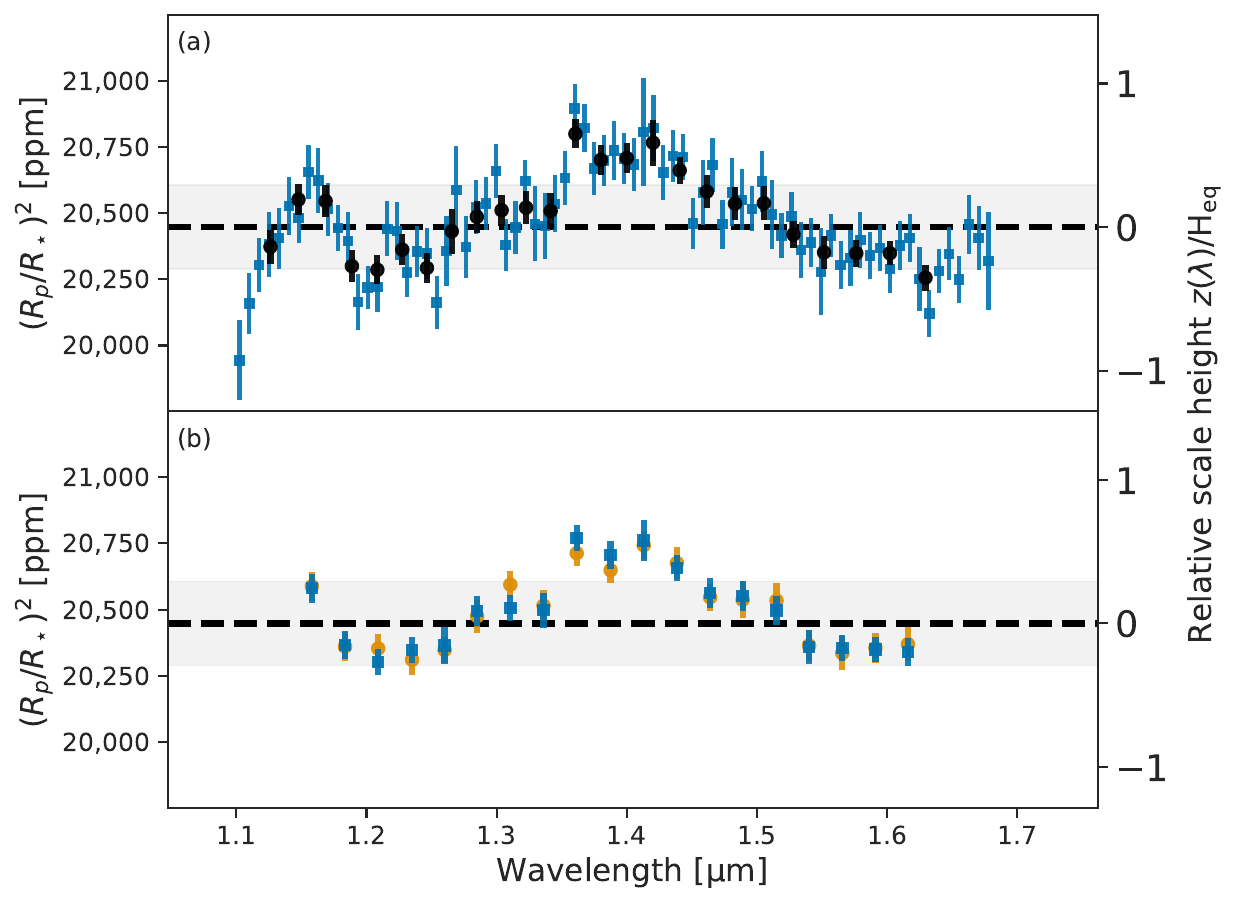}
	\end{center}
	\caption{\textbf{HST/WFC3 transmission spectrum of WASP-107b.} Panel (a) shows the transmission spectrum derived using the \texttt{CASCADe} package (blue squares) together with a slightly lower resolution version of the same spectrum used in the retrieval analysis (black dots). Panel (b) shows the  comparison between the spectrum derived by ref.~\protect\citeSupp{kreidberg_2018} (green dots) and the  \texttt{CASCADe} spectrum (blue squares), binned to the published wavelength resolution of ref.~\protect\citeSupp{kreidberg_2018}. In both panels, the band-averaged transit depth is indicated by the dashed vertical line. The shaded area represents the 95~\% confidence interval of the mean transit depth. The right y-axis gives the planetary spectrum in units of atmospheric
    scale height of the planetary atmosphere assuming it to be  hydrogen dominated. In panel~(b), the ref.~\protect\citeSupp{kreidberg_2018} spectrum was shifted downwards by 145~ppm to the same mean transit depth as found in the \texttt{CASCADe} analysis for better comparison between the two spectra.
    }\label{fig:fig_HST}
\end{figure}

\section{Retrieval analysis}\label{method:retrieval}
To constrain the atmospheric properties of WASP-107b we carried out retrievals with two different codes: \texttt{ARCiS}\,\citeSupp{min_2020} (see Sect.~\ref{Sec:ARCiS}) and \texttt{petitRADTRANS}\,\citeSupp{molliere_wardenier_2019} (see Sect.~\ref{Sec:pRT}). 

\subsection{\texttt{ARCiS} retrieval setup}\label{Sec:ARCiS}

The atmospheric modelling and retrieval code \texttt{ARCiS}\,\citeSupp{min_2020} was used to perform retrievals using a free parameterised retrieval setup. Our nominal model for \texttt{ARCiS} consists of a pressure-temperature profile with a constant value of $d\log T/d\log P$. The temperature at a pressure level of 1\,bar is retrieved. Since this is too deep in the atmosphere to be observable, we report the temperature and uncertainties derived from this profile at $P=10^{-5}\,$bar, which is the pressure level dominating the observed transit spectrum. It is important to realise that the derived temperature gradient is only representative of the uppermost atmosphere that we probe with the observed transit spectrum. We include as absorbing molecular species H$_{2}$O\citeSupp{exo_h2o}, CO\,\citeSupp{rothman_2010}, CO$_2$\,\citeSupp{exo_co2}, CH$_4$\,\citeSupp{exo_ch4}, C$_2$H$_2$\,\citeSupp{exo_c2h2}, SO$_2$\,\citeSupp{exo_so2}, SO\,\citeSupp{Bernath2022JQSRT.29008317B}, H$_{2}$S\,\citeSupp{exo_h2s}, SiO\,\citeSupp{exo_sio}, HCN\,\citeSupp{exo_hcn}, NH$_3$\,\citeSupp{exo_nh3}, and PH$_3$\,\citeSupp{exo_ph3}. The temperature and pressure dependent opacities were computed from the line lists and formatted for \texttt{ARCiS} input\,\citeSupp{chubb_2021}. For each species, a $\log$-uniform prior for the volume mixing ratio from $10^{-12}$ to $1$ was taken. The remaining atmosphere consists of $\rm H_2$ and He with a number density ratio of 0.85:0.15.

The cloud is modelled as a Gaussian layer with a certain width and optical depth at $9\,\mu$m. The specific cloud density as a function of pressure $P$ is given by
\begin{equation}
\label{eq:layerfinf}
f_\mathrm{cloud}=\frac{g\,\tau_\mathrm{cloud}}{\kappa_\mathrm{cloud}\,P\,\sigma_P\sqrt{2\pi}}\exp\left(-\frac{1}{2\sigma_P^2}\left[\log\frac{P}{P_0}\right]^2\right),
\end{equation}
where $g$ is the gravitational acceleration of the planet, $\kappa_\mathrm{cloud}$ is the cloud opacity at $9\,\mu$m. The parameters $\tau_\mathrm{cloud}$ (the cloud optical depth), $P_0$ (the cloud pressure), and $\sigma_P$ (the cloud width) are retrieval parameters. Finally, we consider partial cloud coverage using a retrieval parameter $f_\mathrm{coverage}$ between zero and one.

For the composition of the cloud particles we take a mixture of amorphous MgSiO$_3$\,\citeSupp{jager_2003}, SiO$_2$\,\citeSupp{henningmutschke_1997, palik_2012, kitzmann_2018}, SiO\,\citeSupp{wetzel_2013} and amorphous carbon\,\citeSupp{zubko_1996}. We mix the refractive indices of these materials using effective medium theory. We use the standard multi-component Bruggeman mixing rule. This mixing rule has the benefit that all materials are treated the same and there is no dominant matrix material defined (as is the case in the simpler Maxwell-Garnett mixing rule). Note that amorphous carbon provides a continuum opacity and can be considered a placeholder for any cloud component with a featureless spectrum (like, for example, metallic iron). The size of the particles, $a_\mathrm{cloud}$, is assumed constant throughout the cloud and the optical properties are computed using irregularly shaped particles simulated by the DHS (Distribution of Hollow Spheres) method\,\citeSupp{min_2005} where the irregularity parameter $f_\mathrm{max}$, which describes how far the particle shape deviates from a homogeneous sphere, is another free parameter. The prior for $a_\mathrm{cloud}$ is taken to be $\log$-uniform from 0.01 to 10\,$\mu$m. For $f_\mathrm{max}$ we take a linear prior from $0$ to $1$.

The above setup has 24 free parameters (1 for the radius, 2 for the $T$-$P$ structure, 12 molecules, and 9 for the cloud structure and particle size/shape). We add one additional parameter allowing for scaling of the HST data with respect to the JWST observations with a 0.38\% Gaussian prior corresponding to the uncertainty on the band-averaged transit depth. All parameters and corresponding prior ranges are given in Suppl.~Inf.~Table~\ref{tab:retrieval_parameters}. In the \texttt{ARCiS} retrievals we include the full HST and the MIRI spectrum. In addition to this base model we also perform retrievals where one of the molecular components is removed to test its significance. We convert the natural logarithm of the Bayes factors into a rejection significance using the formalism presented in ref.~\citeSupp{benneke_2013}. To test the significance of the clouds we perform retrievals using no clouds and one where the cloud opacity is replaced with a parameterised opacity. A full corner plot showing the posterior distribution for all retrieval parameters is shown in Suppl.~Inf.~Figure~\ref{fig:cornerplot_ARCiS}.

\begin{figure}[htp]
\centering
\includegraphics[width=\textwidth]{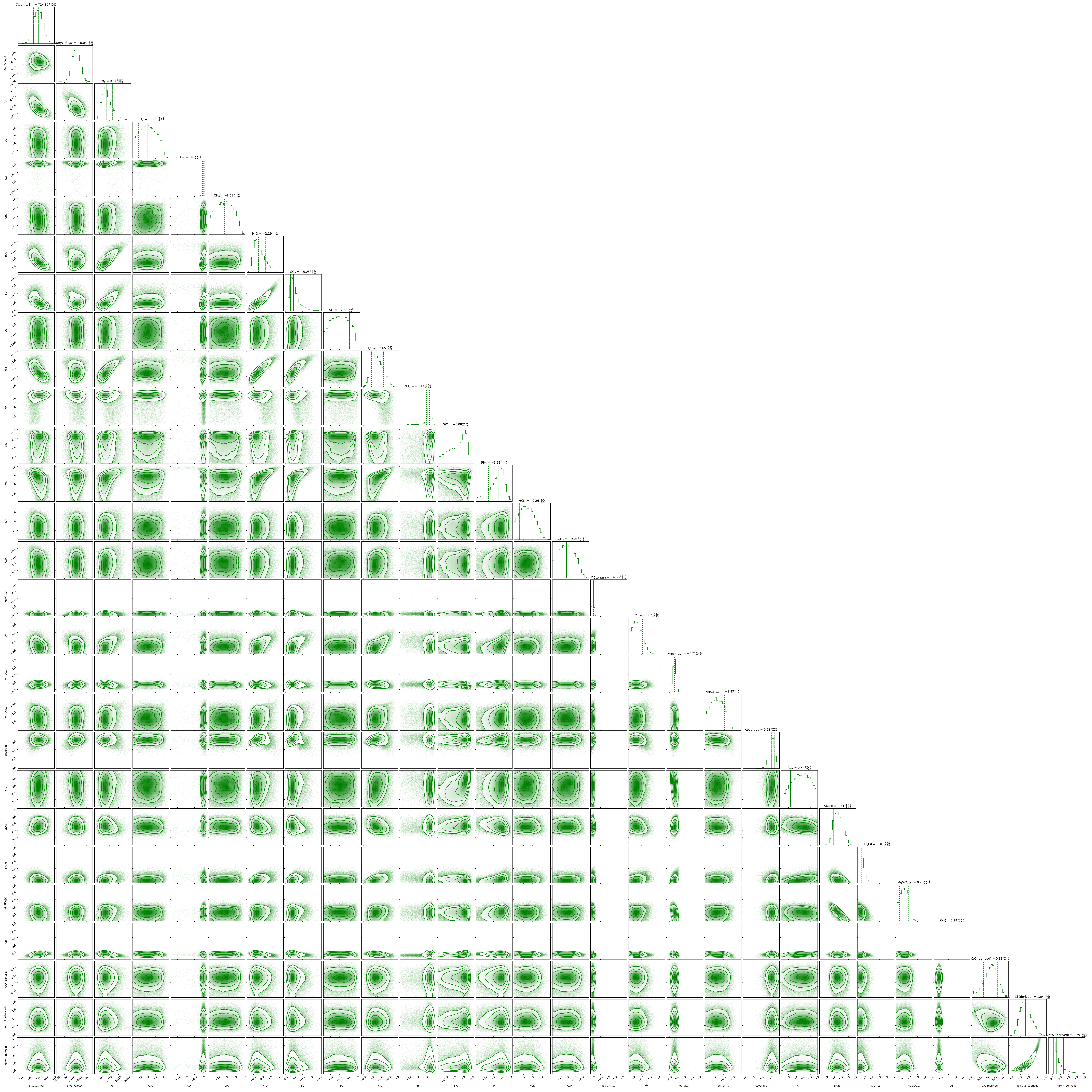}
\caption{\textbf{Full corner plot for the retrieval of the transit spectrum with the \texttt{ARCiS} setup.} The posterior distribution is shown for all retrieval parameters with the addition of the derived parameters metallicity ($Z$), C/O ratio and mean molecular weight (MMW). Gas absorber abundances are shown in logarithms (base 10) of the volume mixing ratios. Note that even though the temperature at 1\,bar was retrieved, we present here the posterior of the temperature at $10^{-5}\,$bar which is closer to the pressure layers determining the shape of the transit spectrum.}
\label{fig:cornerplot_ARCiS}
\end{figure}

\subsection{\texttt{petitRADTRANS} retrieval setup} \label{Sec:pRT}

Our nominal \texttt{petitRADTRANS} (\texttt{pRT}) forward model assumed an isothermal planetary atmosphere with a uniform prior on temperature from 200 to 2,000\,K. The following line absorber species were included: H$_{2}$O and CO\,\citeSupp{rothman_2010}, C$_2$H$_2$\,\citeSupp{exo_c2h2}, CO$_2$\,\citeSupp{exo_co2}, CH$_4$\,\citeSupp{exo_ch4}, SO$_2$\,\citeSupp{exo_so2}, H$_{2}$S\,\citeSupp{exo_h2s}, SiO\,\citeSupp{exo_sio}, HCN\,\citeSupp{exo_hcn}, NH$_3$\,\citeSupp{exo_nh3} and PH$_3$\,\citeSupp{exo_ph3}. The opacities of  all but the first two species were calculated in the \texttt{pRT} format by ref.~\citeSupp{chubb_2021}. The mass fractions of all molecules were retrieved freely, with a $\log$-uniform prior from $10^{-10}$ to 1. The remaining atmospheric gas was assumed to be in the form of $\rm H_2$ and He, at a mass ratio of 0.72:0.28. The retrieved molecular mass fractions were converted to volume mixing ratios for comparison with the \texttt{ARCiS} results. As gas continuum opacities we considered $\rm H_2$-$\rm H_2$ and $\rm H_2$-He collision induced absorption in addition to $\rm H_2$ and He Rayleigh scattering\,\citeSupp{dalgarnowilliams_1962, chandalgarno_1965,borysow_1988,borysow_1989a, borysow_1989b, borysow_2001, borysow_2002, richardgordon_2012}. The planetary gravity was retrieved using tight priors from band-averaged light curve measurements on the planet radius and from radial velocity (RV) measurements on the mass\,\citeSupp{anderson_2017}. The planet radius at the reference pressure (taken to be 0.01~bar) was retrieved as a separate free parameter, using a uniform prior from 0.7 to 2~$R_{\rm J}$. For our `complex' cloud model we included amorphous MgSiO$_3$\,\citeSupp{scottduley_1996}, SiO$_2$\,\citeSupp{henningmutschke_1997, palik_2012, kitzmann_2018} and crystalline KCl\,\citeSupp{palik_2012} clouds, considering them to be irregularly shaped (DHS method\,\citeSupp{min_2005}). The cloud mass fractions at the base of the cloud had $\log$-uniform priors from $10^{-10}$ to 1, and the cloud base pressures $P_{\rm base}$ were retrieved with a $\log$-uniform prior from $10^{-6}$ to 1,000 bar. Above the cloud deck (at lower pressures) the cloud mass fraction was defined as $X_{\rm base}\left(P/P_{\rm base}\right)^{f_{\rm sed}}$, where $X_{\rm base}$ is the mass fraction at the cloud base and $f_{\rm sed}$ is the settling parameter, defined as the cloud particles' mass-averaged ratio of settling and mixing velocities. The prior on $f_{\rm sed}$ was uniform, ranging from 0 to 10. The cloud particle sizes were then found as described in ref.~\citeSupp{ackermanmarley_2001}, namely by assuming a $\log$-normal size distribution, and making use of $f_{\rm sed}$, $K_{\rm{zz}}$, and $\sigma_{\rm g}$, where $K_{\rm{zz}}$ is the vertical eddy diffusion coefficient and $\sigma_{\rm g}$ is the width of the $\log$-normal particle size distribution. We assumed a $\log$-uniform prior from $10^5$ to $10^{13}$~$\rm cm^2 \ s^{-1}$ for $K_{\rm{zz}}$ and a $\log$-uniform prior on $x_\sigma$ from $10^{-2}$ to $1$, where $\sigma_{\rm g} = 1+2 x_\sigma$. For our `simple' cloud model we replaced the cloud extinction opacity by $\kappa(\lambda, P) = \kappa_{\rm base} \left[1 + (\lambda/\lambda_0)^{-p}\right] \left(P/P_{\rm base}\right)^{f_{\rm sed}}$, where $f_{\rm sed}$ and $P_{\rm base}$ have the same meaning and priors as before. The opacity at the cloud base was retrieved with a $\log$-uniform prior from $10^{-20}$ to $10^{20}$ $\rm cm^2 \ g^{-1}$, $\lambda_0$ with a $\log$-uniform prior from 0.01 to 100~$\mu$m, and $P$ with a uniform prior from 0 to 6. For both forward models we allowed for a multiplicative flux scaling by 0.38\% and 0.185\% (Gaussian standard deviation of prior), for the HST and JWST data, respectively, corresponding to the uncertainties on the band-averaged transit depths. In the \texttt{petitRADTRANS} retrievals we include the full HST spectrum and the MIRI spectrum. All parameters and corresponding prior ranges are given in Suppl.~Inf.~Table~\ref{tab:retrieval_parameters}. To convert the natural logarithm of the Bayes factors, $\Delta {\rm log}(Z)$, into a rejection significance we use the formalism presented in ref.~\citeSupp{benneke_2013}.
A full corner plot showing the posterior distribution for all retrieval parameters is shown in Suppl.~Inf.~Figure~\ref{fig:cornerplot_pRT}

\begin{figure}[htp]
\centering
\includegraphics[width=\textwidth]{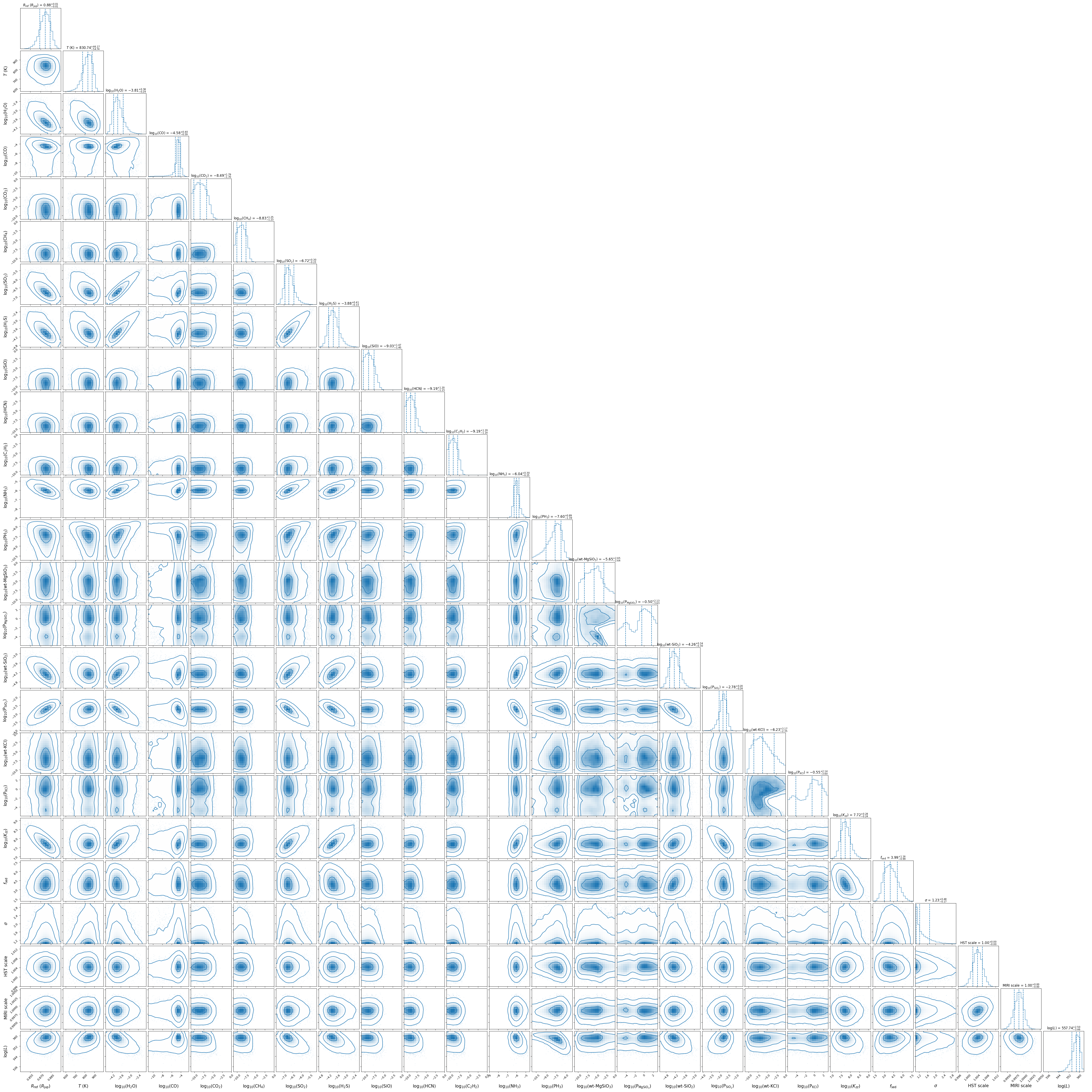}
\caption{\textbf{Full corner plot for the retrieval of the transit spectrum with the \texttt{petitRADTRANS} setup.} The posterior distribution is shown for all retrieval parameters. Gas absorber abundances are shown in logarithms (base 10) of the volume mixing ratios, while the cloud abundance at the cloud deck is given in $\log_{10}$  mass fractions.}
\label{fig:cornerplot_pRT}
\end{figure}

\begin{table*}[!htp]
	\caption{\textbf{Parameters and prior rages used in the \texttt{ARCiS} and \texttt{petitRADTRANS} (\texttt{pRT}) retrieval analysis.}}
	\label{tab:retrieval_parameters}
	\begin{center}
	\begin{tabular}{@{\extracolsep\fill}lccc}
		\toprule%
		& \multicolumn{2}{@{}c@{}}{Prior range} & Prior type
        \\\cmidrule{2-3}%
		Parameter & \texttt{ARCiS} & \texttt{pRT} \\
		\midrule
	T at $1\,$bar [K]	&	$100 - 2000$	&	$200 - 2000$	&	linear \\
	$d\log T/d\log P$	&	$-0.1 - 0.1$	&	0$^{(a)}$		&	linear \\
	$R_p\,[R_\mathrm{Jup} ]$	&$0.7 - 1.14$	&	$0.7 - 2.0$		&	linear \\	
	$\log_{10}(g)$ (cgs units)			&	2.47$^{(a)}$	&	$2.43\,\pm 0.05$	&	Gaussian \\
	Molecular abundances		&	$10^{-12} - 1$$^{(b)}$	&	$10^{-10} - 1$$^{(c)}$	&	logarithmic \\
	\midrule
	Cloud properties \\ 
	\midrule
	$P_0\,[\mathrm{bar} ]$	&	$10^{-5} - 10^{3}$	&	-		&	logarithmic \\
	$\sigma_P$		&	$0.1 - 10$		&	-			&	logarithmic \\
	$\tau_\mathrm{cloud}$&	$10^{-4} - 10^{3}$	&	-		&	logarithmic \\
	$a_\mathrm{cloud} [\mu$m]&	$10^{-2} - 10$	&	-		&	logarithmic \\
	$f_\mathrm{max}$	&	$0 - 1$		&	-			&	linear \\
	Material mass fractions	&	$0 - 1$		&	-			&	linear \\
	$f_\mathrm{coverage}$	&	$0 - 1$		&	-			&	linear \\	
	$P_\mathrm{base}\,$[bar]	per material &		-		&	$10^{-6} - 10^3$	&	logarithmic \\
	$X_{\rm base}$ per material	&	-		&	$10^{-10} - 1$	&	logarithmic \\
	$f_{\rm sed}$			&	-			&	$0 - 10$		&	linear \\
	$K_{\rm{zz}}\,[\rm cm^2 \ s^{-1}]$&	-			&	$10^5 - 10^{13}$	&	logarithmic \\
	$x_\sigma$	&	-			&	$10^{-2} - 1$	&	logarithmic \\
		\bottomrule
	\end{tabular}\\
\end{center}
{ $^{(a)}$~fixed value. $^{(b)}$~volume mixing ratio. $^{(c)}$~mass fraction.}
\end{table*}

\subsection{Silicate cloud detection significance}\label{Sec:clouds}

In order to determine the significance of the silicate cloud contribution to the retrieval we compare the Bayesian evidence to that of a retrieval performed using a parameterised cloud setup and to a retrieval using an atmospheric setup without clouds. The parameterised cloud setup uses exactly the same cloud structure but a wavelength-dependent opacity characterised by
\begin{equation}
    \kappa(\lambda)\propto(1+(\lambda/\lambda_0)^p)^{-1},
    \label{Eq:cloud}
\end{equation}
with the two parameters $\lambda_0$ and $p$ being retrieval parameters. Eq.~(\ref{Eq:cloud}) captures the expected behaviour of cloud opacities, being largely constant at short wavelengths (cut-off set by the wavelength $\lambda_0$) and having a slope at large wavelengths (set by the dimensionless parameter $p$).
The Bayes factor of the silicate cloud model with respect to the parameterised cloud model tells us if the 10\,$\mu$m silicate feature is required to fit the data. The comparison of the silicate cloud model to the model without clouds tells us if clouds are needed at all in the atmosphere. Furthermore, using \texttt{ARCiS}, we compare the cloud setup with only a single cloud component. The results are summarised in Suppl.\ Inf.\ Table\,\ref{tab:silicate_significance}. As can be seen, all setups including any silicate component (either SiO, SiO$_2$ or MgSiO$_3$) are preferred over simplified setups. The cloud containing only carbon opacity acts very similar to our parameterised opacity as it only provides a featureless continuum. It is therefore preferred over no clouds but not preferred over the parameterised setup. We also tested the significance of the silicate cloud setup for the other two data reductions, with \texttt{TEATRO} and \texttt{Eureka!}, and find also these reductions provide strong detections of silicate clouds.

\begin{table*}[!htp]
	\caption{\textbf{Significance of improvement of the fit with \texttt{ARCiS} for various cloud setups with respect to no clouds or with respect to clouds with a parameterised opacity.}}
	\label{tab:silicate_significance}
	\begin{center}
	\begin{tabular}{@{\extracolsep\fill}lccc}
		\toprule%
		Setup & \texttt{CASCADe} & \texttt{TEATRO} & \texttt{Eureka!} \\
        \midrule
	\multicolumn{4}{@{}c@{}}{With respect to no cloud} \\
		\midrule
All cloud components    &   $9.2\sigma$ &   -$^{(a)}$ &   -$^{(a)}$\\
Parameterised opacity   &   $6.0\sigma$ &   -$^{(a)}$ &   -$^{(a)}$\\
Only SiO       &   $9.7\sigma$    &   -$^{(a)}$ &   -$^{(a)}$\\
Only SiO$_2$   &   $7.7\sigma$    &   -$^{(a)}$ &   -$^{(a)}$\\
Only MgSiO$_3$ &   $8.8\sigma$    &   -$^{(a)}$ &   -$^{(a)}$\\
Only carbon    &   $5.8\sigma$    &   -$^{(a)}$ &   -$^{(a)}$\\
		\midrule
	\multicolumn{4}{@{}c@{}}{With respect to a cloud with parameterised opacity} \\
		\midrule
All cloud components    &   $7.2\sigma$ &   $7.6\sigma$ &   $5.2\sigma$\\
Only SiO       &   $7.8\sigma$    &   -$^{(a)}$ &   -$^{(a)}$\\
Only SiO$_2$   &   $5.1\sigma$    &   -$^{(a)}$ &   -$^{(a)}$\\
Only MgSiO$_3$ &   $6.6\sigma$    &   -$^{(a)}$ &   -$^{(a)}$\\
Only carbon    &   $-2.2\sigma$    &   -$^{(a)}$ &   -$^{(a)}$\\
        \bottomrule
	\end{tabular}\\
\end{center}
{ $^{(a)}$~not tested.}
\end{table*}

To investigate how sensitive our silicate cloud detection is to specific wavelength regions in the MIRI spectrum, we performed the analysis described above for either the full MIRI wavelength range and additionally using only the MIRI spectrum up to a given wavelength $\lambda_\mathrm{max}$. With this exercise we aim to establish the robustness of our retrieved cloud setup acknowledging ongoing discussions in the community on potential higher systematic errors for MIRI transit depths at longer wavelengths. We emphasise here that in our data we see no indications of any shadowed region that would increase the systematic errors for wavelengths between 10 and $\sim$12\,$\mu$m.

\begin{figure}[htp]
\centering
\includegraphics[width=\textwidth]{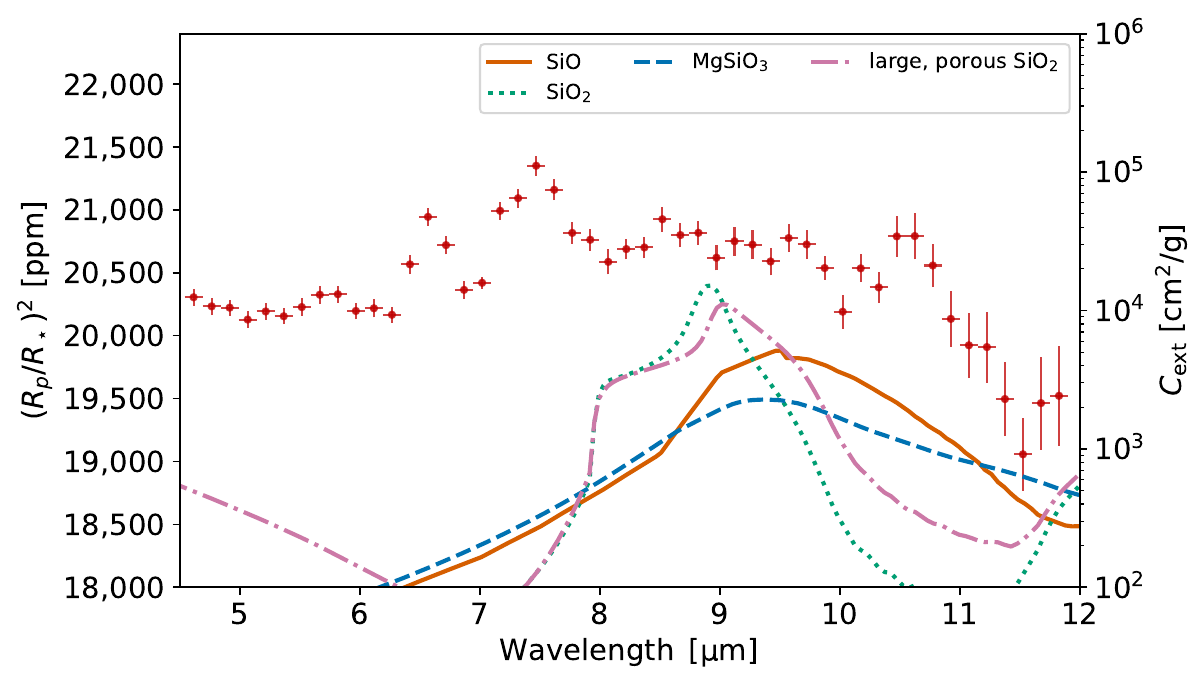}
\caption{\textbf{Comparison of the extinction coefficient of the silicate cloud particles with the transit spectrum of WASP-107b.} The extinction curves are computed for $0.01\,\mu$m solid particles for SiO (red), SiO$_2$ (green) and MgSiO$_3$ (blue), representative of the particle size found by the \texttt{ARCiS} retrievals. The pink curve is computed using a size distribution of particles between $0.1$ and $2\,\mu$m and a porosity of 0.25 (representative of the particles found by the \texttt{pRT} retrievals).}
\label{fig:opacities}
\end{figure}

It is expected that the significance of the silicate detection drops quickly if we exclude all wavelengths longer than 10\,$\mu$m because this is where the silicate feature is present (see Suppl.~Inf.~Figure~\ref{fig:opacities}). In Extended Data Figure~\ref{fig:silicate_detection} we show the resulting detection significance as a function of the maximum wavelength used in the analysis. It is clear that clouds are required no matter what wavelength range we choose. As expected, if we remove the entire wavelength range where the silicate feature is present (so wavelengths above 9.5\,$\mu$m) the silicate clouds are no longer detected. In these cases it is seen that the model prefers the setup with fewer parameters, which is the parameterised cloud setup. 

Silicate clouds are preferred with a significance of 5.7$\sigma$ even if we remove all MIRI observation with wavelengths longer than 10\,$\mu$m. This significance quickly increases if we increase the maximum wavelength used in the analysis showing that the detection of silicate clouds is a robust result.

\section{(Photo)chemical models}\label{Sec:photochemical}

\subsection{(Photo)chemical model setup}\label{Ref:photochemical_setup}
The goal of the forward (photo)chemical models is to understand the gas-phase formation of molecules in the atmosphere of WASP-107b and to derive the sensitivity of the predicted molar fractions on the model's input parameters. Since the primary focus is on the gas-phase formation of SO$_2$, CH$_4$, and H$_2$O, no cloud-formation processes have been included in these models.

In the case of WASP-107b, a tidally-locked planet with an equilibrium temperature of $\sim$740\,K, and orbiting a K6 dwarf host star \citeSupp{piaulet_2021, hejazi_2023}, it is anticipated that there will be no significant spatial gradients in the temperature structure and zonal wind speeds\,\citeSupp{Baeyens2021MNRAS.505.5603B}. Consequently, we adopt a one-dimensional configuration to examine the chemical abundance distribution within the atmosphere of WASP-107b. 

The forward chemical models\,\citeSupp{Agundez_2014ApJ...781...68A} for WASP-107b were computed considering a host star radius, R$_\star$, of 0.676\,R$_\odot$, a planet radius R$_\mathrm{p}$ of 0.94\,R$_{\rm{J}}$, and a planet mass M$_\mathrm{p}$ of 30.51\,M$_\oplus$\,\citeSupp{hejazi_2023}.
The temperature-pressure profile ($T$-$P$) has been computed using the analytical equation derived by ref.~\citeSupp{Guillot2010A&A...520A..27G}, assuming an infrared (IR) atmosphere opacity $\kappa_\mathrm{IR}= 0.01$, a ratio between  optical and IR opacity $\gamma=0.4$, an equilibrium temperature $T_\mathrm{eq} = 740$ K, and an intrinsic temperature, $T_{\rm{int}}$, in the range of 250\,--\,600\,K. Vertical mixing in 1D chemical models is commonly parameterized by eddy diffusion. However, for exoplanets, the eddy diffusion coefficient $K_{\rm{zz}}$ is loosely defined\,\citeSupp{Baeyens2021MNRAS.505.5603B}. For the 1D photochemical models used in this work, we assume a constant $K_{\rm{zz}}$, with values varying between 10$^8$\,--\,10$^{11}$\,cm$^2$ s$^{-1}$. We explore a range of C/O ratios, from solar (0.55) to sub-solar (0.1), the lower limit informed by planet formation models\,\citeSupp{Khorshid2022A&A...667A.147K} that predict a C/O ratio for the planet above $\sim$0.15. Our base model used in Extended Data Figure~\ref{fig:chemistry_Tint_COratios} has an intrinsic temperature of 400\,K, a solar C/O ratio, a metallicity of 10$\times$ solar, and a $\log_{10}$($K_{\rm{zz}}$, \rm{cgs})\,=\,10.

Our 1D chemical kinetics model treats thermochemical and photochemical reactions. The thermochemical network is based on the C–H–N–O–S network from VULCAN\,\citeSupp{tsai_2021} for reduced atmospheres containing 89 neutral C-, H-, O-, N-, and S-bearing species and 1028 total thermochemical reactions (i.e., 514 forward-backward pairs)\,\citeSupp{Baeyens2023arXiv230900573B}.  The photo-absorption cross sections are taken from the KIDA database\,\citeSupp{Venot2020A&A...634A..78V} and complemented with additional sulphur photo-absorption cross sections (O.\ Venot, private communication).  The full network cross sections were benchmarked against WASP-39b\,\citeSupp{tsai_photochemically_2023}.  

The chemical model predictions are sensitive to the flux impinging the outer atmosphere. To simulate the spectral energy distribution (SED) of  the host star WASP-107, we take the stellar spectrum of HD~85512, which is of similar spectral type (K6\,V) and for which a panchromatic SED was constructed in the MUSCLES survey\,\citeSupp{Loyd2016ApJ...824..102L}. Being both K6 dwarf stars, the bolometric luminosity of both SEDs is similar, but the chromospheric and coronal activity can differ between both stars. To assess that difference, we observed contemporaneously with the JWST observations, the Near-Ultraviolet (NUV) emission of the host star WASP-107 with {\it Neil Gehrels Swift}. We also reanalysed the X-ray emission observed with \textit{XMM-Newton} in 2018. The measured flux densities incident on WASP-107b yields a NUV flux value that is $\sim$6.4\,erg\ cm$^{-2}$ s$^{-1}$\ \AA$^{-1}$ and an X-ray flux value that is $\sim$1\,$\times$\,10$^3$\,erg\ cm$^{-2}$ s$^{-1}$; see Sect.~\ref{Sec:NUV}\,--\ref{Sec:Xray}. The folding of the MUSCLES HD~85512 spectrum with the \textit{Swift} filter transmission curve yields a value that is lower by only $\sim$30\% compared to WASP-107,  while the X-ray emission of HD~85512 is lower by a factor of $\sim$20. The rotation period of $\sim$47 days\,\citeSupp{France2016ApJ...820...89F} implies an age of $\sim$5.6\,Gyr for HD~85512, hence considerably older than WASP-107 with an estimated age of $\sim$3.4\,Gyr\,\citeSupp{piaulet_2021}. Therefore, it is not unexpected that HD~85512 is significantly less magnetic and/or has less coronal activity than WASP-107.
However, for our photochemical models mainly the NUV and FUV flux is of importance, since the X-ray emission primarily impacts photoionization which is not included in our models. We therefore can use the MUSCLES HD~85512 spectrum as representation of WASP-107's SED.

Each chemical model was executed with a vertical resolution comprising 130 layers spanning the pressure range from 10$^{-7}$ to 100\,bar. Subsequently, the hydrodynamical input and the abundances resulting from the chemical kinetics simulations are used to compute a synthetic transmission spectrum, using the radiative transfer package \texttt{petitRADTRANS}\,\citeSupp{molliere_wardenier_2019} (see above). Next to the line absorption opacities described in the \texttt{petitRADTRANS} retrieval setup, we also include the line absorption opacities listed in ref.~\citeSupp{Baeyens2021MNRAS.505.5603B}. Since the primary goal of the forward model computations is to understand the gas-phase formation of SO$_2$, CH$_4$ and H$_2$O in this  planet independent of cloud-formation, no condensate opacity was added in this last post-processing setup. For each pressure layer, the mean molecular weight is calculated based on the mixture that resulted from the disequilibrium chemistry models. The planetary radius at reference pressure (0.01\,bar) was set to 0.9\,R$_\mathrm{p}$. Finally, the predicted synthetic spectra are rebinned to the spectral resolution of the WASP-107b JWST MIRI data.

\subsection{(Photo)chemical model predictions}\label{Ref:photochemical_output}
 
Figure~\ref{fig:wasp107_model} provides evidence that only models incorporating photochemistry in combination with a super-solar metallicity predict a detectable level of SO$_2$ in WASP-107b. The large atmospheric scale height of WASP-107b enables highly efficient photochemical processes to operate within the $\sim$740\,K temperature regime of this low-density planet, resulting in SO$_2$ volume mixing ratios being $>$5$\times$10$^{-7}$ at pressures between 10$^{-7}$\,--\,10$^{-4}$\,bar.

We explored the sensitivity of SO$_2$ to both the metallicity and the C/O ratio itself. Extended Data Figure~\ref{fig:chemistry_Tint_COratios} shows that the SO$_2$ molar fraction in the upper atmosphere of WASP-107b displays a mild sensitivity to the explored C/O ratio, increasing by a few factors as the C/O decreased from solar (0.55) to sub-solar (0.10).  In contrast, the SO$_2$ molar fraction is highly sensitive to the metallicity (see Figure~\ref{fig:wasp107_model}) owing to the fact that both the sulphur and oxygen abundance scale with metallicity.  Our photochemical models show that SO$_2$ becomes detectable at super-solar metallicities, an effect already noted for higher temperature atmospheres\,\citeSupp{Polman2023A&A...670A.161P}.

Two critical parameters influencing the detectability of SO$_2$ within a planetary atmosphere are the UV irradiation and the gravity ($g$), which in turn determines the atmospheric scale height (see Extended Data Figure~\ref{fig:chemistry_UV_g}). Although the atmospheric scale height for both WASP-107b and WASP-39b is roughly equivalent (estimated at $\sim$1$\times$10$^6$\,m), their gravity differs, with WASP-107b at $\sim$260\,cm/s$^2$ and WASP-39b at $\sim$430\,cm/s$^2$. It is important to note that simulations of WASP-39b in previous studies were conducted at higher gravity values of 1,000\,cm/s$^2$ \citeSupp{tsai_photochemically_2023} and 2,140\,m/s$^2$ \citeSupp{Polman2023A&A...670A.161P}. Extended Data Figure~\ref{fig:chemistry_UV_g} juxtaposes the SO$_2$  predictions under $g$\,=\,260\,cm/s$^2$ and $g$\,=\,430\,cm/s$^2$ (similar to WASP-39b) and 1,000\,cm/s$^2$, where the gravity has been adapted by scaling the mass of the planet. While the increase of gravity from 260\,cm/s$^2$ to 430\,cm/s$^2$ only slightly alters the SO$_2$ abundance profile, a gravity of 1,000\,cm/s$^2$ significantly decreases the SO$_2$ abundance at pressures between $\sim$10$^{-5}$\,--\,1\, bar. This  is attributed to the reduced efficiency of photochemistry in deeper layers of atmospheres with high gravity. Consequently, this reduction diminishes the reservoir of OH radicals necessary for the synthesis of SO$_2$.

A last simulation employs a gravitational force of 260\,cm/s$^2$ as well, but uses the SED of HD 85512 - as a proxy for WASP-107 - scaled by a factor 100 (brown line in Extended Data Figure~\ref{fig:chemistry_UV_g}) or the WASP-39 spectrum from Ref.~\citeSupp{tsai_photochemically_2023} (purple line in Extended Data Figure~\ref{fig:chemistry_UV_g}) as the input stellar spectrum. The flux density originating from the host star, incident at the planet, is approximately 200 times greater for WASP-39b than for WASP-107b in the near-ultraviolet (NUV) range, and exhibits a factor of $\sim$100\,--\,1,000 in the far-ultraviolet (FUV), with comparable EUV and X-ray fluxes (see Suppl.\ Inf.\ Figure~\ref{fig:SED_crossections}). Photodissociation of SO$_2$ and H$_2$S mainly operates in the FUV, with absorption cross sections reaching around 10$^{-16}$\,cm$^2$. While the NUV absorption cross sections for SO$_2$ are about two orders of magnitude lower than in the FUV, it's worth noting that the H$_2$S cross sections are only available up to $\sim$250\,nm; see Suppl.\ Inf.\ Figure~\ref{fig:SED_crossections}.

\begin{figure}[htp]
    \centering
    \includegraphics[width=.8\textwidth]{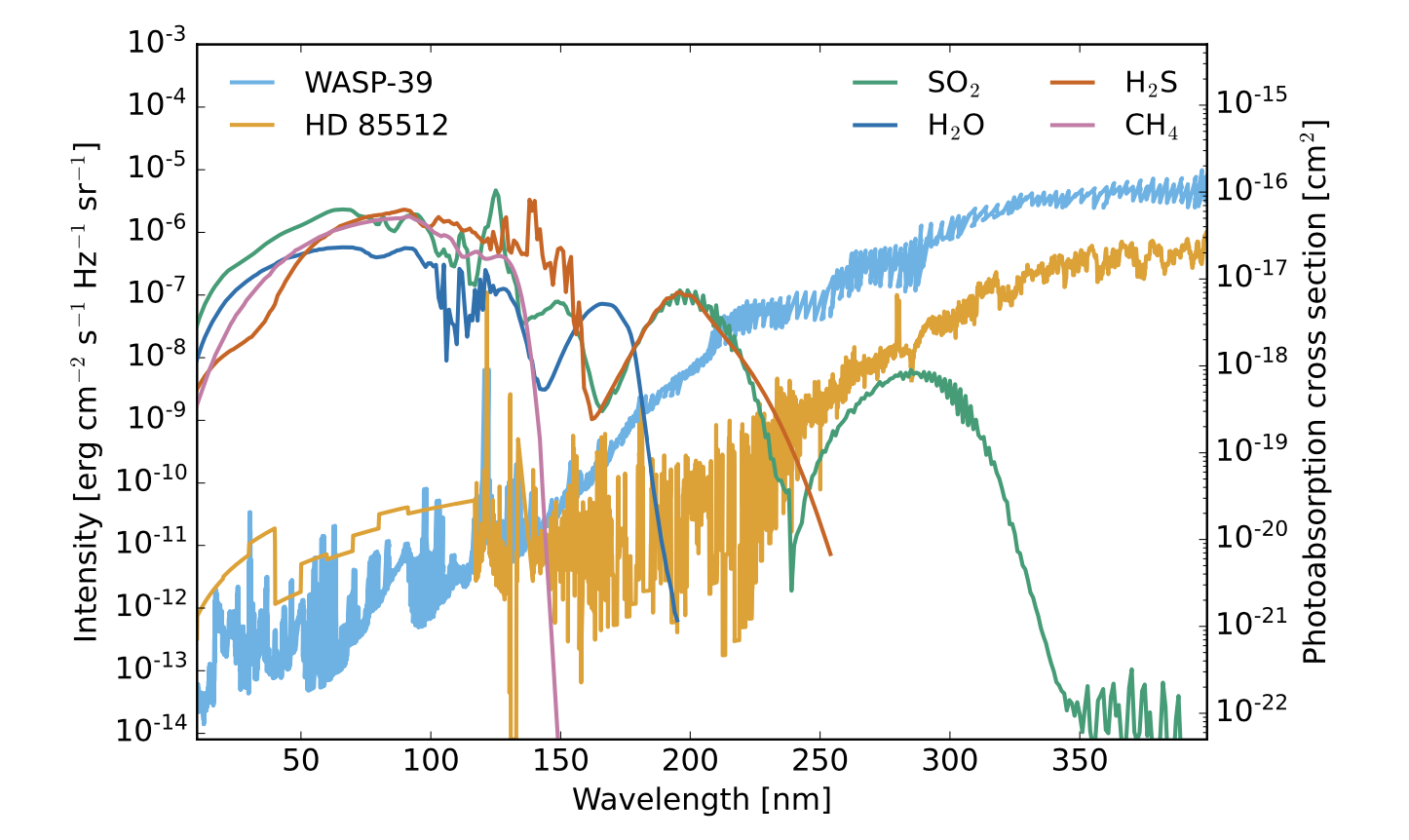}
    \caption{\textbf{Input stellar spectral energy distributions (SED) and photo-absorption cross sections.}
    The orange and light blue curve show the SED of HD\,85512 -- used as proxy for WASP-107 -- and of WASP-39, with corresponding intensity values given on the left y-axis. The photo-absorption cross sections of SO$_2$, H$_2$O, H$_2$S, and CH$_4$ are shown in green, pink, dark blue, and brown, respectively, with corresponding values given on the right y-axis.}
    \label{fig:SED_crossections}
\end{figure}

Increasing the UV irradiation with a factor 100 triggers the direct photodissociation of SO$_2$ at altitudes near 10$^{-4}$\,bar. At higher altitudes, this process is somewhat counteracted by H$_2$O photolysis,  generating additional OH radicals that react with S and SO to form SO$_2$ (see brown line in Extended Data Figure~\ref{fig:chemistry_UV_g}). However, when adopting the WASP-39 spectrum with its more extreme FUV/NUV ratio, an interesting observation emerges (purple line in Extended Data Figure~\ref{fig:chemistry_UV_g}): around 10$^{-3}$\,bar, direct FUV-driven photodissociation of SO$_2$ takes place, while at altitudes near 10$^{-4}$\,bar, the additional destruction of H$_2$S liberates sulphur radicals. These sulphur atoms are subsequently oxidized into SO$_2$, partially offsetting the SO$_2$ loss at deeper levels. At the uppermost atmospheric levels, approximately several times 10$^{-7}$\,bar, SO$_2$ undergoes photodissociation across all simulations. Hence, a low gravity together with modest UV irradiation and FUV/NUV ratio are the key ingredients for the formation of SO$_2$ in detectable amounts.

Extended Data Figure~\ref{fig:chemistry_Tint_COratios} shows that the  eddy diffusion and the intrinsic temperature have a  minor impact on the abundance of SO$_2$  at those pressure levels where the MIRI SO$_2$ features predominantly emerge, i.e.\ at pressures below a few times 10$^{-5}$\,bar (see Extended Data Figure~\ref{fig:contribution_functions_ARCiS}). Even when excluding vertical transport in the disequilibrium models ($K_{\rm{zz}}$\,=\,0\,cm$^2$ s$^{-1}$) a significant abundance of SO$_2$ is still predicted at pressures below a few times 10$^{-4}$\,bar (see panel~(b) in Suppl.\ Inf.\  Figure~\ref{fig:chem_only}), proving the crucial role of photolysis in establishing the chemical composition in WASP-107b's atmosphere. The increase in SO$_2$ formation around 10$^{-3}$ bar (for $K_{zz}$\,=\,0\,cm$^2$ s$^{-1}$, purple line) is caused by the breaking up of H$_2$S yielding sulphur radicals that are subsequently oxidised. 
When including eddy diffusion, these sulphur atoms are redistributed through the atmosphere, resulting in a SO$_2$ molar fraction depicted with the full black line.

\begin{figure}
    \centering
    \includegraphics[width=\textwidth]{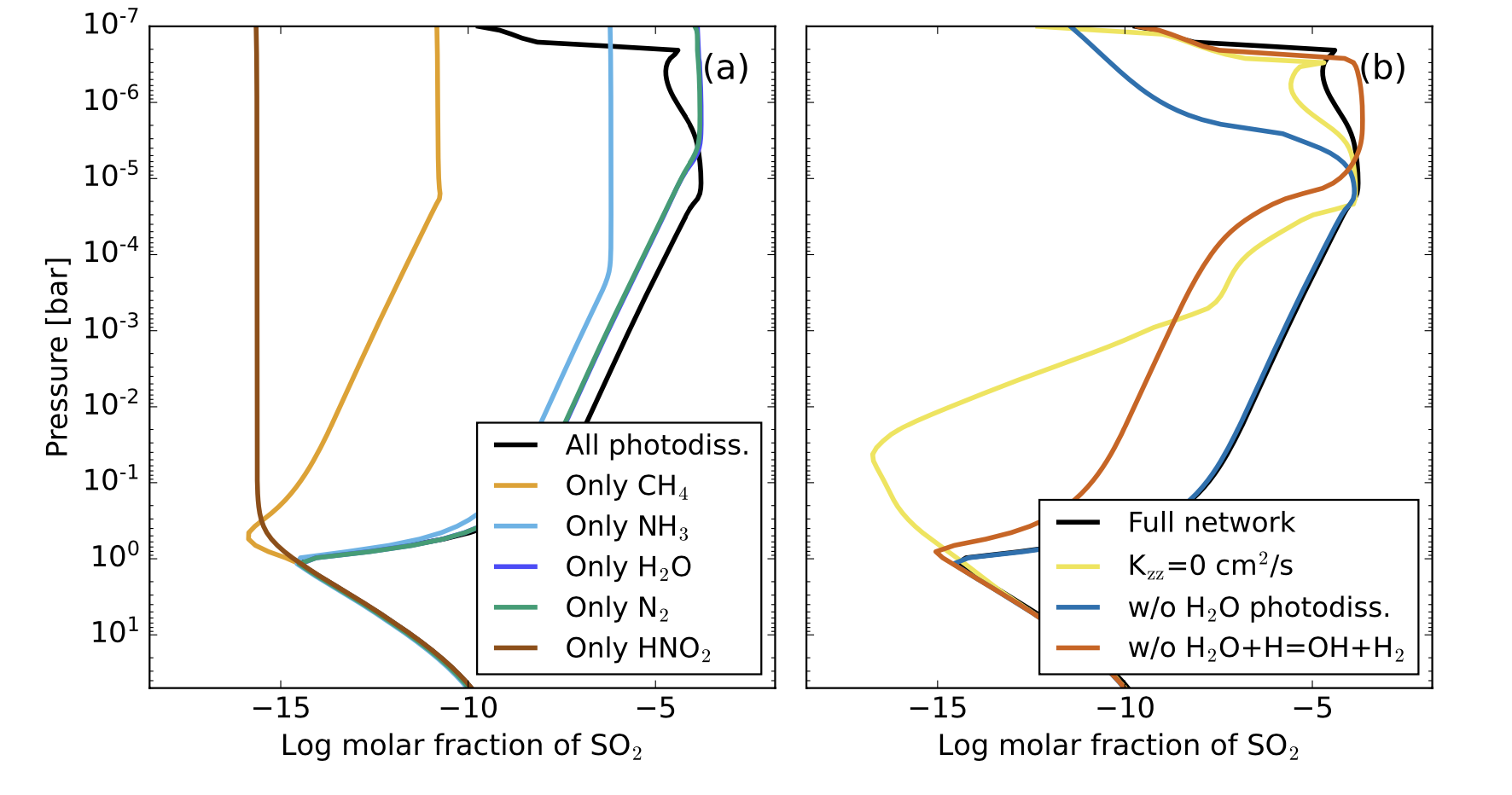}
    \caption{\textbf{SO$_2$ molar fraction predictions for WASP-107b for different set-ups of the photochemical network.}
    The base model (shown in black in each panel) has an intrinsic temperature of 400\,K, a solar C/O ratio, a metallicity of 10$\times$ solar, a $\log_{10}(K_{\rm{{zz}}}, \rm{cgs})$\,=\,10, and uses the SED of HD 85512 - used as a proxy for WASP-107 - as input stellar spectrum. 
    Panel~(a): Predicted SO$_2$ molar fractions when all  photo-absorption cross sections (black) are taken into account during the chemistry simulation, compared to predictions where only CH$_4$ (orange), NH$_3$ (light blue), H$_2$O (dark blue), N$_2$ (green) or HNO$_2$ (brown) are used. 
    Panel~(b): Predicted SO$_2$ molar fractions without vertical mixing, i.e.\ $K_{\rm{{zz}}}$\,=\,0\,cm$^2$/s (yellow), without photodissociation of H$_2$O (blue), and without including the thermochemical reaction H$_2$O+H$\rightleftharpoons$H$_2$+OH (orange).
    }
    \label{fig:chem_only}
\end{figure}

A chemical network analysis indicated that the primary trigger for the formation of SO$_2$ in the atmosphere of WASP-39b is water photolysis\,\citeSupp{tsai_photochemically_2023}. However, at first sight, it seems that water photolysis only plays a minor role for the production of SO$_2$ in WASP-107b. This conclusion is drawn from  panel~(b) in Suppl.\ Inf.\ Figure~\ref{fig:chem_only} where we exclude water photodissociation from our photochemical models (green line). It can be seen that  the influence on  the predicted SO$_2$ abundance is only confined to  pressures below 10$^{-5}$\,bar. The reason for this behaviour is that H$_2$O is predominantly photodissociated in the uppermost atmospheric layers. This is also shown in panel~(b) of Suppl.~Inf.~Figure~\ref{fig:OH} where we compare the [OH]/[H] ratio under equilibrium and disequilibrium conditions. While the omission of H$_2$O photodissociation explains the difference between both curves for pressures around a few times 10$^{-7}$\,bar, the vertical transport is the main reason for the difference between equilibrium and disequilibrium predictions for pressures between $\sim$10$^{-4}$\,--\,1\,bar. The thermochemical reaction of main importance for establishing the [OH]/[H] ratio in that pressure regime (see panel~(b) in Suppl.\ Inf.\  Figure~\ref{fig:chem_only}) is 
\begin{align}
\text{H$_2$O+H} &\underset{k_r}{\stackrel{k_f}{\rightleftharpoons}} \text{H$_2$ + OH.}
\label{reac:H2O}
\end{align} 
The reverse reaction rate, $k_r$, is given in the VULCAN network in its Arrhenius form 
\begin{equation}
    k_r = A_r\, T^B_r \,\exp(-C_r/T)\ \   \rm{[cm^3\ s^{-1}]}\,,
\end{equation} 
with $T$ the temperature (in Kelvin), and the corresponding parameters being the pre-exponential factor $A_r$\,= 3.57$\times$10$^{-16}$\,cm$^3$ s$^{-1}$, the temperature-dependent exponent $B_r$\,=\,1.52, and the activation energy $C_r$\,=\,1740\,K. Using the NTRS-NASA thermodynamic data\,\citeSupp{McBride2001tdff.rept.....M}
and assuming thermodynamic equilibrium, the Gibbs free energy of formation of the forward reaction, and the corresponding equilibrium constant can be calculated\,\citeSupp{Gail2013pccd.book.....G}. This allows the calculation of the forward reaction rate $k_f$. Fitting these results with the Arrhenius form yields $A_f$\,=\,1.54$\times$10$^{-14}$\,cm$^3$ s$^{-1}$, $B_f$\,=\,1.245, and a high energy barrier of $C_f$\,=\,9468\,K. It can be seen that at the temperatures relevant for planet atmospheres, the forward reaction rate is much lower than the reverse rate.

\begin{figure}[htp]
    \centering
    \includegraphics[width=\textwidth]{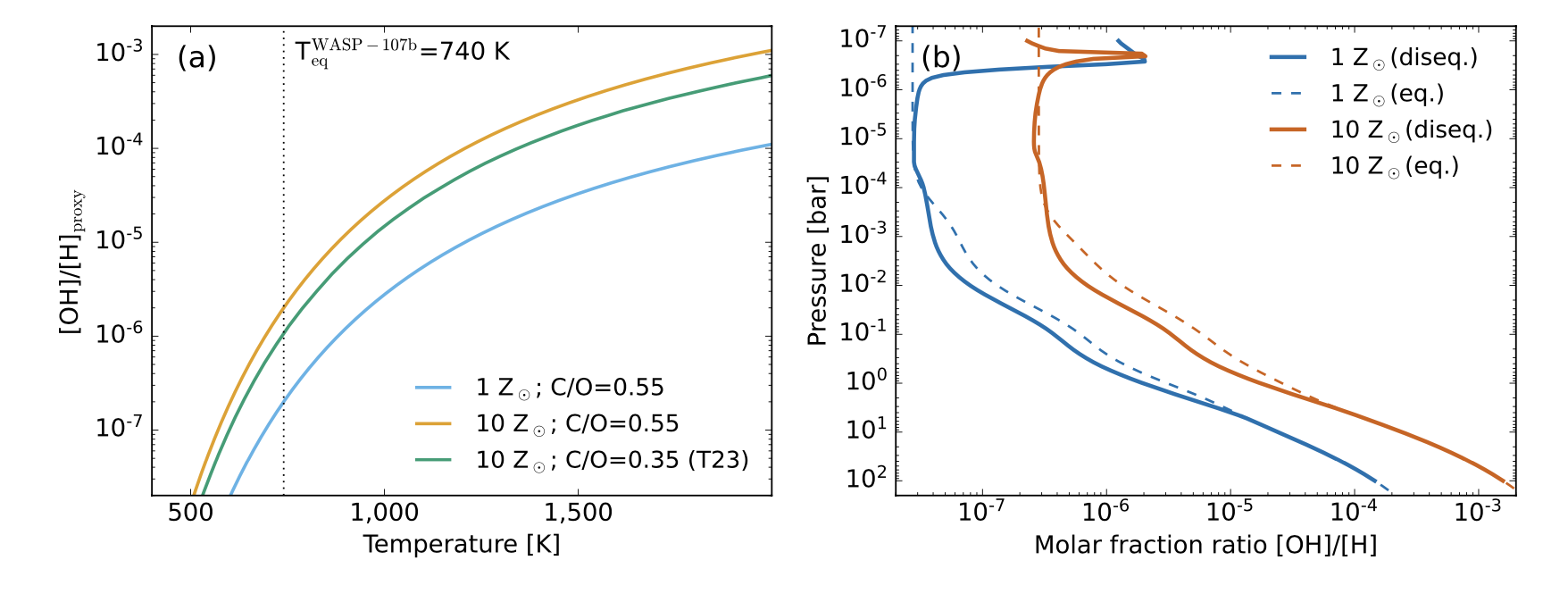}
    \caption{\textbf{[OH]/[H] ratio for equilibrium and disequilibrium predictions.} Panel~(a): Proxy for the [OH]/[H] ratio assuming thermodynamic equilibrium and that most of the O is in H$_2$O at solar metallicity (blue) and at $Z=10\,Z_\odot$ (orange). The prediction by Ref.~\protect\citeSupp{tsai_photochemically_2023} is shown as a green line for comparison. Panel~(b): The molar fraction of [OH]/[H] under two conditions: solar metallicity (blue) and 10$\times$ solar metallicity (orange), is depicted in both chemical equilibrium (dashed line) and disequilibrium (solid line) calculations.}
    \label{fig:OH}
\end{figure}

Similar to Ref.~\citeSupp{tsai_photochemically_2023}, we then can calculate the [OH]/[H] ratio assuming thermodynamic equilibrium and that most of the O is in H$_2$O. This yields panel~(a) in Suppl.\ Inf.\  Figure~\ref{fig:OH}, to be compared with Extended Data Figure~10 of Ref.~\citeSupp{tsai_photochemically_2023}. At temperatures below $\sim$750\,K, the [OH]/[H] ratio drops below $\sim$2$\times$10$^{-6}$ for $Z$\,=\,10\,$Z_\odot$, and hence a factor 10 lower at solar metallicity. This scarcity of OH has been used as an argument for the lack of  SO$_2$ formation at equilibrium temperatures below approximately $\sim$1,000\,K for a planet with WASP-39b parameters, favouring instead the prevalence of sulphur allotropes S$_x$\,\citeSupp{tsai_photochemically_2023}.

The central inquiry that emerges is how SO$_2$ can be created  within the atmosphere of WASP-107b if the aforementioned argument stands. The solution becomes evident through Suppl. Inf. Figure~\ref{fig:chem_only}, where it is demonstrated that the photodissociation of various specific abundant molecules sparks the generation of SO$_2$. This assertion is exemplified in panel~(a) of Suppl.\ Inf.\  Figure~\ref{fig:chem_only}, where the photodissociation of either only H$_2$O (or only N$_2$ or NH$_3$) leads to the emergence of SO$_2$. But the photodissociation of only CH$_4$ (only acting at wavelengths $\la$140\,nm; see Suppl.\ Inf.\ Figure~\ref{fig:SED_crossections})
yields negligible amounts of SO$_2$, while the photodissociation of the rare molecule HNO$_2$ yields outcomes consistent with chemical equilibrium predictions, wherein all photodissociation is thus excluded (see dotted line in panel~(b) of Figure~\ref{fig:wasp107_model}). 
This phenomenon arises from the fact that the photodissociation of various specific abundant molecules releases atoms and radicals that induce a very active photochemistry even down to pressure layers of approximately 1\,bar. Reactions involving the liberated atoms and radicals often display temperature-independent behaviour without energy barriers (i.e., $B$\,=\,$C$\,=\,0) and possess pre-exponential factors typically on the order of 10$^{-11}$\,--\,10$^{-7}$\,cm$^3$\,s$^{-1}$. Consequently, a significant amount of H atoms and OH radicals is formed, leading to the oxidisation of sulphur into SO$_2$. Hence, although the photolysis of H$_2$O can initiate the production of SO$_2$ in WASP-107b, it is not the sole molecule whose photodissociation holds the potential to induce SO$_2$ formation.

In summary, the overarching scenario that unfolds reveals that the primary pathways initiating the formation of SO$_2$ in the low-density atmosphere of WASP-107b are twofold. Firstly, through the photodissociation of H$_2$O in the upper atmospheric layers at pressures below a few times 10$^{-6}$\,bar, yielding atomic H and OH radicals. These OH radicals are key for oxidising sulphur that is liberated from H$_2$S. Secondly, in the pressure range of 10$^{-5}$\,--\,1\,bar, the prevailing determinant of the chemical composition is the interplay of photochemical processes acting upon various abundant molecules, not limited to H$_2$O. These processes can generate a sufficiently substantial quantity of free atoms and radicals, that can be redistributed through eddy diffusion. This initiates  a cascade of barrierless thermochemical reactions that progressively culminate in the formation of SO$_2$. Given the fact that a large ensemble of those reactions are temperature-independent, the equilibrium temperature stands as just one among several factors dictating the formation (or not) of SO$_2$. As long as the UV irradiation and FUV/NUV ratio remain moderate and the gravity is low, these processes will lead to the formation of SO$_2$ in sufficient amounts to be detectable even within a $\sim$740\,K temperature planet.

\afterpage{\clearpage}

\longrefs=1
\bibliographystyleSupp{naturemag}
\bibliographySupp{sn-bibliography}


\end{document}